\documentclass[twocolumn,times]{aastex62}
\usepackage{graphicx}
\usepackage[flushleft]{threeparttable}
\usepackage{blindtext}
\usepackage{amsmath}
\usepackage{mathtools}
\usepackage{multirow}
\usepackage{comment}
\usepackage{booktabs}
\turnoffeditone

\newcommand{\beq}{\begin{equation}}
\newcommand{\eeq}{\end{equation}}
\newcommand{\bea}{\begin{eqnarray}}
\newcommand{\eea}{\end{eqnarray}}
\DeclareUnicodeCharacter{0301}{\'{e}}

\begin{document}
\shortauthors{Park et al.}
\def\nar{New Astron.}
\def\na{New Astron.}
\title{\large \textbf{Crucial Factors for L\MakeLowercase{y}$\alpha$ Transmission in the Reionizing Intergalactic Medium: Infall Motion, HII Bubble Size, and Self-shielded Systems}}

\correspondingauthor{Hyunbae Park}
\email{hyunbae.park@lbl.gov}

\correspondingauthor{Intae Jung}
\email{intae.jung.2013@gmail.com}

\author[0000-0002-7464-7857]{Hyunbae Park}
\affil{Lawrence Berkeley National Laboratory, CA 94720, USA}
\affil{Berkeley Center for Cosmological Physics, UC Berkeley, CA 94720, USA}
\affil{Kavli IPMU (WPI), UTIAS, The University of Tokyo, Kashiwa, Chiba 277-8583, Japan}

\author[0000-0003-1187-4240]{Intae Jung}
\affil{Astrophysics Science Division, Goddard Space Flight Center, Greenbelt, MD 20771, USA}
\affil{Department of Physics, The Catholic University of America, Washington, DC 20064, USA}
\affil{Center for Research and Exploration in Space Science and Technology, NASA/GSFC, Greenbelt, MD 20771}

\author{Hyunmi Song}
\affil{Department of Astronomy and Space Science, Chungnam National University, Daejeon 34134, Republic of Korea}

\author{Pierre Ocvirk}
\affil{Universit́e de Strasbourg, CNRS, Observatoire astronomique de Strasbourg, UMR 7550, F-67000 Strasbourg, France}

\author{Paul R. Shapiro}
\affil{Department of Astronomy, University Texas, Austin, TX 78712-1083, USA}

\author{Taha Dawoodbhoy}
\affil{Department of Astronomy, University Texas, Austin, TX 78712-1083, USA}

\author{Ilian T. Iliev}
\affil{Astronomy Centre, Department of Physics \& Astronomy, Pevensey III Building, University of Sussex, Falmer, Brighton, BN1 9QH, UK}

\author{Kyungjin Ahn}
\affil{Department of Earth Sciences, Chosun University, Gwangju 61452, Republic of Korea}

\author{Michele Bianco}
\affil{Astronomy Centre, Department of Physics \& Astronomy, Pevensey III Building, University of Sussex, Falmer, Brighton, BN1 9QH, UK}

\author{Hyo Jeong Kim}
\affil{Department of Earth Sciences, Chosun University, Gwangju 61452, Republic of Korea}



\begin{abstract} 
 Using the CoDa II simulation, we study the Ly$\alpha$ transmissivity of the intergalactic medium (IGM) during reionization. At $z>6$, a typical galaxy without an active galactic nucleus fails to form a proximity zone around itself due to the overdensity of the surrounding IGM. The gravitational infall motion in the IGM makes the resonance absorption extend to the red side of Ly$\alpha$, suppressing the transmission up to roughly the circular velocity of the galaxy. In some sight lines, an optically thin blob generated by a supernova in a neighboring galaxy results in a peak feature, which can be mistaken for a blue peak. Redward of the resonance absorption, the damping-wing opacity correlates with the global IGM neutral fraction and the UV magnitude of the source galaxy. Brighter galaxies tend to suffer lower opacity because they tend to reside in larger HII regions, and the surrounding IGM transmits redder photons, which are less susceptible to attenuation, owing to stronger infall velocity. The HII regions are highly nonspherical, causing both sight-line-to-sight-line and galaxy-to-galaxy variation in opacity. Also, self-shielded systems within HII regions strongly attenuate the emission for certain sight lines. All these factors add to the transmissivity variation, requiring a large sample size to constrain the average transmission. The variation is largest for fainter galaxies at higher redshift. The 68\% range of the transmissivity is similar to or greater than the median for galaxies with $M_{\rm UV}\ge-21$ at $z\ge7$, implying that more than a hundred galaxies would be needed to measure the transmission to 10\% accuracy.
\end{abstract}

\section{Introduction}

How was the intergalactic medium (IGM) reionized at $z \gtrsim 6$? This question is connected to the unknown properties of early galaxies and is, therefore, one of the key questions of modern astronomy. 
Cosmic microwave background radiation observed by the Planck satellite constrains the midpoint of reionization to be at $z\sim 8$ \citep{2020A&A...641A...6P}\footnote{We note that this statement is somewhat model dependent as the electron-scattering optical depth measured by Planck only constrains the integrated column density of free electrons. For a more detailed analysis on this matter, we refer the readers to, e.g., \cite{2021PhRvD.104f3505H}.}. Another powerful probe of cosmic history is the Ly$\alpha$ emission from high-$z$ star-forming galaxies. Those galaxies are expected to generate strong emission near Ly$\alpha$, which the neutral gas in the IGM can easily scatter. Thus, the statistics of the Ly$\alpha$ line strength is closely related to the global neutral fraction of the IGM \citep[e.g.,][]{2004ApJ...617L...5M,2014PASA...31...40D}. 

Today, numerous Ly$\alpha$ emitters (LAEs) have been observed up to $z\sim 9$ by narrowband imaging \citep[e.g.,][]{2006ApJ...648....7K,2008ApJS..176..301O,2010ApJ...723..869O,2010ApJ...725..394H,2011ApJ...734..119K,2014ApJ...797...16K,2015MNRAS.451..400M,2016MNRAS.463.1678S,2017ApJ...842L..22Z,2017ApJ...844...85O,2018PASJ...70...55I,2018PASJ...70S..13O} and spectroscopic surveys \citep[e.g.,][]{2008ApJ...680.1072D,2011ApJS..192....5A,2013Natur.502..524F,2014ApJ...795...20S,2015A&A...573A..24C,2015ApJ...804L..30O,2015ApJ...810L..12Z,2016ApJ...826..113S,2017A&A...606A..12H,2017MNRAS.464..469S,2017NatAs...1E..91H,2017MNRAS.472..772M,2018ApJ...864..103J,2018PASJ...70S..15S,2020ApJ...904..144J,2020A&A...634A..97K,2021ApJ...919..120M,2021ApJ...914...79T}. The observed samples indeed show a steep decline in the number of LAEs found per galaxy above $z=6$, contrary to the much more gradual evolution at lower redshifts. This decline appears more prominent for relatively fainter galaxies ($M_{\rm UV}\gtrsim -20$) \citep[e.g.,][Jung, I. et al. 2021 in preparation]{2010ApJ...725L.205F,2011ApJ...728L...2S,2011ApJ...743..132P,2012ApJ...744...83O,2012ApJ...760..128M, 2012MNRAS.422.1425C,2013ApJ...775L..29T,2014ApJ...794....5T,2017ApJ...842L..22Z,2021MNRAS.502.6044E}. On the other hand, the number of bright LAEs tend to change less with redshift and are often found clustered in overdense regions \citep{2018ApJ...863L...3C,2019ApJ...877..146J,2019ApJ...883..142H,2019ApJ...879...28H,2020ApJ...904..144J,2020ApJ...891L..10T,2021NatAs...5..485H}. In most recent surveys, several LAEs have been detected at $z\sim 7$, constraining the neutral fraction at the highest redshifts ever \citep{2019ApJ...878...12H,2019MNRAS.485.3947M,2020ApJ...904..144J,2021ApJ...919..120M} although the results still have a large uncertainty and have not converged yet.

In order to constrain the reionization history from observed Ly$\alpha$ emission statistics precisely, one crucially needs to understand the IGM opacity and how it evolves during reionization. Transmission of Ly$\alpha$ photons in the IGM depends on the neutral hydrogen density, velocity, and temperature of the intergalactic gas. Ultraviolet (UV) radiation from star-forming galaxies forms expanding HII regions around them, where some Ly$\alpha$ emission can be transmitted owing to low HI density. Thus, capturing the statistics of these HII regions, which can be up to tens of comoving megaparsecs (cMpc) large, is a key to the modeling the IGM transmissivity during the reionization era \citep[e.g.,][]{2005MNRAS.363.1031F}. Additionally, small-scale velocity and density structures below the megaparsec scale around the source galaxy are also important factors for the IGM transmissivity \citep{2013MNRAS.429.1695B,2015MNRAS.446..566M,2016MNRAS.463.4019K}. 

The IGM opacity of Ly$\alpha$ photons can be classified into two components: resonance and damping-wing opacity. At the line resonance, the cross section is so high that even a tiny trace of neutral fraction in highly ionized gas (well below $1 \%$) can make the IGM opaque to Ly$\alpha$ photons. Thus, photons on the blue side of Ly$\alpha$ typically face high opacity at some point in the rest frame of the IGM and are scattered off the line of sight. On the other hand, photons on the red side of the resonance are much more likely to be transmitted. The IGM opacity for these photons mostly comes from the damping-wing cross section, which extends far away from the resonance. This damping-wing cross section is much smaller than the resonance cross section, but the HI density in neutral regions can be high enough to create a substantial optical depth. Therefore, this damping-wing opacity is expected to depend on the global neutral fraction of the IGM during reionization. 

Our goal in this study is to clarify what aspects of the IGM density/velocity field modulate the resonance and damping-wing absorption and quantify how they depend on the progress of reionization. To this end, we shall analyze the Cosmic Dawn II (CoDa II) simulation of \citet{2020MNRAS.496.4087O}, which covers a suitable dynamic range to compute the IGM opacity during reionization and the relevant physics. This work parallels \cite{2021MNRAS.508.3697G} in that both works are based on similar IGM transmission calculations from the same dataset. Whereas \cite{2021MNRAS.508.3697G} mainly focused on the transmission of the resonance to understand the recently observed blue peaks from some LAEs at $z\gtrsim 6.5$ \citep{2018ApJ...859...91S, 2018A&A...619A.136M,2021MNRAS.500..558M}, we focus on the red side that is relevant to the evolution of the LAE number density during reionization.

The rest of the paper is as follows. In Section~\ref{sec:method}, we describe the CoDa II simulation and how we calculate the IGM transmission. In Section~\ref{sec:results}, we present our main results. In Section~\ref{sec:discussion}, we discuss our results in light of the current understanding of LAEs. For the numerical calculations, we adopt the cosmology parameters used by the CoDa II simulation for self-consistency. The parameters are $\Omega_{m,0}=0.307$, $\Omega_{b,0}=0.048$, and $h=0.678$ \citep{2014A&A...571A..16P}. 

\section{Method} \label{sec:method}

\begin{figure*}
\begin{center}
\includegraphics[scale=0.55]{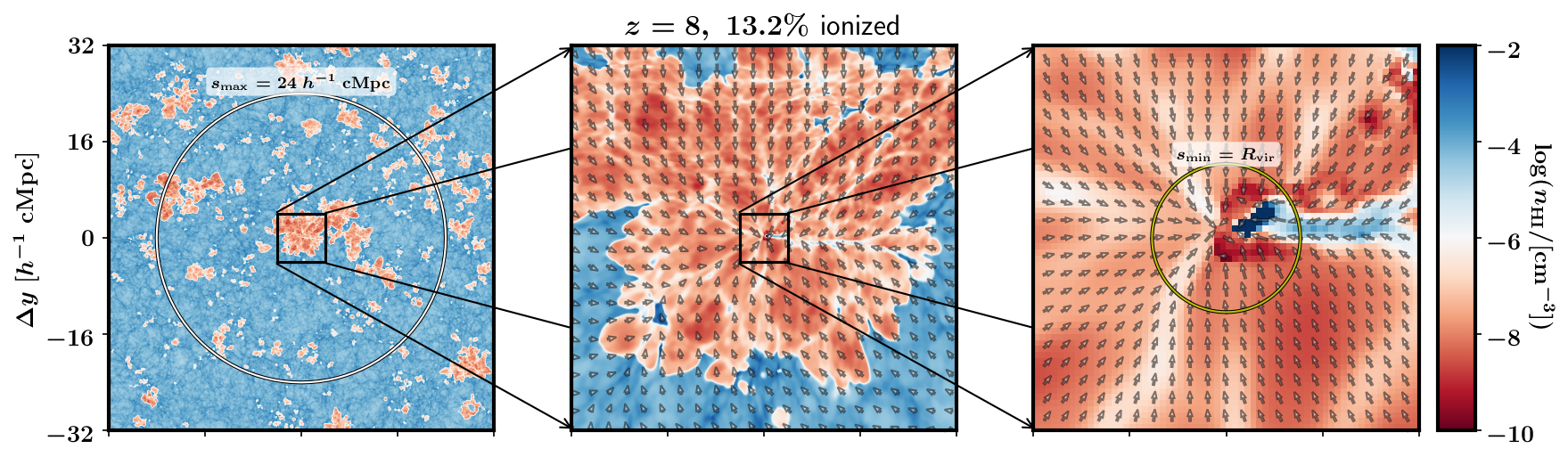}
\includegraphics[scale=0.55]{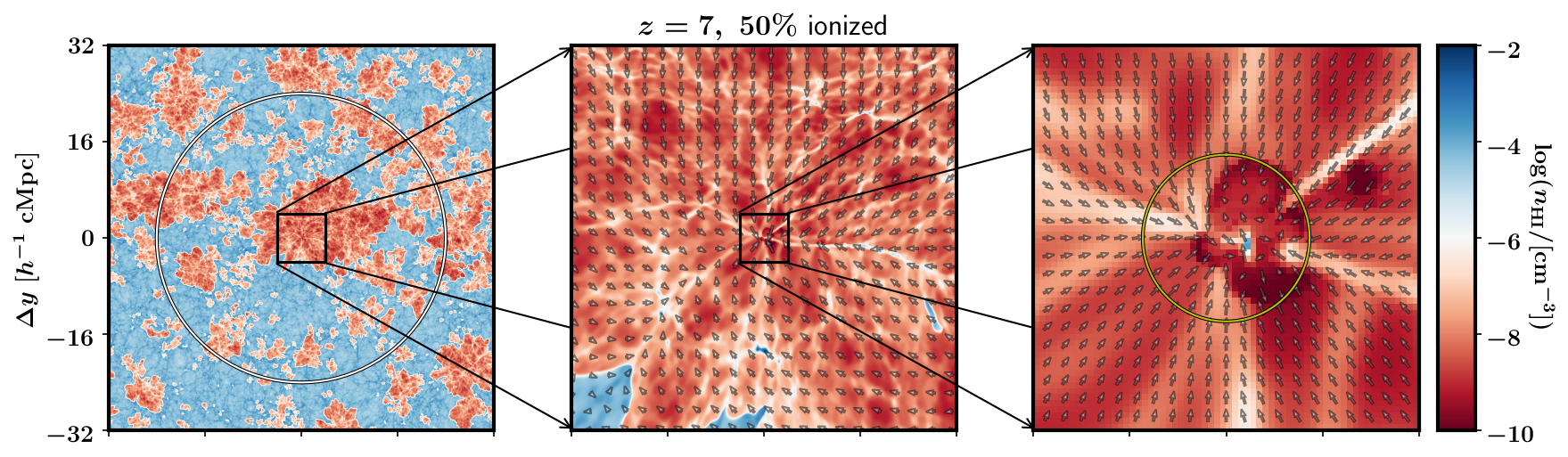}
\includegraphics[scale=0.55]{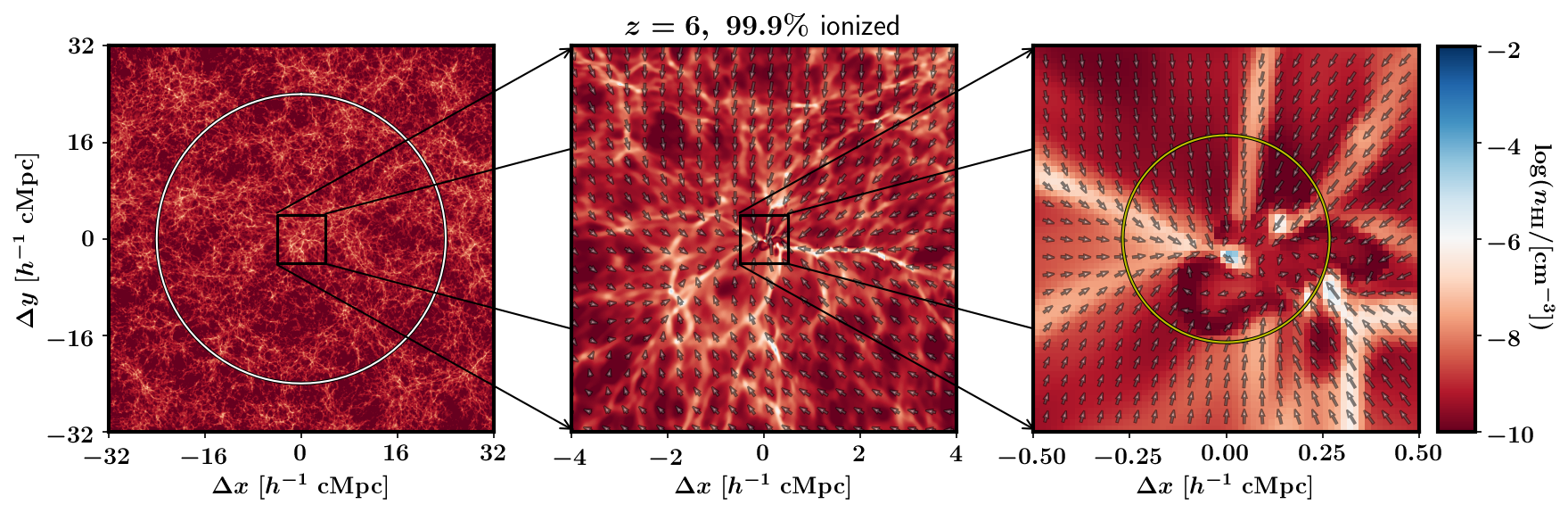}
\caption{$xy$ plane of the neutral hydrogen density map centered on the most massive galaxy in the CoDa II simulation at $z = 8$ (top), 7 (middle), and 6 (lower panels). We zoom in from the entire simulation box of $(64~h^{-1}~{\rm cMpc})^2$ to the $(8~h^{-1}~{\rm cMpc})^2$ and $(1~h^{-1}~{\rm cMpc})^2$ subboxes from left to right. The map is color-coded according to the neutral hydrogen density, but the blue/red color coincides with neutral/ionized regions in most cases. In the middle and right column panels, the partially transparent arrows describe the peculiar motion of gas with respect to the galaxy. The white circles in the left panels describe the maximum integration distance in Equation~(\ref{eq:tau}), $s_{\rm max}=24~h^{-1}~{\rm Mpc}$,  for the local optical depth $\tau_{\rm L}$ described in Section~\ref{sec:CalcT}. The yellow circles in the right panels show the virial radius of the galaxies at the center, which is the minimum integration distance, $s_{\rm min}=R_{\rm vir}$.}
\label{fig:nHmap} 
\end{center}
\end{figure*}

\subsection{Cosmic Dawn II simulation}

We compute the IGM transmission of Ly$\alpha$ photons during the Epoch of Reionization based on the simulated galaxies and IGM of the CoDa II simulation. CoDa II is a fully coupled radiation-hydrodynamic simulation of the reionization era in a cubic volume of $[64~h^{-1}~{\rm cMpc}]^3$ on a $4096^3$ regular grid, which resolves atomically cooling halos down to $10^8~M_\odot$ with $~200$ dark matter particles. The simulation was initialized at $z=150$ and run down to $z=5.8$ using the RAMSES-CUDATON code \citep{2016MNRAS.463.1462O}. The global ionization fraction of the IGM reaches $99\%$ at $z\approx 6.2$ in the simulation. The simulated galaxies reproduced the observed UV luminosity function at $z\gtrsim 6$. The details of this simulation are elaborated on in \citet{2020MNRAS.496.4087O}. Here, we briefly list the features that are relevant to our work. 

In CoDa II, photoionization is computed self-consistently by casting rays from stellar particles, which form from gas cells that satisfy certain star-formation conditions. The CoDa II simulation captures a statistically meaningful number of HII regions with its large volume and resolves small-scale structures around galaxies needed to compute the transmission accurately. Figure~\ref{fig:nHmap} illustrates this with the HI density in the IGM around the UV-brightest galaxies at $z=6$, 7, and 8 from the full-box scale in the left column down to the virial radius scale in the right column. We also note that CoDa II uses the exact speed of light in its radiative transfer calculation and, thus, is free from possible problems arising from using the reduced speed-of-light approximation \citep{2019A&A...622A.142D, 2019A&A...626A..77O}. 


The UV luminosity of each galaxy is computed from the age of the stellar particles based on stellar synthesis models. We name galaxies according to their ranking in UV luminosity at $1600$ \AA (i.e., $M_{\rm UV}\equiv M_{{\rm AB}1600}$) within the snapshot to which it belongs. For example, we refer to the 100th UV-brightest galaxies as galaxy \#0100. The brightest sample galaxy (galaxy \#0001) of the $z=7$ snapshot is $M_{\rm UV}=-23.1$ in the UV magnitude, for example, and there are hundreds of galaxies up to $M_{\rm UV}\sim -19$, which makes CoDa II suitable for understanding LAEs observed at $z\gtrsim 7$. Note that galaxy \#$n$ at $z=7$ and galaxy \#$n$ at $z=8$ are generally not the same because the ranking in UV magnitude changes over time.

\subsection{IGM Transmission Calculation} \label{sec:CalcT}

\begin{figure}
\begin{center}
\includegraphics[scale=0.48]{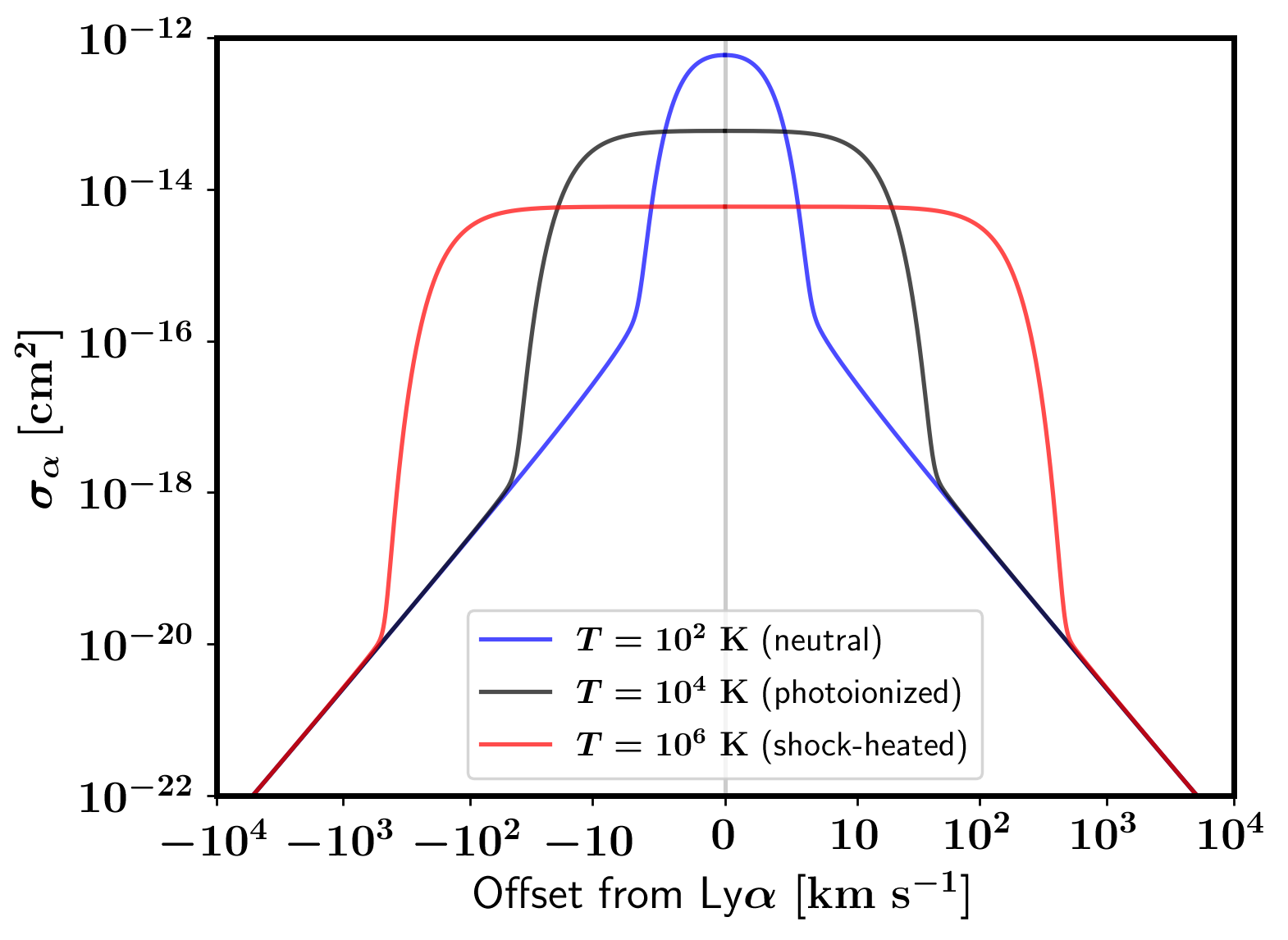}
\caption{ Ly$\alpha$ scattering cross section as a function of wavelength offset in velocity units ($v_\alpha$). The blue, black, and red lines describe the cross-section for three different gas temperatures, $T=10^2,~10^4,~$ and $10^6$ K, respectively.}
\label{fig:sig_a} 
\end{center}
\end{figure}

We calculate the IGM transmissivity for photons emitted around the Ly$\alpha$ frequency from galaxies in the $z=6$, 7, and 8 snapshots. In the simulation, the IGM is almost fully ionized at $z=6$, while it is $50\%$ and $13.2\%$ ionized at $z=7$ and $8$, respectively. Throughout this work, we assume neutral hydrogen is the only opacity source given that the dust and metals from star-formation are unlikely to exist in meaningful amount in the IGM during reionization. 

The transmissivity, $\mathcal{T}\equiv e^{-\tau}$, for a frequency $\nu_{\rm e}$ at the rest frame of a source is given by the optical depth
\bea \label{eq:tau}
\tau(\nu_{\rm e})=\int^{s_{\rm max}}_{s_{\rm min}} n_{\rm HI}(\mathbf{r}) \sigma_\alpha( \nu_a(\mathbf{r}),T(\mathbf{r}))a(z)ds,
\eea 
where $s$ is the comoving distance between the source and the IGM, $a(z)$ is the scale factor at redshift $z$, $\mathbf{r}=\mathbf{r}_s+s\hat{\gamma}$ is the comoving coordinate of the IGM along the line-of-sight direction $\hat{\gamma}$ from the source location $\mathbf{r}_s$, $n_{\rm HI}$ is the neutral hydrogen number density, $\sigma_\alpha(\nu,T)$ is the ensemble-averaged Ly$\alpha$ cross section of gas with temperature $T$, and
\bea
\nu_a(\mathbf{r})=\nu_{\rm e}
\left[1-\frac{v_{\rm pe}(\mathbf{r})+sH(z)a(z)}{c}\right]  \nonumber
\eea
describes the frequency shift in the rest-frame of the IGM due to the IGM peculiar motion $v_{\rm pe}$ and the cosmic expansion rate $H(z)$. 
Here, $v_{\rm pe}$ is calculated with respect to the peculiar motion of the galaxy, which we obtain by taking the density-weighted average of the velocity field with the virial radius of the host halo. We define the virial radius as the distance within which the mean density becomes 200 times the cosmic mean ($R_{\rm vir} \equiv r_{200}$). 

The Ly$\alpha$ cross-section is given by
\bea \label{eq:sigma}
\sigma_\alpha(\nu,T) = 5.889\times10^{-14}~T_4^{-0.5} \phi(x) ~{\rm cm}^2,
\eea
where $T_4\equiv T/[10^4~{\rm K}]$ and $\phi(x)$ is the Voigt function as a function of the dimensionless frequency $x\equiv[\nu - \nu_\alpha]/\Delta\nu_D$. Here, $\nu_\alpha\approx2.46\times 10^{15}~{\rm Hz}$ is the resonance frequency of Ly$\alpha$, and 
\bea
\Delta \nu_D&=&\nu_\alpha\sqrt{\frac{2k_BT}{m_pc^2}}= 4.28\times 10^{-5}~T_4^{0.5} \nu_\alpha \nonumber
\eea
is the thermal broadening frequency (Doppler width), $k_B$ is the Boltzmann constant, and $m_p$ is the proton mass. The Voigt function is given by
\bea
\phi(x)=
\frac{a_V}{\pi}\int^{\infty}_{-\infty} dy \frac{e^{-y^2}}{[y-x]^2+a_V^2},
\eea
where $a_V\approx4.7\times10^{-4}~T_4^{-0.5}$ is the Voight parameter. For fast computation, it is convenient to use the analytic fitting formula provided by \citet{2006ApJ...645..792T}.

The shape of $\sigma_\alpha(\nu,T)$ is shown in Figure~\ref{fig:sig_a} for $T=10^2,~10^4,$ and $10^6~{\rm K}$. The profile of $\sigma_\alpha(\nu)$ has a core at the center with the damping wing extending outward. The FWHM is roughly 2, 20, and 200 ${\rm km}~{\rm s}^{-1}$ for $T=10^2,~10^4,$ and $10^6~{\rm K}$, respectively. The gas temperature is approximately $T=10^4~{\rm K}$ for the ionized IGM and $T\lesssim 10^2~{\rm K}$ for the neutral IGM. Some gas is shock-heated to $\sim 10^6~{\rm K}$ by supernova feedback, but such gas rarely exists outside the virial radius, where we shall calculate the IGM transmissivity. 

For each galaxy, we calculate $\tau(\nu_{\rm e})$ for 2000 random lines of sight to account for the sight-line variation. For each sight line, we evaluate Equation~(\ref{eq:tau}) in two segments to obtain $\tau=\tau_{\rm L}+\tau_{\rm DW}$. Here, $\tau_{\rm L}$ is the local contribution within a distance of $24~h^{-1}~{\rm cMpc}$, which is far enough from the source galaxy to include the surrounding HII bubble. $\tau_{\rm DW}$ is the damping-wing absorption by distant IGM beyond that distance. 

For the local contribution $\tau_{\rm L}$, we capture the local IGM dynamics and the inhomogeneity of reionization by sampling $n_{\rm HI}(\mathbf{r})$, $T(\mathbf{r})$, and $v_{\rm pe}(\mathbf{r})$ along each sight line by interpolating the mesh data of the simulation. In this work, we regard the IGM as the gas outside the virial radius ($R_{\rm vir}$) and integrate to $24~h^{-1}~{\rm cMpc}$ from the galaxy by setting $s_{\rm min}=R_{\rm vir}$ and $s_{\rm max}=24~h^{-1}~{\rm cMpc}$ in Equation~(\ref{eq:tau}). We describe this integration range by white circles in Figure~\ref{fig:nHmap}. Within this range, we do not consider the time evolution of IGM and calculate the opacity from the still snapshot of the target redshift. The circles in the left panels of Figure~\ref{fig:nHmap} describe the integration range for $\tau_{\rm L}$,  which is large enough to capture the surroundings of the ionized bubble at $z=7$ and $8$ created by the central galaxy.

For the large-scale opacity contribution from the neutral gas in the IGM, $\tau_{\rm DW}$, we integrate in Equation~(\ref{eq:tau}) from $s_{\rm  min}=24~h^{-1}~{\rm cMpc}$ to $z=6$, where the reionization has ended in the simulation. For this large-scale contribution, we only consider the damping-wing opacity from neutral IGM and ignore the peculiar motion. For the neutral hydrogen density, we use the globally averaged value given by
\bea
&&\bar{n}_{\rm HI}=\bar{x}_{\rm HI}(z)[1+z]^3 \Omega_b \frac{3H_0^2}{8\pi G}\frac{X}{m_p}\nonumber\\
&=&9.62\times 10^{-5}\bar{x}_{\rm HI}(z)\left[\frac{\Omega_b}{0.048}\right]\left[\frac{h}{0.678}\right]^2\left[\frac{z+1}{8}\right]^3~{\rm cm}^{-3}, 
\eea
where $\bar{x}_{\rm HI}$ is the global average neutral fraction and $X=0.76$ is the mass fraction of hydrogen in baryons. In the damping wing, the Voigt profile can be approximated as $\phi(x)=a_V\pi^{-0.5}x^{-2}$ giving
\bea \label{eq:tau_D}
&&\tau_{\rm DW}(v_\alpha)= 5.7
\left[\frac{h}{0.678}\right] 
\left[\frac{\Omega_b}{0.048}\right]
\left[\frac{\Omega_m}{0.307}\right]^{-1}
\nonumber\\
&&
\int_{s_{\rm min}}^{s_{\rm max}} \left[\frac{ds}{h^{-1}~{\rm cMpc}}\right]  ~\left[\frac{z+1}{8}\right] ~\bar{x}_{\rm HI}(z) \left[\frac{s+\Delta s(v_\alpha)}{h^{-1}~{\rm cMpc}}\right]^{-2},
\eea
where 
\bea
&& \Delta s(v_\alpha) = 0.64~[h^{-1}~{\rm cMpc}]  \nonumber\\
&& \times \left[\frac{v_\alpha}{100~{\rm km}~{\rm s}^{-1}}\right] \left[\frac{\Omega_m}{0.307} \right]^{-0.5} \left[\frac{1+z}{8} \right]^{-0.5}
\eea
accounts for the initial frequency offset of emitted photons from the Ly$\alpha$ in the velocity unit: $v_\alpha\equiv-c\Delta\nu/\nu_\alpha$.  Here, we have assumed $H\approx H_0\sqrt{\Omega_m (1+z)^3}$ given that $z\gg 1$ during reionization. We calculate $\bar{x}_{\rm HI}$ by interpolating in redshift the globally averaged values within each snapshot. In the case of $\tau_{\rm DW}$, the integration starts from $s_{\rm min}=24~h^{-1}~{\rm cMpc}$ and ends at $s_{\rm max} = 263~ (482)~h^{-1}~{\rm cMpc}$ for galaxies at $z=7~(8)$. We find that roughly 10\% (20\%) of the photons are scattered due to this large-scale damping-wing opacity at $z=7~(8)$.

\section{Results} \label{sec:results}

\begin{figure*}
\begin{center}
\includegraphics[scale=0.4]{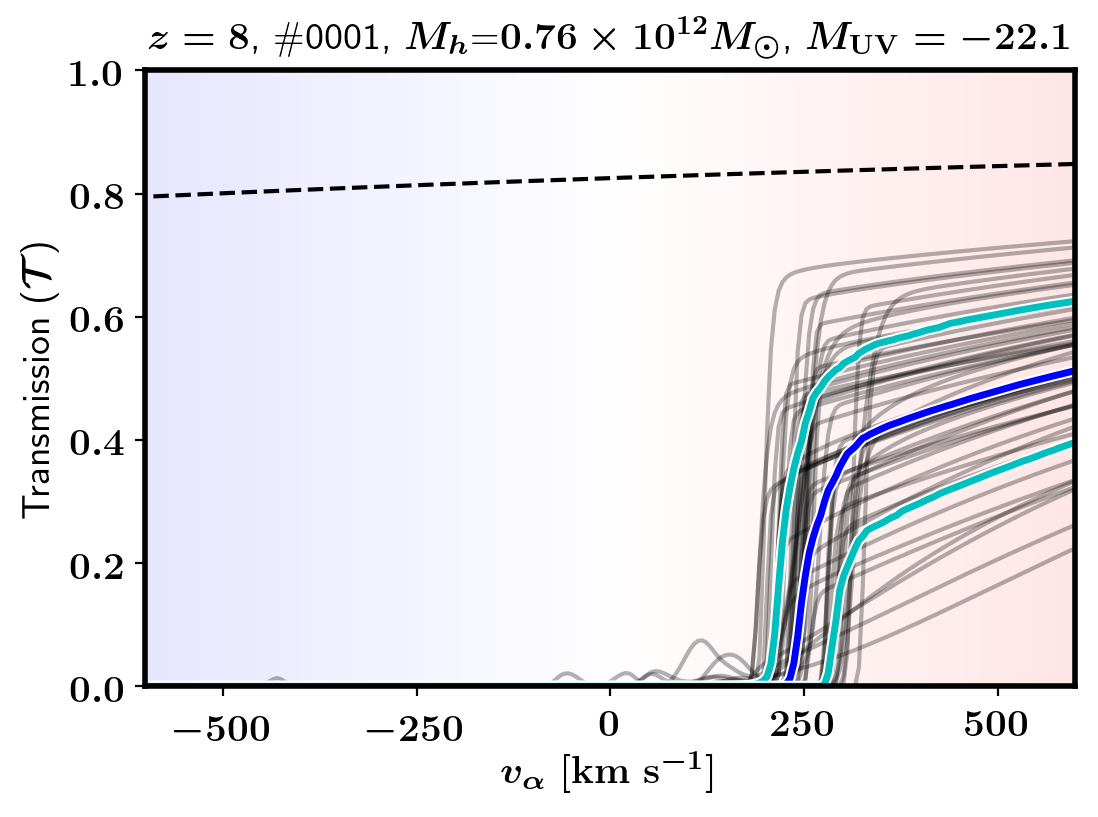}
\includegraphics[scale=0.4]{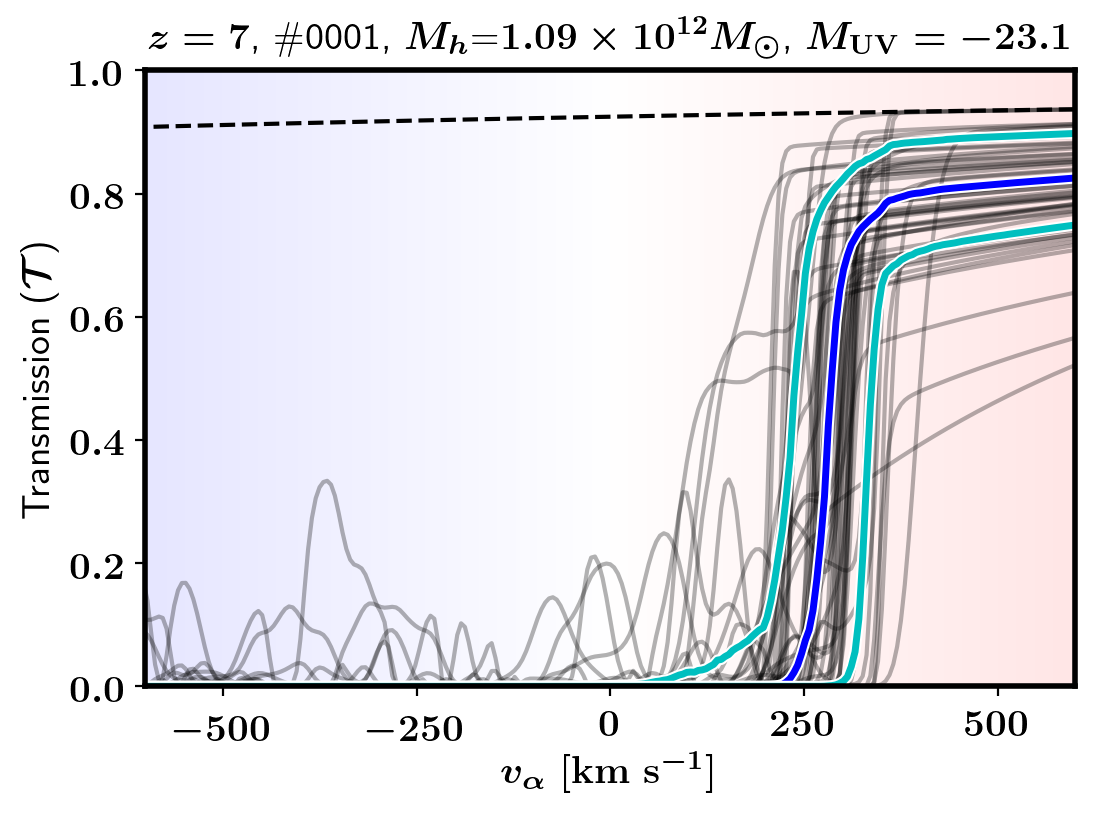}
\includegraphics[scale=0.4]{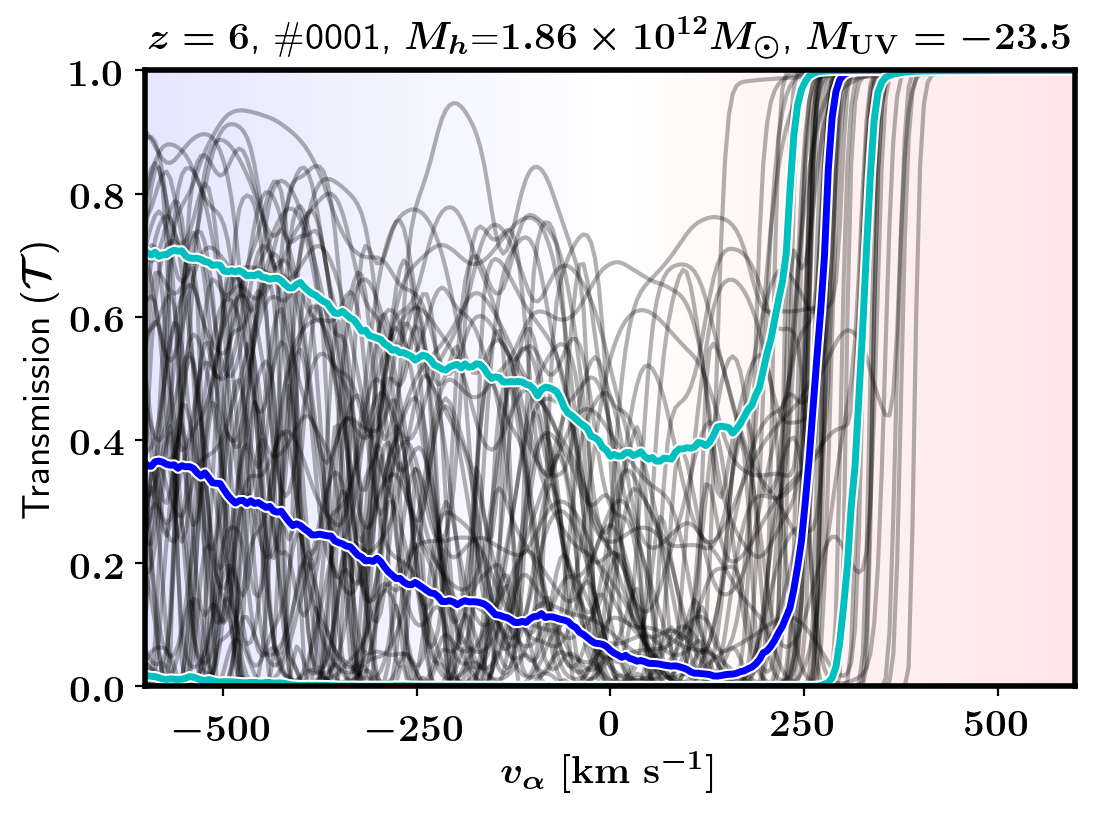}
\includegraphics[scale=0.4]{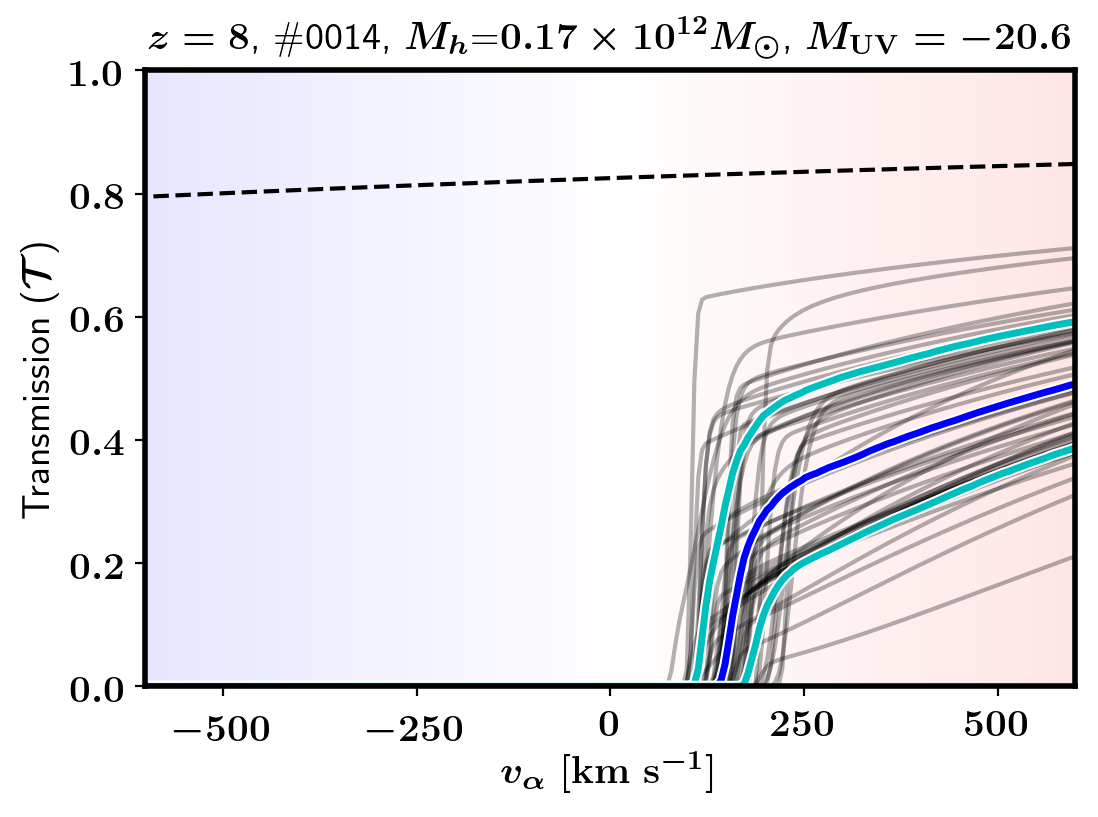}
\includegraphics[scale=0.4]{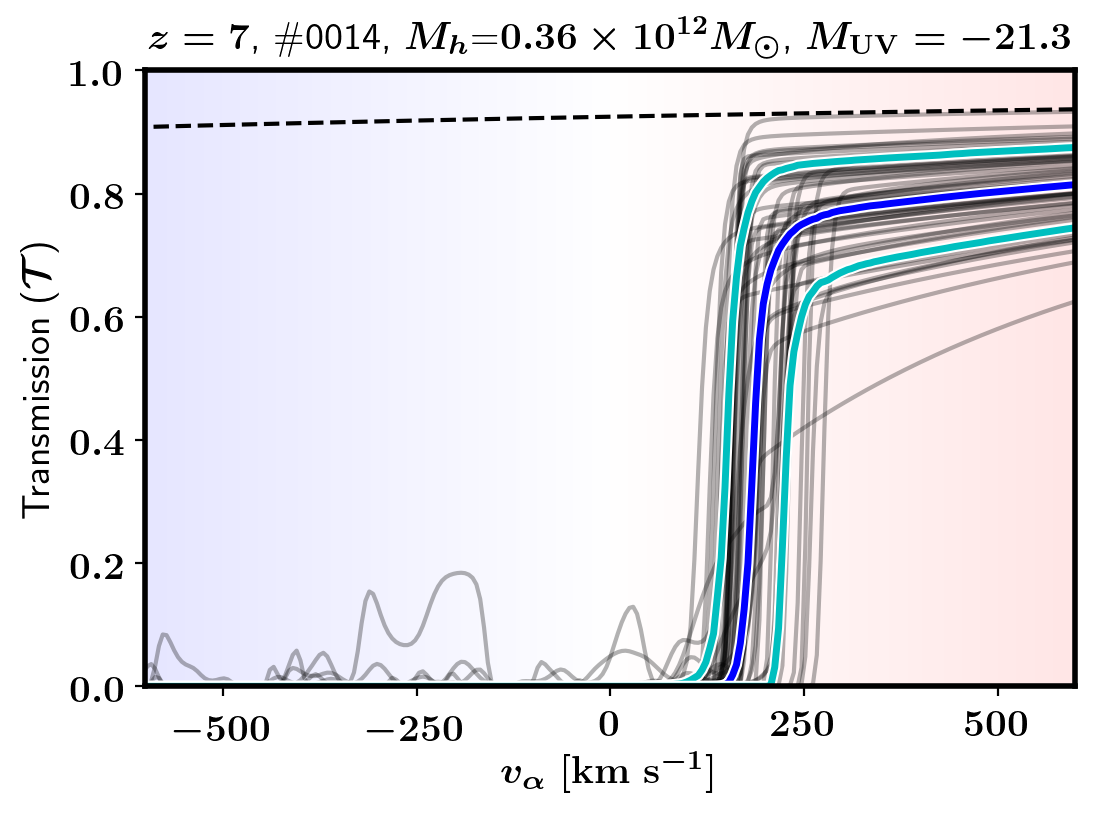}
\includegraphics[scale=0.4]{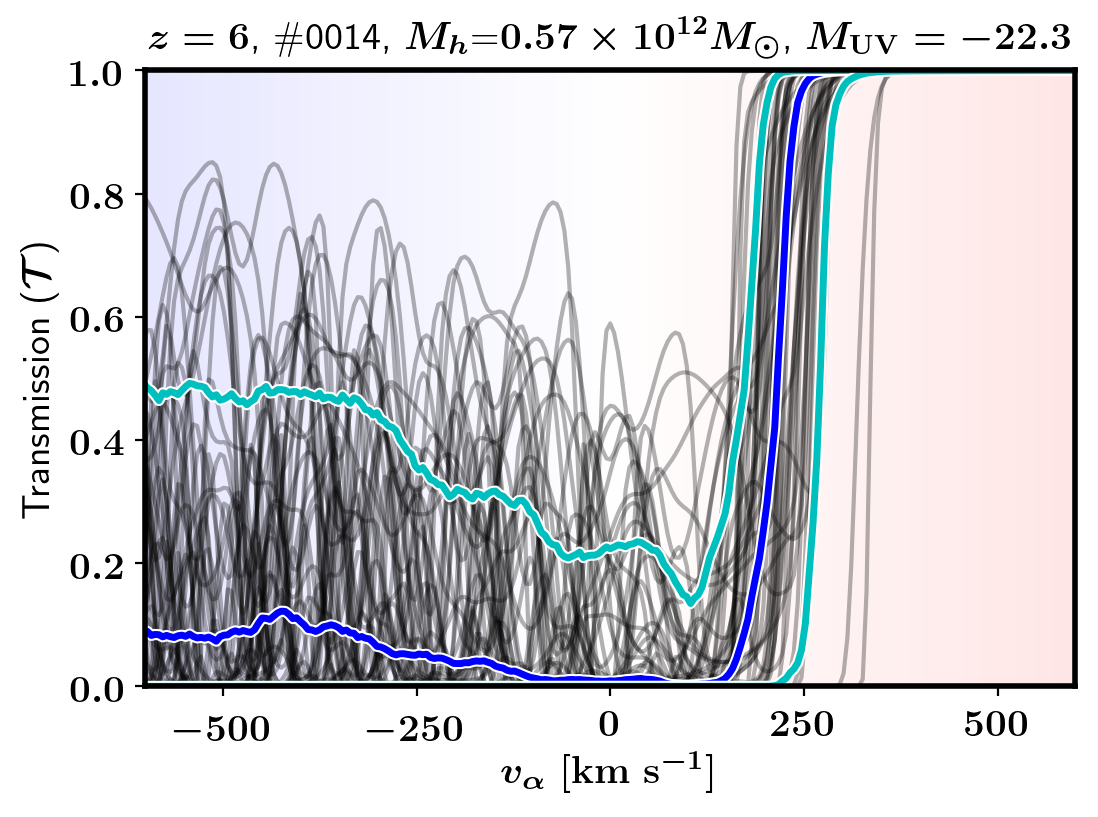}
\includegraphics[scale=0.4]{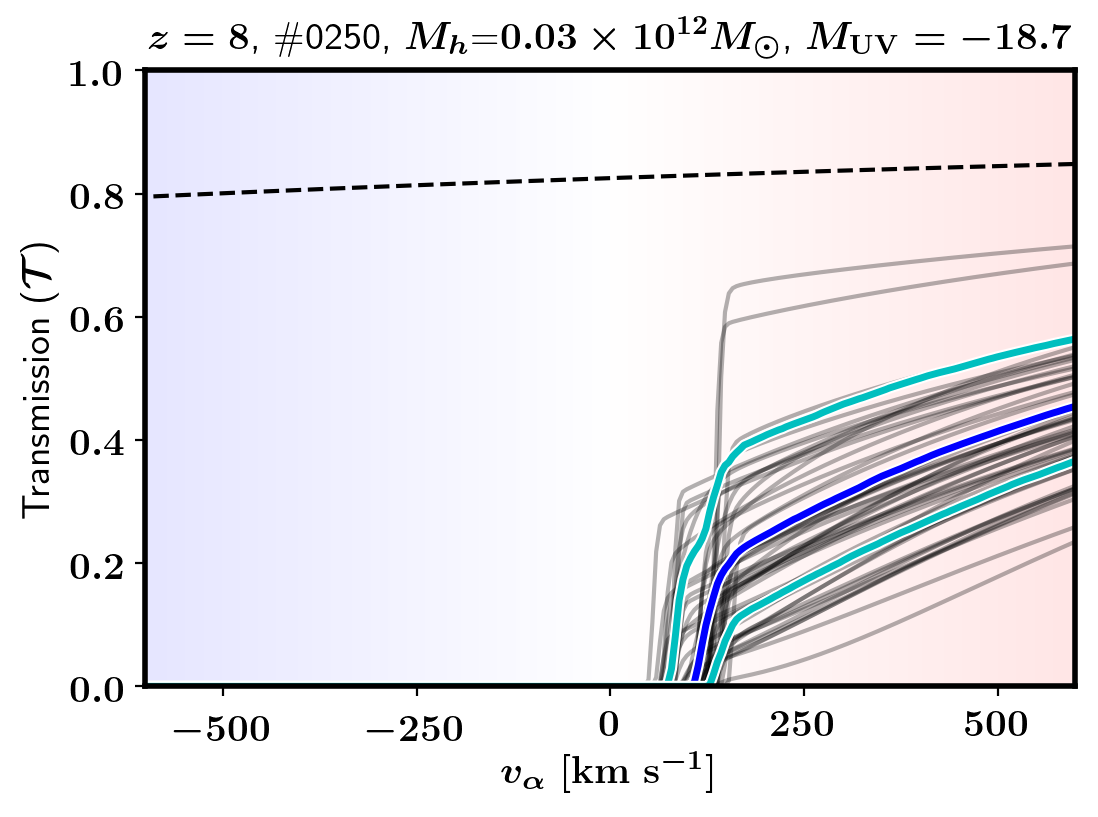}
\includegraphics[scale=0.4]{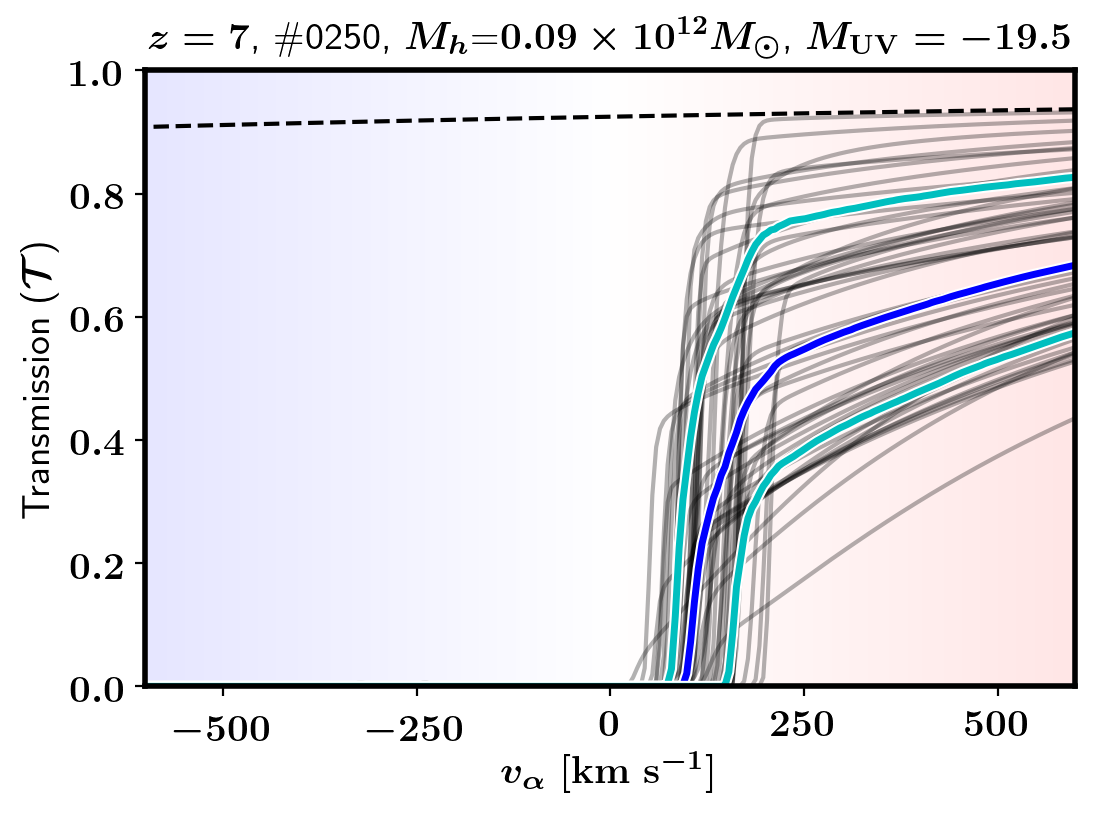}
\includegraphics[scale=0.4]{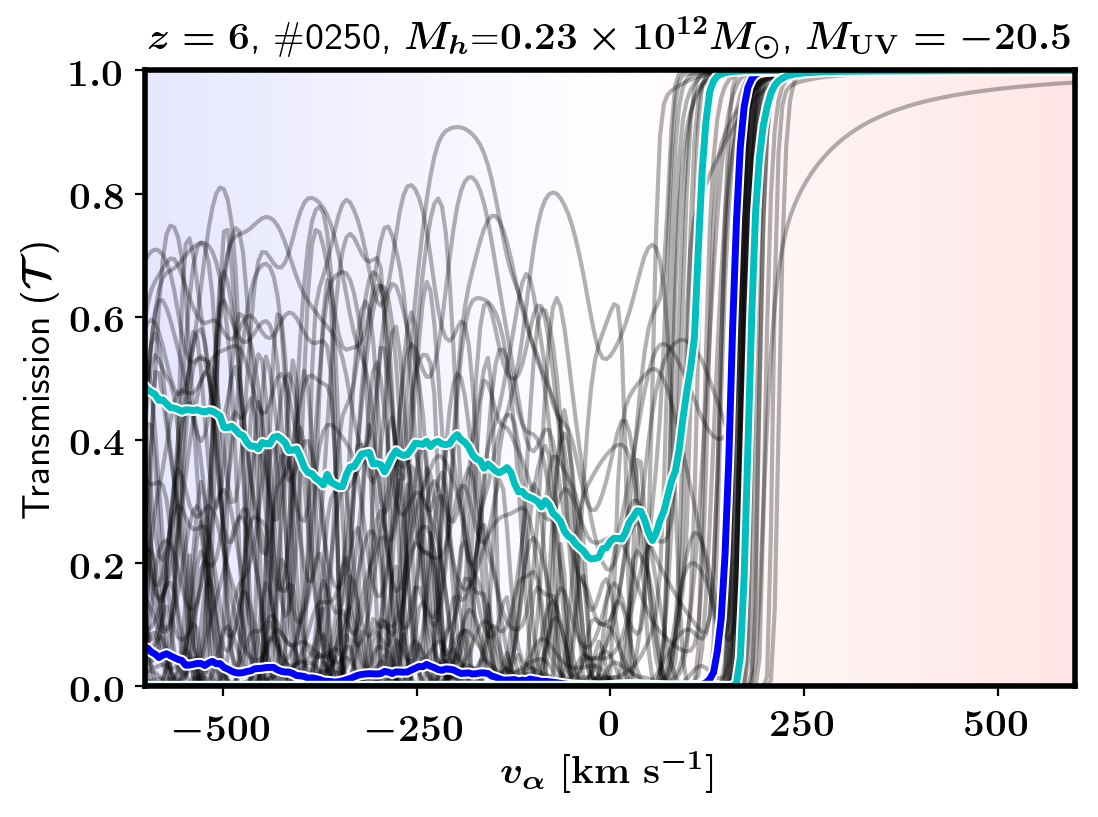}
\caption{ IGM transmission as a function of frequency. The upper, middle, and lower panels show the results for galaxy \#0001, \#0014, and \#0250, respectively, in the $z=8$ (left), $7$ (center), and $6$ (right) snapshots. The dashed lines denote the large-scale absorption ($\tau_{\rm DW}$) coming from the  IGM  farther than $s>24~h^{-1}~{\rm cMpc}$ from the source galaxy. Fifty out of 2000 individual sight lines are shown as partially transparent black solid lines. The blue solid line marks the median of the results from 2000 sight lines binned for each $v_\alpha$, and the cyan lines bracket the $68\%$ range.}
\label{fig:TrH1}
\end{center}
\end{figure*}

Figure~\ref{fig:TrH1} shows the IGM transmissivity $\mathcal{T}=e^{-\tau}$ as a function of frequency in the velocity unit ($v_\alpha$) for galaxies \#0001, \#0014, and \#0250 in the $z=6,~7,$ and $8$ snapshots. Note that the positive side of the $x$-axis ($v_\alpha>0$) is the red side of Ly$\alpha$. The transmissions for different sight lines are shown as multiple partially transparent curves to depict the sight-line variation. The blue solid line shows the median value at each frequency, and the cyan lines bracket the 68\% range around the median. 

\begin{figure*}
\begin{center}
\includegraphics[scale=0.5]{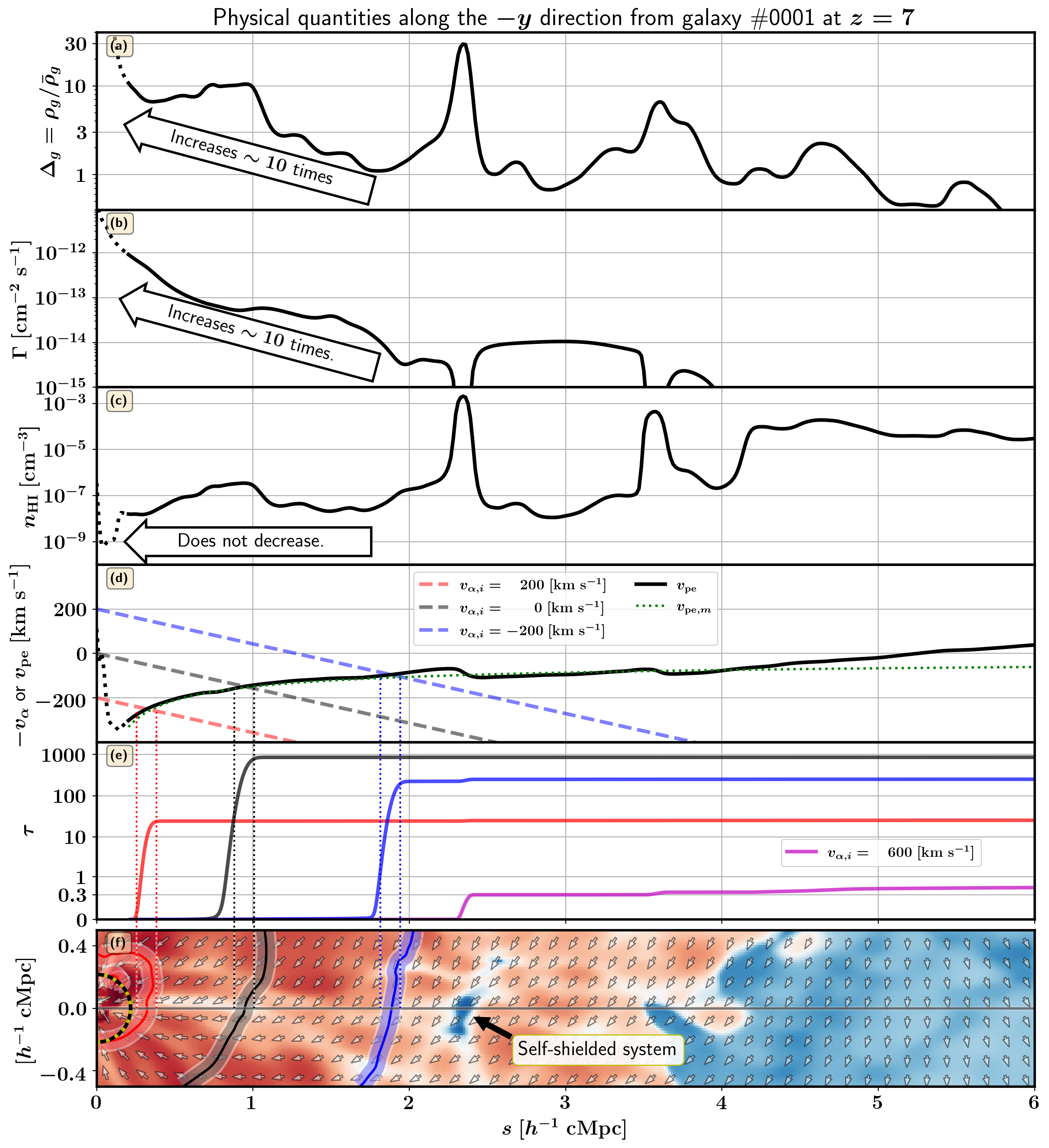}
\caption{Physical quantities of the IGM relevant to Ly$\alpha$ transmission along the $-y$ direction from the center of galaxy \#0001 at $z=7$. Panel $a$: gas density normalized by the cosmic mean. The range within the virial radius of the source galaxy is plotted as a dotted line and is not included in the optical depth calculation. This also applies to panel $b$, $c$, and $d$. Panel $b$: ionization photon density. Panel $c$: neutral hydrogen number density. Panel $d$: peculiar velocity of IGM w.r.t. to the galaxy is shown as the black solid line, and the green dotted lines are the model fit, $v_{{\rm pe},m}=\sqrt{GM_h/s}$. The blue, black, and red dashed lines show the frequency of the photons that were emitted at $v_{\alpha,i}=-200$, $0$, and $200~{\rm km}~{\rm s}^{-1}$, respectively. We plot minus the photon frequency, $-v_\alpha$, so that the intersection between the dashed and solid lines (i.e. $ v_{\alpha}+v_{\rm pe}=0$) is where the photons encounter resonance in the rest frame of the IGM. Panel $e$: optical depth of the IGM for four different frequencies. The line colors correspond to the same frequencies as for the dashed lines of panel $d$ plus the magenta line corresponds to $ v_{\alpha,i}=600~{\rm km}~{\rm s}^{-1}$. Panel $f$: the neutral hydrogen density map around the line of sight. The color scheme is the same as the one used in Figure~\ref{fig:nHmap}. The half-circle in the black dashed line indicates the virial radius. The contours lines describe the location where $v_{\alpha}+v_{\rm pe}=0$ for the frequencies plotted in panel $d$ and $e$. The thickness of the contours depicts the thermally broadened line-absorption profile of photoionized gas with $T=10^4$ K. The vertical dotted lines extending across panel $d$, $e$, and $f$ connect the locations where $v_{\alpha}+v_{\rm pe}=0$. } 
\label{fig:LOS}
\end{center}
\end{figure*}

\subsection{Resonance Absorption}

The cross section of Ly$\alpha$ is large enough to make even the mostly ionized IGM opaque at the resonance. Photons that escape the circumgalactic medium (CGM) at a frequency bluer than the resonance are redshifted to the line center in the rest frame of the IGM at a certain point and encounter this high opacity of the resonance absorption. Notably, all the transmission curves in Figure~\ref{fig:TrH1} show a rapid drop at $v_\alpha= 100-300~{\rm km}~{\rm s}^{-1}$ on the red side of the Ly$\alpha$, implying that the photons are blueshifted to the resonance in the rest frame of IGM.

This red-side resonance absorption occurs for two reasons. First, the peculiar velocity field of the IGM is dominated by the gravitational infall motion toward the source galaxy. Second, the rising IGM density toward the galaxy keeps the IGM optically thick at the resonance frequency despite the rising ionizing intensity from the source galaxy. Without the infall motion and the overdensity of the IGM, the transmission cutoff should appear at or on the blue side of Ly$\alpha$ \citep[e.g.,][]{2020MNRAS.499.1395M}.

The velocity map around galaxies in Figure~\ref{fig:nHmap} clearly shows the infall motion from all directions. The infall motion extends to at least several comoving Mpc from the galaxy (see middle column panels) and can extend up to tens of Mpc (see Figure~6 of \cite{2008MNRAS.391...63I}). There also are outflow motions generated by star-formation feedback in hot low-density blobs shown as dark red spots. However, the outflowing gas is mostly confined within the virial radius, which we take as the CGM in this work. 

In order to describe how the infall motion affects the IGM transmission, we plot gas density, HI density, peculiar velocity, and optical depth toward the $-y$ direction from galaxy \#0001 at $z=7$ in Figure~\ref{fig:LOS}. The plotted quantities are baryon density, ionizing photon density, HI number density, peculiar velocity, the optical depth of IGM to photons emitted at three different frequencies at the galaxy center, and the local density map around the line of sight. 

The peculiar velocity shown as the black solid line in panel $d$ is negative (i.e., infalling) up to $\sim 5~h^{-1}~{\rm cMpc}$ from the source. The infall velocity becomes as large as $\approx300~{\rm km}~{\rm s}^{-1}$ when it peaks near $s=R_{\rm vir}$ and gradually decreases toward larger distances. The infall velocity at a distance $s$ is similar to the circular velocity at that distance:
\bea
v_{{\rm pe},m}(s) \equiv \sqrt{\frac{G M_h}{s} },
\eea
where $M_h$ is the total galaxy mass (see the green dotted line). This indicates that the gravitational force from the galaxy is the dominant cause of the infall motion. The peculiar velocity profile is also similar to the analytical models of \cite{2004MNRAS.349.1137S} and \citet{2007MNRAS.377.1175D}.

The blueshift in the rest frame of the IGM due to the infall motion results in a high resonance opacity for some photons on the red side of Ly$\alpha$. The dashed lines in panel d show that the frequency change of the photons emitted at $ v_{\alpha,i}=-200,~0,$ and $200~{\rm km}~{\rm s}^{-1}$ due to cosmic expansion along the line of sight. When the dashed line intersects with the solid line (i.e., $v_{\alpha}+v_{pe}=0$), the photons face a sudden increase in the optical depth due to resonance opacity (see panel $e$). The photons that started at $v_{\alpha,i}=0$ and $200~{\rm km}~{\rm s}^{-1}$ would never have encountered the resonance cross-section without the infall motion.

Another reason for the red-side absorption is that the infalling gas near the virial radius remains opaque to Ly$\alpha$ photons despite the intense ionizing radiation from the source. From $s=10R_{\rm vir} \approx 1.8~h^{-1}~{\rm cMpc}$ to $s= R_{\rm vir} \approx0.18~h^{-1}~{\rm cMpc}$, the gas density in in panel $a$ rises by 10 times the cosmic mean while the ionizing radiation intensity ($\Gamma$) rises roughly by a factor of a hundred (panel $b$). The gas temperature remains close to 20,000 K throughout the interval. The 10 times increase in density cancels the hundred times increase in the intensity because the neutral hydrogen density of highly ionized gas ($n_{\rm HI}=x_{\rm HI}n_{\rm H}$) goes as $n_{\rm H}(n_{\rm HII}/\Gamma) \approx n_{\rm H}^2/\Gamma$ in the ionization equilibrium for constant gas temperature.  Therefore, the neutral hydrogen density does not fall toward the galaxy as shown in panel $c$, keeping the IGM optically thick to Ly$\alpha$. 

\begin{figure}
\begin{center}
\includegraphics[scale=0.5]{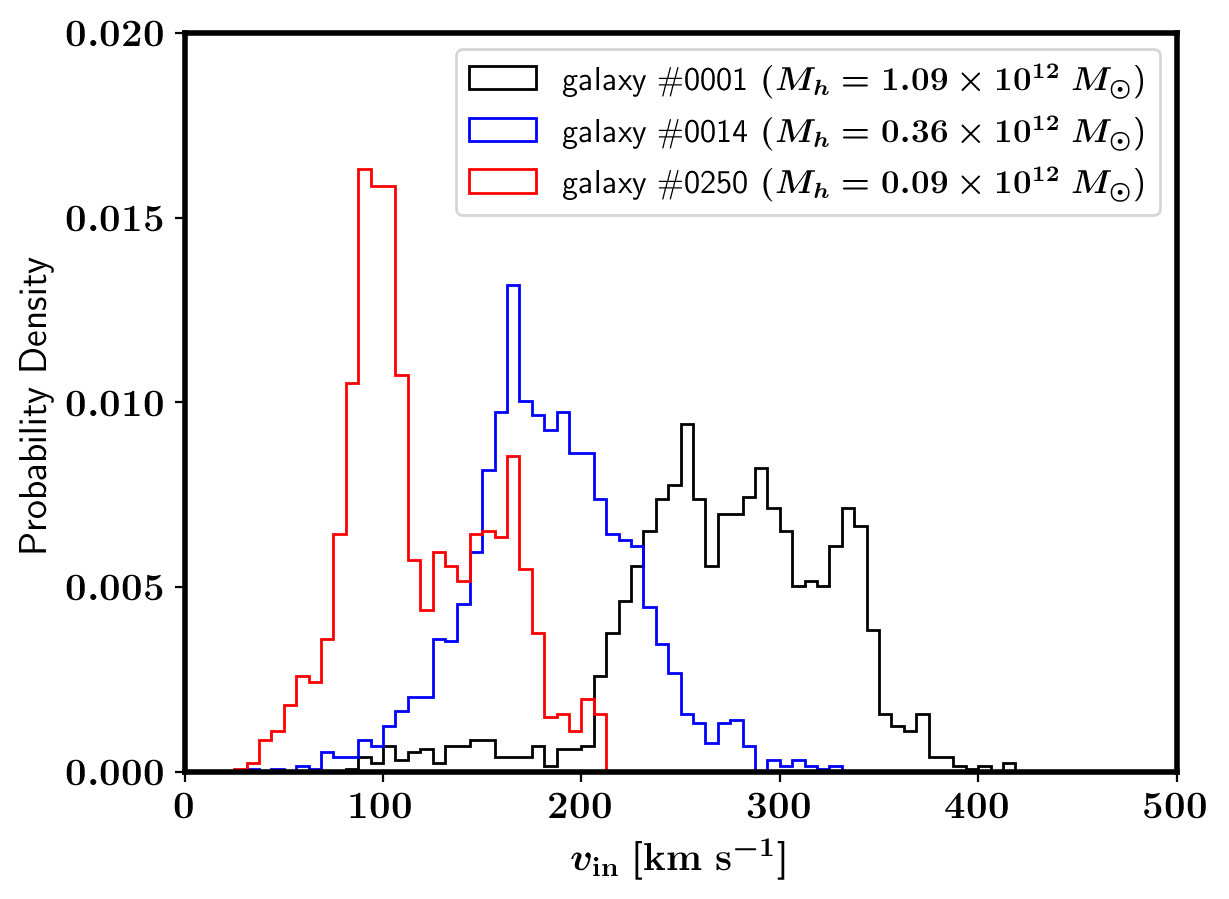}
\caption{Probability distribution of the maximum infall velocity $v_{\rm in}$ for 2000 sight lines for galaxies \#0001, \#0014, and \#0250 shown in black, blue, and red, respectively. $v_{\rm in}$ is obtained by identifying the location of the cutoff in the Ly$\alpha$ transmission. The galaxy masses are shown in the legend.} 
\label{fig:vin_dist}
\end{center}
\end{figure}

The transmission cutoff frequency is set by the maximum infall velocity outside the virial radius $R_{\rm vir}$. In most cases, the infall velocity peaks around $s=R_{\rm vir}$ (panel $d$), where the infall velocity is close to the circular velocity at the virial radius, $V_c\equiv \sqrt{GM_h/R_{\rm vir}}$. Here, the galaxy mass, $M_h$, is the total mass that includes both baryonic and dark matter within $R_{\rm vir}$. For each transmission curve, we measure the maximum infall velocity, $v_{\rm in}$, by identifying the location of the transmission cutoff, which coincides with the peak of the transmission curve slope. We plot in Figure~\ref{fig:vin_dist} the probability distribution of $v_{\rm in}$ based on the results from the 2000 sight lines. The $v_{\rm in}$ distribution has a fairly large scatter but shows a clear galaxy mass dependence. $v_{\rm in}$ is distributed around 300, 200, and 100 ${\rm km}~{\rm s}^{-1}$ for galaxies with their masses $1.09$, $0.36$, and $0.09\times 10^{12}~M_\odot$, respectively. 

\begin{figure*}
\begin{center}
\includegraphics[scale=0.5]{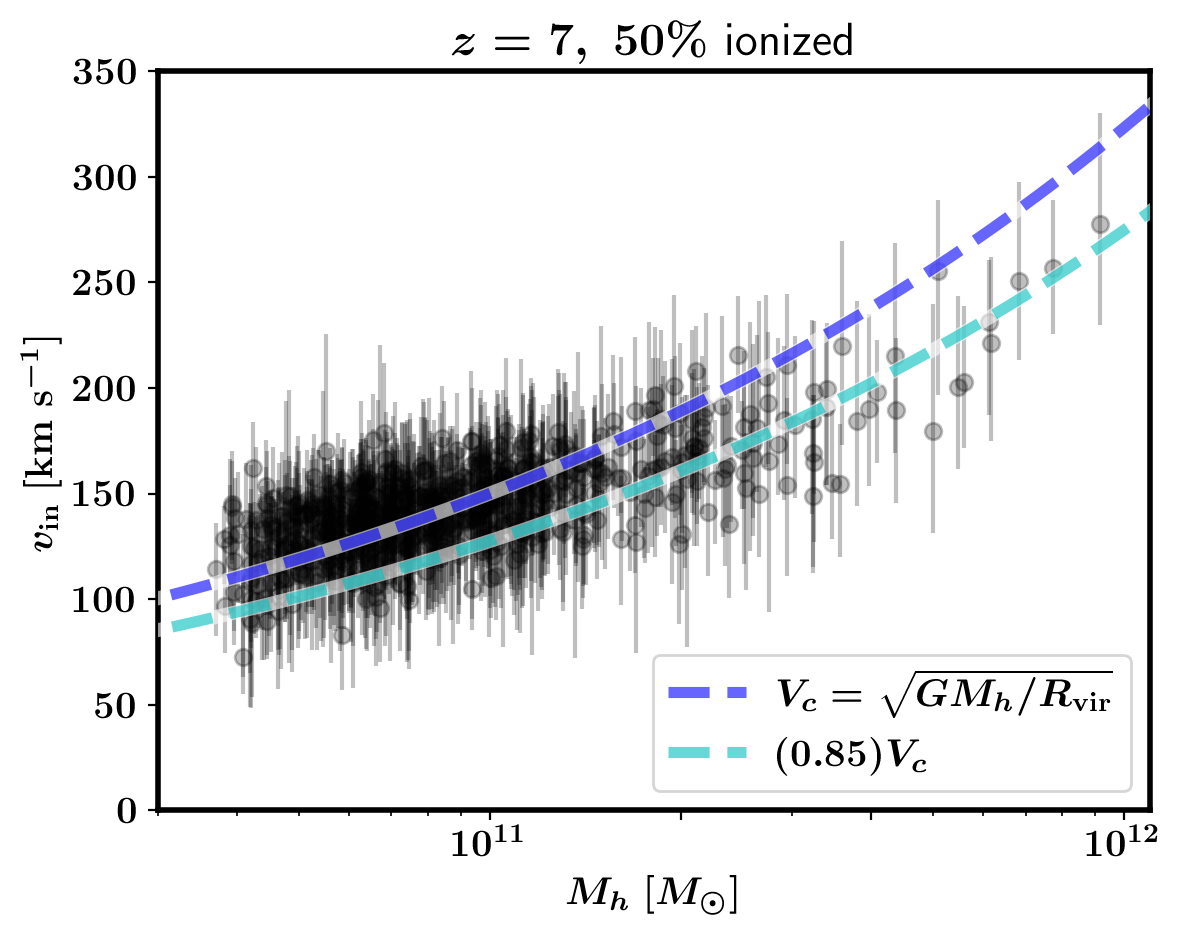}
\includegraphics[scale=0.5]{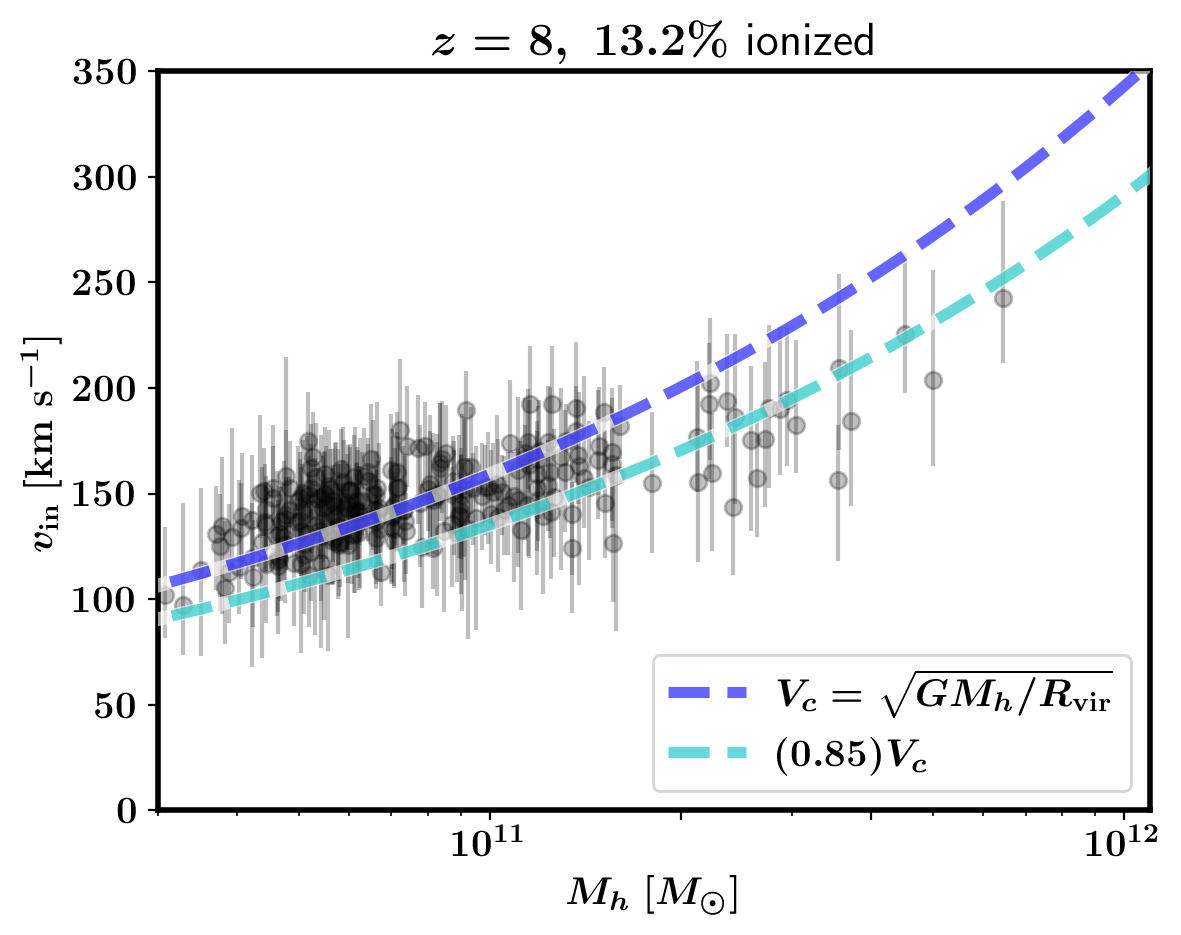}
\caption{Maximum infall velocity $v_{\rm in}$ as a function of galaxy mass $M_h$. Error bars for the data points show the 1$\sigma$ range computed for individual galaxies from 2000 sight lines.} 
\label{fig:dv_Mh}
\end{center}
\end{figure*}

\begin{figure*}
\begin{center}
\includegraphics[scale=0.5]{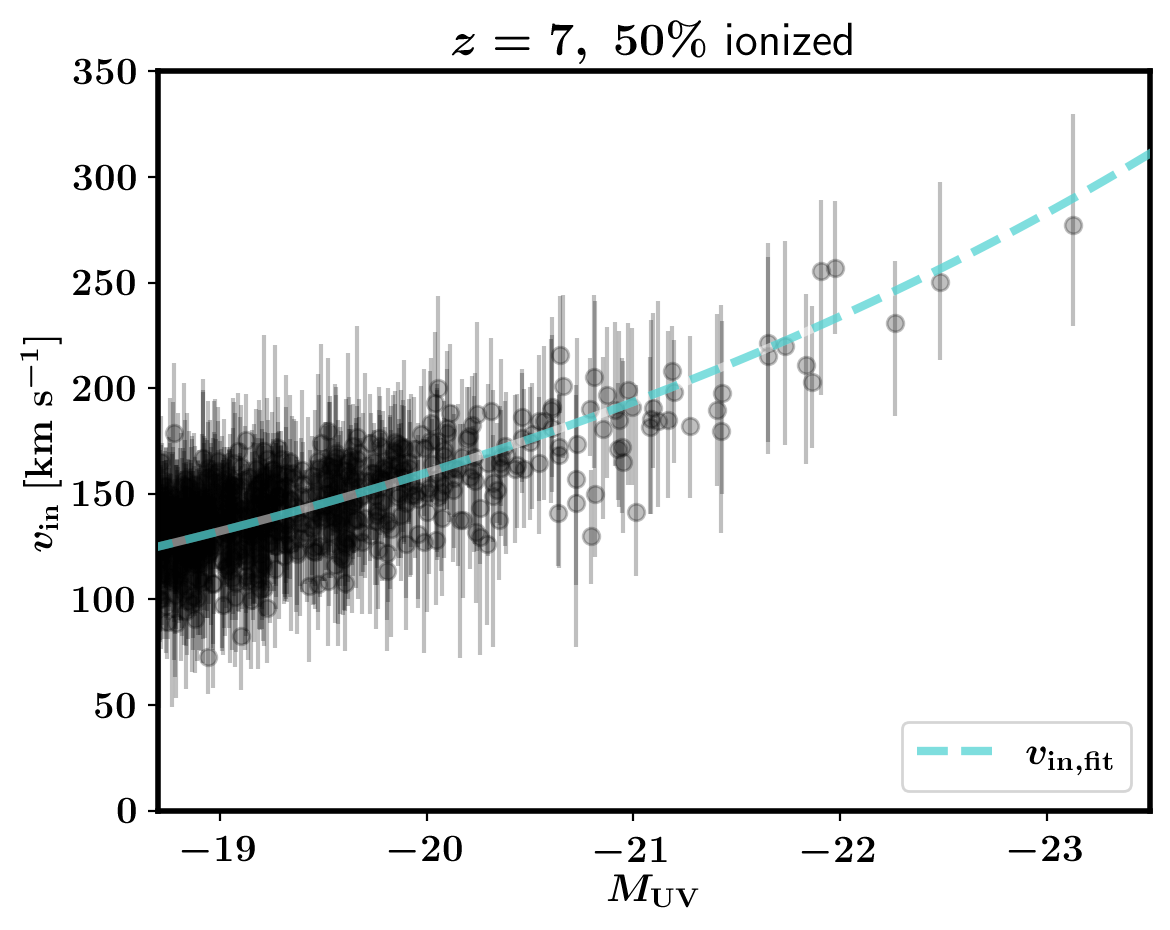}
\includegraphics[scale=0.5]{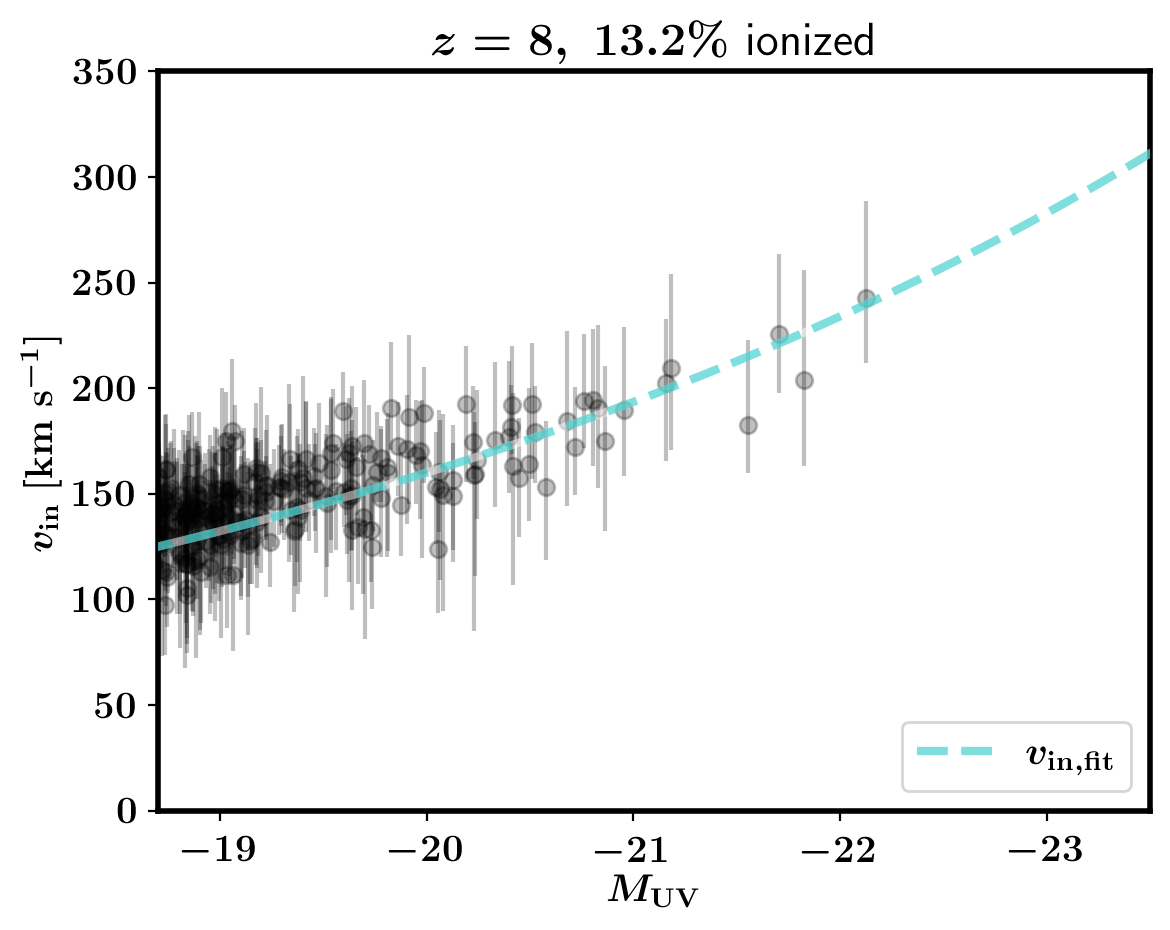}
\caption{Maximum infall velocity $v_{\rm in}$ as a function of UV magnitude $M_{\rm UV}$. Error bars for each data point show the 1$\sigma$ range computed for individual galaxies from 2000 sight lines. The cyan dashed line is a fitting function given by Eq.~\ref{eq:vinfit}.} 
\label{fig:dv_MUV}
\end{center}
\end{figure*}

\begin{figure}
\begin{center}
\includegraphics[scale=0.5]{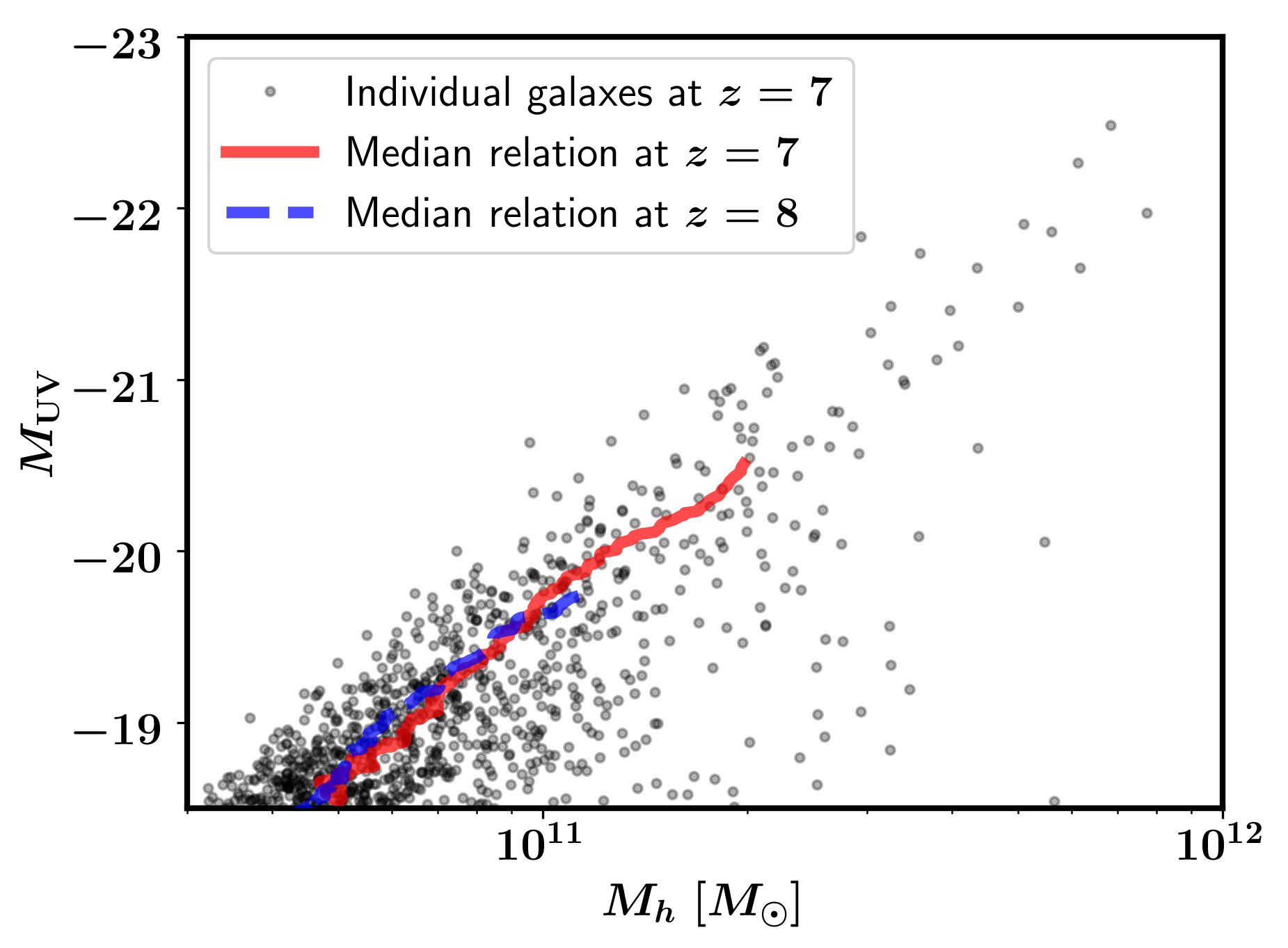}
\caption{Trend of $M_{\rm UV}$ with $M_h$. Black dots are for individual galaxies in the $z=7$ snapshot. The red solid and blue dashed lines are the median trend for $z=7$ and 8, respectively.} 
\label{fig:Mh_vs_Mab}
\end{center}
\end{figure}

We show the complete statistics of $v_{\rm in}$ from the entire galaxy sample as a function of halo mass in Figure~\ref{fig:dv_Mh} and as a function of UV magnitude in Figure~\ref{fig:dv_MUV}. The error bar for each galaxy denotes the 1$\sigma$ sight-line variation calculated from 2000 sight lines for each galaxy. $v_{\rm in}$ is similar to $V_c$ throughout the mass range of our sample, although it is slightly lower at high mass. For example, $v_{\rm in}$ prefers $V_c$ at $M_h=10^{11}~M_\odot$ while  it prefers $0.85 V_c$ at $M_h=10^{12}~M_\odot$.

We show in Figure~\ref{fig:dv_MUV} that $v_{\rm in}$ also correlates with the UV magnitude. This is due to the strong correlation between $M_{\rm UV}$ and $M_h$, which hardly evolves in time (see Figure~\ref{fig:Mh_vs_Mab}). We find that the correlation coefficient between $M_{\rm UV}$ and $M_h$ is above 95\% for the galaxy samples used in this work. The relation between $v_{\rm in}$ and $M_{\rm UV}$ is well described by a fitting function
\bea \label{eq:vinfit}
v_{\rm in,fit}\approx [160~{\rm km}~{\rm s}^{-1}]e^{-0.19[M_{\rm UV}+20]}
\eea
with a 1$\sigma$ scatter of $\sim 40~{\rm km}~{\rm s}^{-1}$ (see the cyan dashed line in Figure~\ref{fig:dv_MUV}).

Nonvanishing transmission on the blue side of Ly$\alpha$ often appears at $z=6$, while it is mostly zero at $z=7$ and $8$. At $z=7$, only bright galaxies with $M_{\rm UV}\lesssim-23$ can transmit a small fraction of blue-side photons. According to \cite{2021MNRAS.508.3697G}, this is because the typical HI density in ionized IGM is too high ($\gtrsim 10^{-9}~{\rm cm}^{-3}$) to transmit the blue-side photons at $z>6.5$. We refer readers to their work for a detailed analysis on this subject.

We note that the residual IGM neutral fraction at the postoverlap phase ($z<6.2$) is a factor of $\sim 5$ lower than the estimate of \cite{2006ARA&A..44..415F}. We checked for a subsample of halos that increasing the calculated optical depths by a similar factor did not affect the transmission redward of the cutoff.

\subsection{Damping-wing absorption}\label{sec:DW}

On the red side of the transmission cutoff, photons do not encounter resonance absorption, leading to a much higher transmissivity. Here, the opacity comes from the damping-wing absorption from highly neutral segments of sight lines. The opacity contribution from each segment depends sensitively on the distance to the segment from the source: $d\tau_{\rm DW}\propto s^{-2}$, as shown in Equation~(\ref{eq:tau_D}). 




The red side of the transmission curves in Figure~\ref{fig:TrH1} show the damping-wing opacity for different sight lines, halos, and redshifts. This red-side transmission decreases toward high redshift due to increasing damping-wing opacity. For example, the median transmission at $v_\alpha=400~{\rm km}~{\rm s}^{-1}$ for galaxies \#0001, \#0016, and \#0150 is nearly 100\% at $z=6$, while it drops to 80, 53, and 62\% at $z=7$ and to 44, 42, and 37\% at $z=8$, respectively. We find a substantial sight-line-to-sight-line variation in $\tau_{\rm DW}$ for individual galaxies at $z>6$. The 1$\sigma$ range of the transmission bracketed by the cyan lines is up to 30\%, implying that the ionization field surrounding the source galaxies is highly anisotropic.

\subsubsection{HII Bubble Size Estimation}

\begin{figure*}
\begin{center}
\includegraphics[scale=0.7]{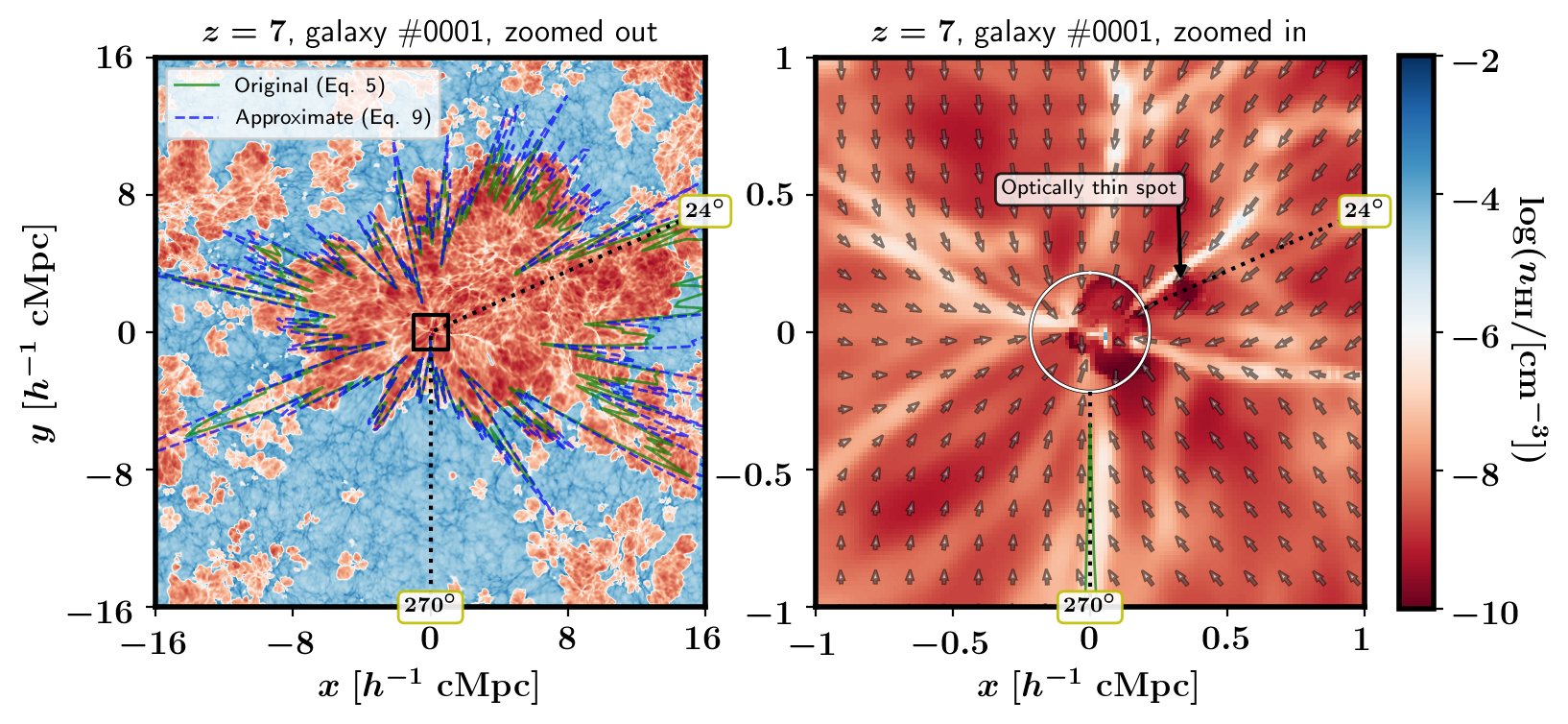}
\includegraphics[scale=0.7]{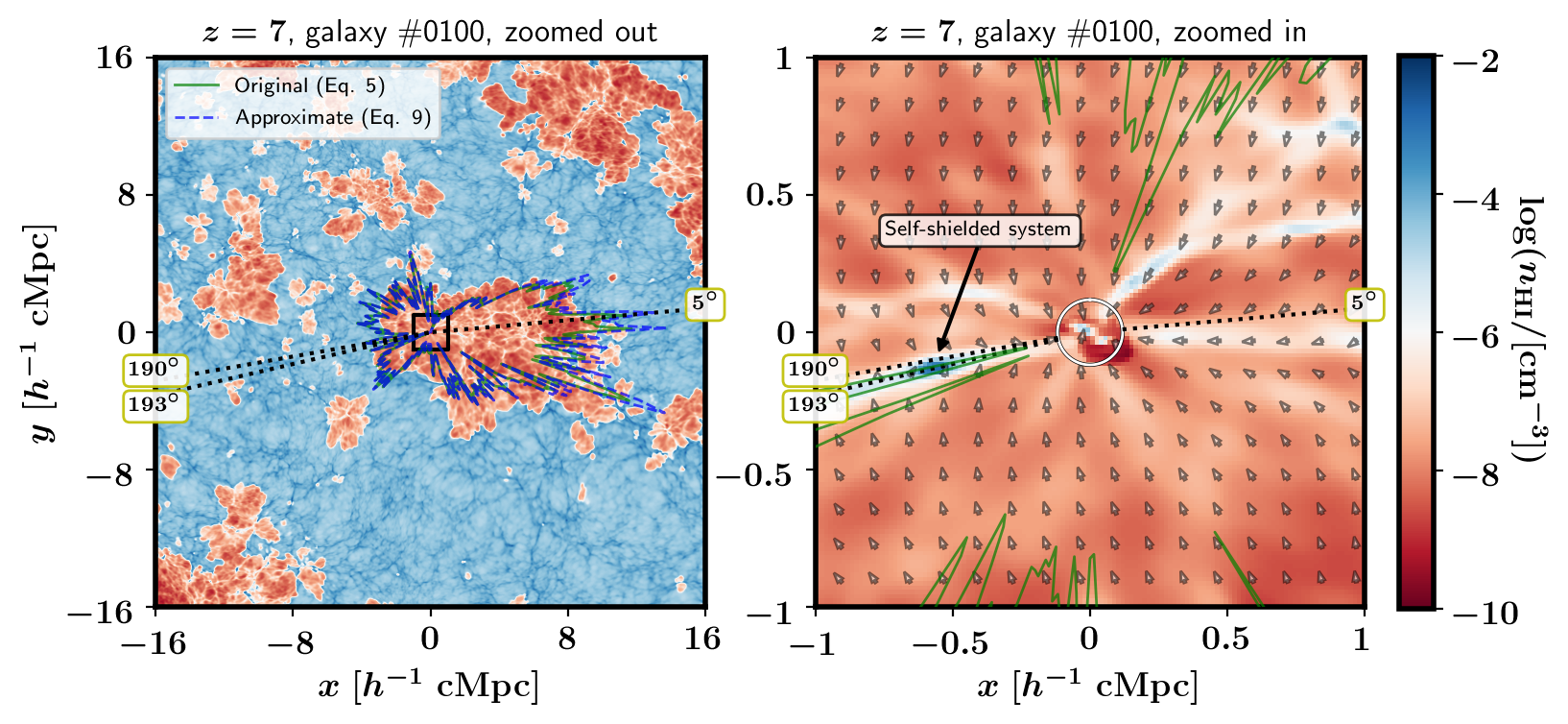}
\caption{Effective HII bubble size $R_b$ for directions on the $xy$ plane is shown on top of the HI density field around galaxies \#0001 (upper panel) and \#0100 (lower panel) in the $z=7$ snapshot. The green solid line contour describes the $R_b$ calculated from the original expression (Eq.~\ref{eq:tau_D}), while the blue dashed line is from the approximate expression (Eq.~\ref{eq:tau_D_app}). The left panels show the maps in a $32~h^{-1}~{\rm cMpc}$ box, while the right panels show in a $2~h^{-1}~{\rm cMpc}$ box. The black dotted lines with angles in degree mark the sight lines described in Figure~\ref{fig:FluxH100}.}
\label{fig:nHmap2}
\end{center}
\end{figure*}

If IGM is fully ionized inside the HII bubble and fully neutral outside, the damping-wing opacity, $\tau_{\rm DW}$, is given by integrating Equation~(\ref{eq:tau_D}) from the bubble size $s_{\rm min}=R_b$ to the end of reionization $s_{\rm max}=s(z=6)$. Then, $\tau_{\rm DW}$ would depend only on the bubble size $R_b$ and the reionization history $\bar{x}_{\rm HI}$, and one can constrain $\bar{x}_{\rm HI}$ and $R_b$ by measuring $\tau_{\rm DW}$ from observations. 

Often, an approximate expression for Equation~(\ref{eq:tau_D}) is used for convenience. In that case, one assumes that the opacity comes entirely from the neighborhood of the source galaxy before the redshift term in the integral changes significantly \citep{2014PASA...31...40D}. The redshift-dependent terms can then be taken out from the integral, giving
\bea \label{eq:tau_D_app}
&& \tau_{{\rm DW,~app}}(v_\alpha)=
5.7
\left[\frac{z_g+1}{8}\right]
\bar{x}_{\rm HI}(z_g) 
\nonumber\\
&&
\times
\left[\frac{R_b+\Delta s(v_\alpha)}{h^{-1}~{\rm cMpc}}\right]^{-1}
\left[\frac{h}{0.678}\right] 
\left[\frac{\Omega_b}{0.048}\right]
\left[\frac{\Omega_m}{0.307}\right]^{-1},
\eea
where $z_g$ is the redshift of the source galaxy. In this case, the opacity does not depend on the reionization history along the line of sight, and one can trivially convert the measured $\tau_{\rm DW}$ to $\bar{x}_{\rm HI}$, given a relation between $\bar{x}_{\rm HI}$ and $R_b$ computed from reionization models \citep[e.g.,][]{2005MNRAS.363.1031F}.

In this work, we define $R_b$ to be the value that matches the damping-wing optical depth $\tau_{\rm DW}$ with the computed IGM transmissivity at $v_\alpha=400~{\rm km}~{\rm s}^{-1}$ (hereafter $\mathcal{T}_{400}$). $v_{\rm in}$ is smaller than $400~{\rm km}~{\rm s}^{-1}$ for all sight lines considered in this work. Thus, the optical depth at $400~{\rm km}~{\rm s}^{-1}$ is purely from the damping-wing opacity.

In Figure~\ref{fig:nHmap2}, we show $R_b$ for galaxies \#0001 and \#0100 at $z=7$ as the line contours overplotted on the HI density maps. We computed $R_b$ for 360 equally spaced directions ($\hat{n}$) on the $xy$ plane and connected the locations of $R_b \hat{n}$ to create a closed contour. The green solid line contour is from the original expression of Equation~(\ref{eq:tau_D}), while the dashed line contour is from the approximate expression of Equation~(\ref{eq:tau_D_app}). 

We note that the original and approximate expressions generally give similar results, although the approximate value occasionally overshoots in some directions. The overshoot tends to be larger when $R_b$ is larger (i.e., $\tau_{\rm DW}$ is small). This is because assuming the $[1+z]\bar{x}_{\rm HI}(z)$ term in Equation~(\ref{eq:tau_D}) to be constant overestimates $\tau_{\rm DW}$ by $\sim 0.1$ at $z=7$, due to the decline of $\bar{x}_{\rm HI}$ toward low redshift. For large HII bubbles like the one surrounding galaxy \#0001, the error in $\tau_{\rm DW}$ can bias $R_b$ more because $\tau_{\rm DW}$ is generally smaller. However, galaxy \#0001 is a rare extreme case, and most of the galaxies are in smaller HII bubbles like galaxy \#0100, where the error is insignificant.

\begin{table}[]
\centering
\caption{Median IGM transmissivity at $v_\alpha=400~{\rm km}~{\rm s}^{-1}$}
\label{tab:Trdw}
\begin{tabular}{@{}c|c|c|c@{}}
\toprule
$ M_{\rm UV}$ & -21 & -20 & -19 \\  \midrule
$\mathcal{T}_{400}(z=7)$ &  $73.1\%$ & $70.6\%$ & $70.3\%$ \\  \midrule
$\mathcal{T}_{400}(z=8)$ &  $37.2\%$ & $33.3\%$ & $30.0\%$ \\ \bottomrule
\end{tabular}
\end{table}



\begin{figure*}
\begin{center}
\includegraphics[scale=0.5]{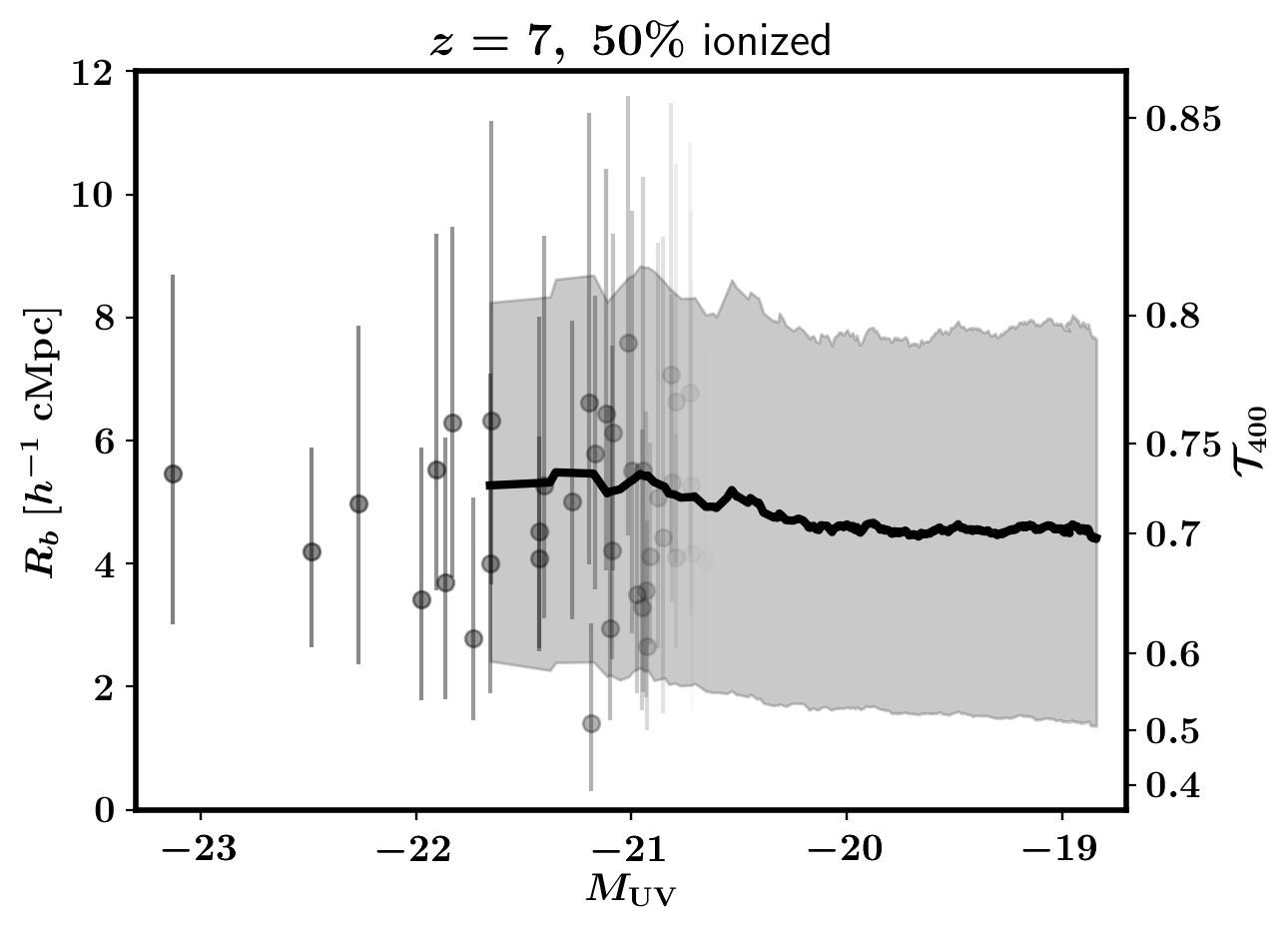}
\includegraphics[scale=0.5]{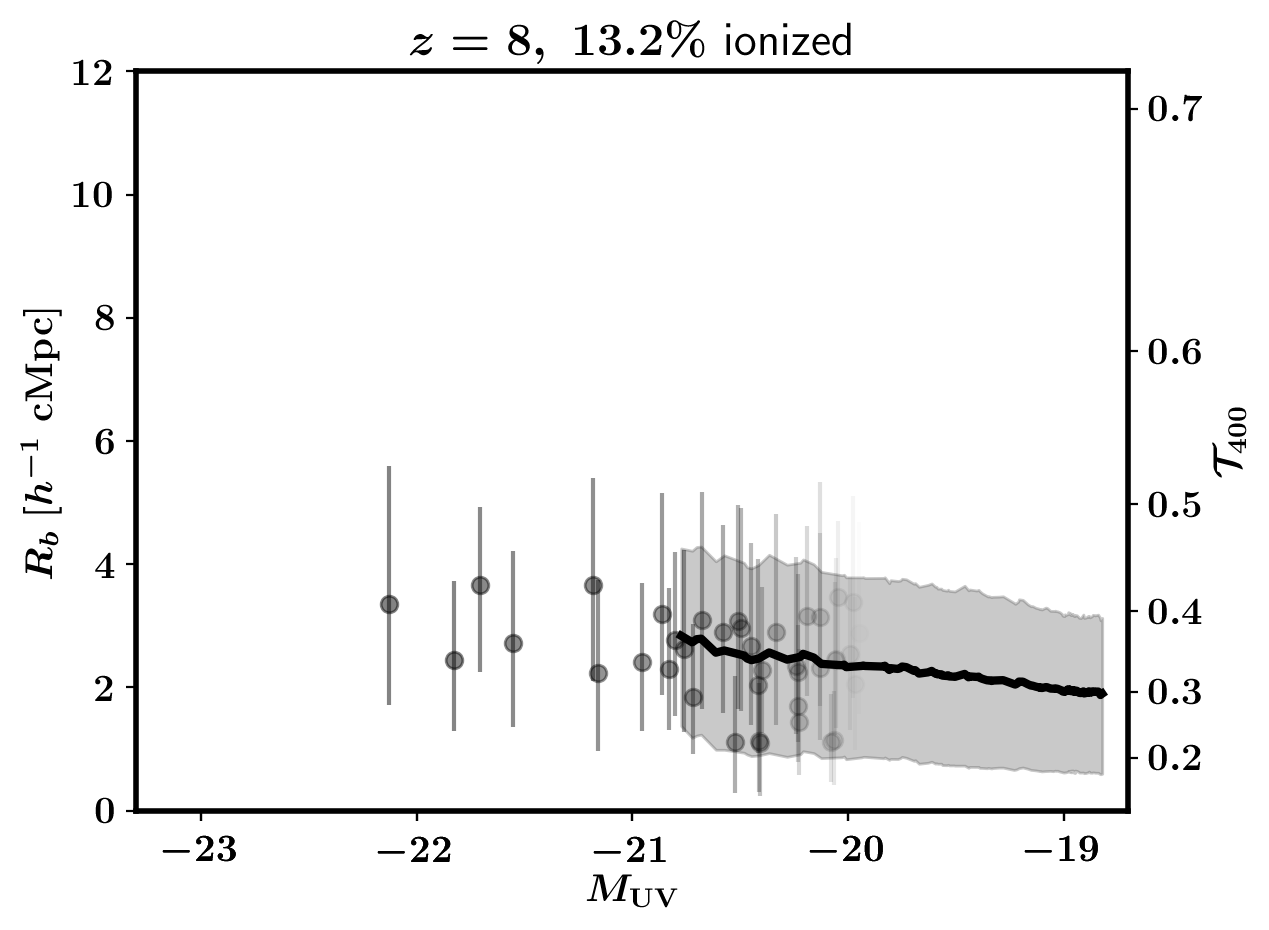}
\caption{Effective HII bubble size $R_b$ for sample galaxies in the $z=7$ (left) and $8$ (right) snapshots. Vertical bars on the data points describe the sight-line variation of individual galaxies. On the right $y$-axis, we show the corresponding IGM transmission at $v_\alpha=400~{\rm km}~{\rm s}^{-1}$. The black solid line marks the median of $R_b$ each at given $M_{\rm UV}$'s. The grey shade brackets the 1$\sigma$ range of the combined sight-line-to-sight-line and galaxy-to-galaxy variation. Individual data points are shown only for relatively rare bright galaxies to avoid crowding the figure with too many data points.} 
\label{fig:Rb_stat}
\end{center}
\end{figure*}

The $R_b$ contour generally follows the HII bubble outline with some deviations that make the contour spikier than the actual bubble shape. The deviation between the apparent and effective bubble sizes largely owes to the bubble geometry being more complex than what is assumed in Equation~(\ref{eq:tau_D}): ``fully ionized inside the HII bubble and fully neutral outside." When another large HII bubble is located in the vicinity of the HII bubble surrounding the source galaxy, for example, $R_b$ would be larger than what Equation~(\ref{eq:tau_D}) gives toward the direction of the neighboring bubble. 

Another important factor ignored in Equation~(\ref{eq:tau_D}) is self-shielded systems, which are small and dense pockets of neutral gas in mini halos within HII regions. When a sight line encounters such a clump, it can pick up a substantial amount of opacity even inside a HII region. The $-y$ direction ($\phi=270^\circ$) from galaxy \#0001 (upper panels of Figure~\ref{fig:nHmap2}) is an example that is affected by a clump. This clump is described in detail in Figure~\ref{fig:LOS}, where it is shown as a blue spot in panel~$f$ at $s\approx 2.3~h^{-1}~{\rm cMpc}$. The baryon density in this neutral clump reaches 30 times the cosmic mean at the peak (panel~$a$). The optical depth for the photon emitted at $v_{\alpha,i} =600~{\rm km}~{\rm s}^{-1}$ rises to 0.3 at this clump (magenta line in panel $e$). $R_b$ for this sight line is $0.33~h^{-1}~{\rm cMpc}$, although the neutral region actually starts at $s\approx 4~h^{-1}~{\rm cMpc}$. 

We present the $R_b$ statistics of the entire sample with $M_{\rm UV}<-18.7$ at $z=7$ and $8$ in Figure~\ref{fig:Rb_stat}. We also give the corresponding transmissivity at $v_\alpha=400~{\rm km}~{\rm s}^{-1}$ on the right-hand-side $y$-axis. We also tabulate the median values of $\mathcal{T}_{400}$ for $M_{\rm UV}=-21,~-20,$ and, $-19$ in Table~\ref{tab:Trdw}. The scattered data points with vertical bars in Figure~\ref{fig:Rb_stat} highlight the large galaxy-to-galaxy and sight-line-to-sight-line variation in the damping-wing opacity; the 1$\sigma$ range depicted by the gray shape shows is larger than the median value. 

$R_b$ is expected to evolve with redshift, which is considered to be the main cause of reduced LAE visibility at high redshifts. The data points of Figure~\ref{fig:Rb_stat} shows this expected redshift dependence of $R_b$ and $\mathcal{T}_{400}$ clearly: the median values of $R_b$ and $\mathcal{T}_{400}$  are around $5~h^{-1}~{\rm Mpc}$ and $73\%$ at $z=7$, while $2.5~h^{-1}~{\rm Mpc}$ and $37\%$ at $z=8$, respectively. 

The $M_{\rm UV}$ dependence is not easily visible from the data points, but the median curve reveals a weak dependence. At $z=7$, the median $R_b$ is $5.5~h^{-1}~{\rm Mpc}$ for galaxies $M_{\rm UV}=-21$, whereas it is about $1~{\rm Mpc}$ smaller for the galaxies with $M_{\rm UV}=-19$. Similarly, at $z=8$, $R_b$ is about $3~h^{-1}~{\rm Mpc}$ for $M_{\rm UV}=-21$ and $2~h^{-1}~{\rm Mpc}$ for $M_{\rm UV}=-19$. The corresponding $\mathcal{T}_{400}$ drops from $73\%$ to $70\%$ at $z=7$ in the same $M_{\rm UV}$ interval, while from $37\%$ to $30\%$ at $z=8$. 

\subsection{Self-shielded Systems}\label{sec:SSS}

\begin{figure*}
\begin{center}
\includegraphics[scale=0.4]{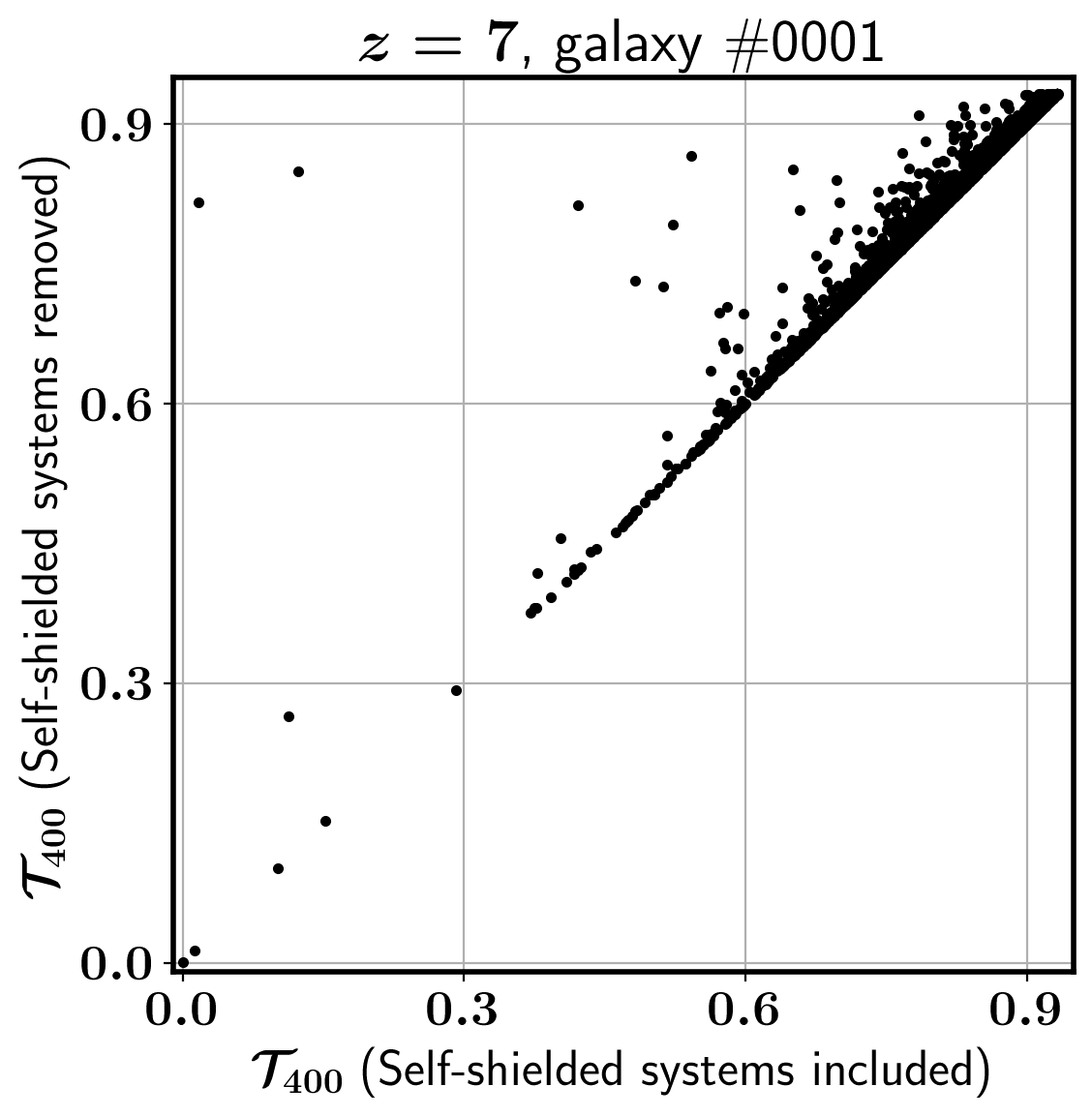}
\includegraphics[scale=0.4]{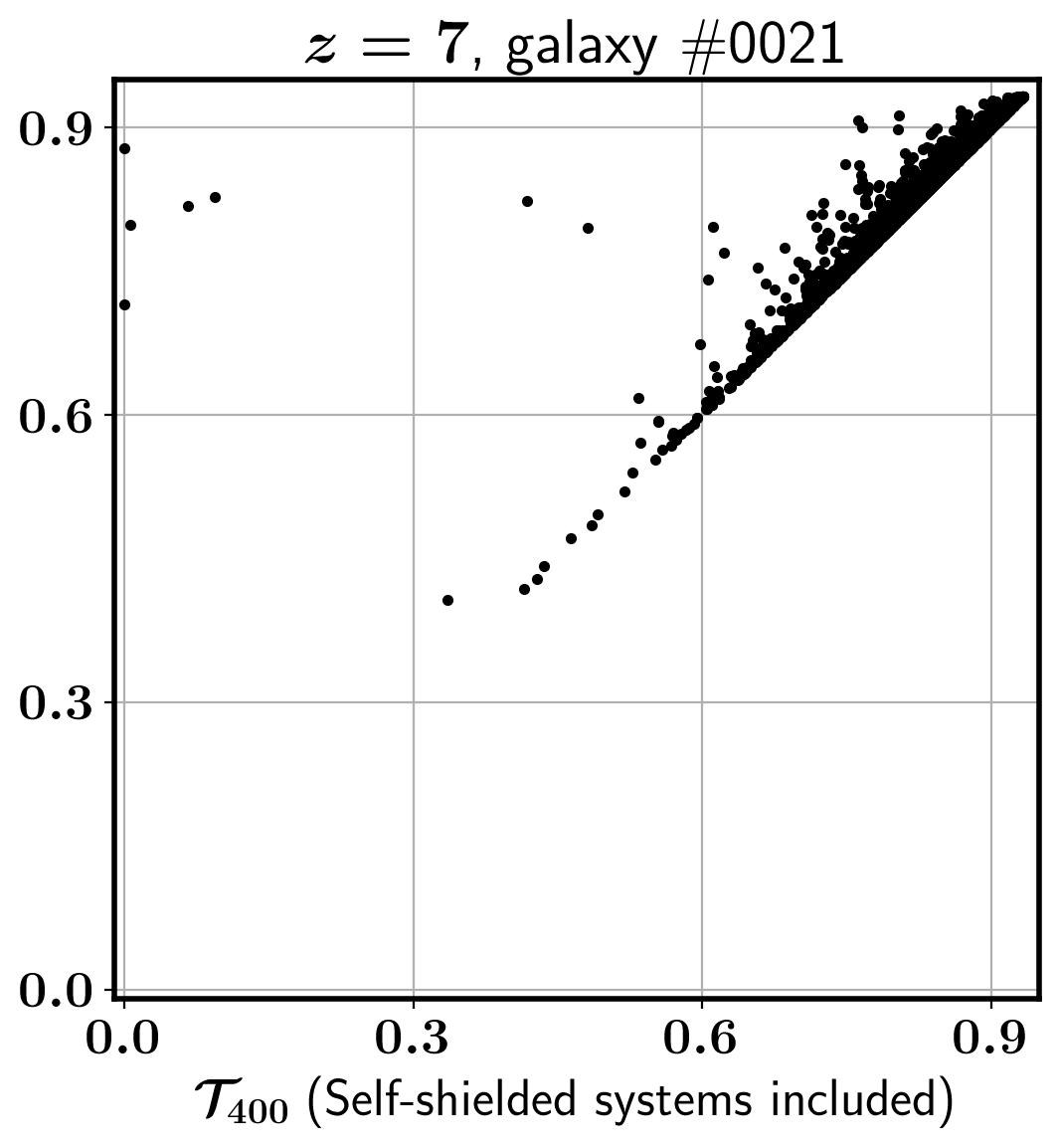}
\includegraphics[scale=0.4]{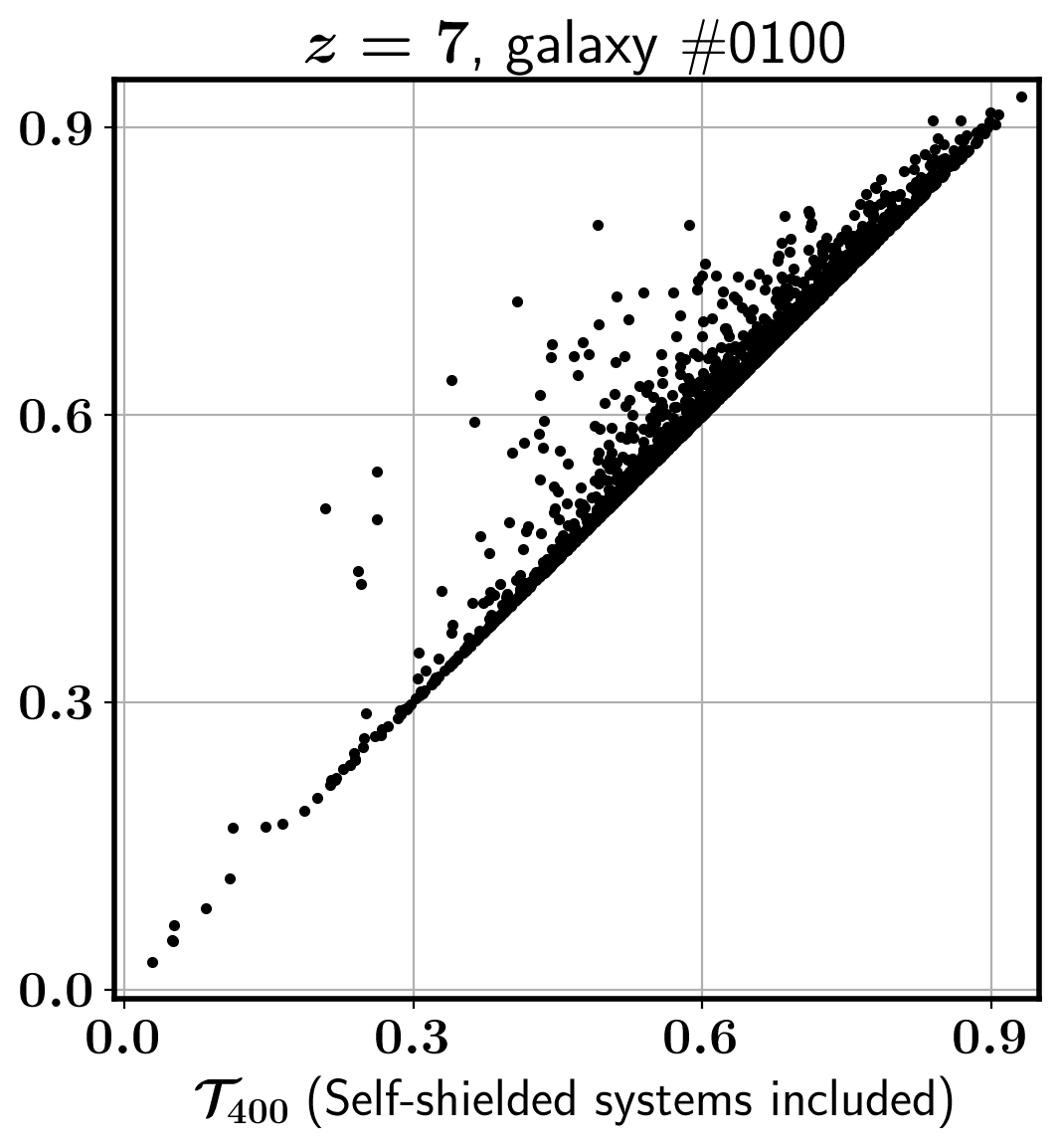}
\caption{Transmission at $v_\alpha=400~{\rm km}~{\rm s}^{-1}$ before and after removing the self-shielded systems in the opacity integral of Equation~(\ref{eq:tau}) as explained in Section~\ref{sec:SSS}. The results for 2000 sight lines of galaxies \#0001, \#0021, and \#0100 are shown in the left, middle, and right panels, respectively.}
\label{fig:T400NS}
\end{center}
\end{figure*}

Self-shielded systems can add a large opacity to certain sight lines to create a large discrepancy between the effective and actual bubble size as mentioned in Section~\ref{sec:DW}. In this section, we estimate the quantitative impact of these systems on the transmissivity by removing these systems in the transmissivity calculation. For this, we first identify highly neutral segments of slight lines with $n_{\rm HI}>10^{-5}~{\rm cm}^{-3}$. We then assume all those segments shorter than $0.3~h^{-1}~{\rm cMpc}$ are the compact neutral clumps within the HII region and lower the HI density of those segments to $10^{-5}~{\rm cm}^{-3}$ when evaluating Equation~(\ref{eq:tau}). This way, we exclude the damping-wing opacity of the self-shielded system like the one in Figure~\ref{fig:LOS}. 

In Figure~\ref{fig:T400NS}, we compare $\mathcal{T}_{400}$ before and after removing the self-shielded systems for 2000 sight lines of galaxies \#0001, \#0021, and \#0100. The data points that are above the $x=y$ line are the sight lines that are affected by the neutral clumps. The figure shows that some sight lines have substantially higher transmission when clumps are removed. 

We find that removing the self-shielded clumps has a small impact on the global statistics of the transmission. Removing the clumps increases the median value of $\mathcal{T}_{400}$ for galaxies \#0001, \#0021, and \#0100 increase from 80.1\% to 80.9\%, from 82.9\% to 83.5\%, and from 62.6\% to 63.8\%, respectively. Also, only 0.78\%, 0.69\%, and 2.3\% of the sight lines experience more than a 10\% increase in $\mathcal{T}_{400}$, respectively. 

However, if we limit the statistics to low-transmission sight lines with $\mathcal{T}_{400}<0.3$, the impact appears much stronger, finding that $\mathcal{T}_{400}$ increases by more than 10\% for 33\%, 100\%, and 12\% of the sight lines, respectively. This result suggests that unusual nondetections of Ly$\alpha$ in UV bright galaxies can be partially attributed to the self-shielded systems blocking the sight line.

We note that our method of removing self-shielded systems is not perfectly accurate. For example, if the edges of two adjacent HII bubbles are closer than $0.3~h^{-1}~{\rm cMpc}$, our method can erroneously regard the thin HI region between the edges as an isolated neutral clump. However, such complications on the bubble edge would have small impacts on the opacity compared to the impact of dense neutral clumps within the edge of a HII region, which is generally closer to the source. We leave more rigorous and comprehensive analyses of the self-shielded systems for future studies.

\subsection{Transmissivity of the L\MakeLowercase{y}$\alpha$ emission line} \label{sec:Xa}

We adopt a model spectral energy distribution for the Ly$\alpha$ emission line profile from the CGM to estimate the quantitative impact of IGM absorption. The adopted model is 
\bea \label{eq:intrinsic}
F_{\rm in}(v_\alpha)\propto \exp{\left( -\left[ \frac{ v_\alpha-V_c}{V_c/2.355}\right]^2 \right) }. 
\eea
This model assumes that the profile is given by a Gaussian which the peak location and the FWHM are given by the circular velocity of the galaxy $V_c=\sqrt{GM_h/R_{\rm vir}}$ as motivated by recent observations of LAEs at $z<6$ \citep{2016ApJ...820..130Y,2018MNRAS.478L..60V}.\footnote{ We discuss how varying the intrinsic profile affects our results in Section~\ref{sec:Muv-dependence} based on the experiment results in the Appendix.}
Here, we only model on the ratio between the intrinsic and transmitted flux for an arbitrary normalization for the intrinsic flux. Also, we do not consider any emission from the blue side of Ly$\alpha$ as the transmission is nearly zero there at $z=7$ and $8$. We note that, in reality, the spectral shape of the Ly$\alpha$ emission entering the IGM is expected to vary from sight line to sight line due to the complex dynamics of the CGM \citep[e.g.,][Song, H. et al. in preparation]{2012A&A...546A.111V}. We take the product of the transmission curve and the intrinsic spectral shape to obtain the IGM transmitted flux: $ F_{\rm out}(v_\alpha)= \mathcal{T}(v_\alpha) F_{\rm in}(v_\alpha)$. Then, the transmitted fraction of the line emission is given by
\bea
X_{{\rm Ly}\alpha} \equiv \frac{\int dv_\alpha \mathcal{T}(v_\alpha) F_{\rm in}(v_\alpha) }{\int dv_\alpha F_{\rm in}(v_\alpha)}.
\eea

\subsubsection{Notable Cases}

\begin{figure*}
\begin{center}
\includegraphics[scale=0.5]{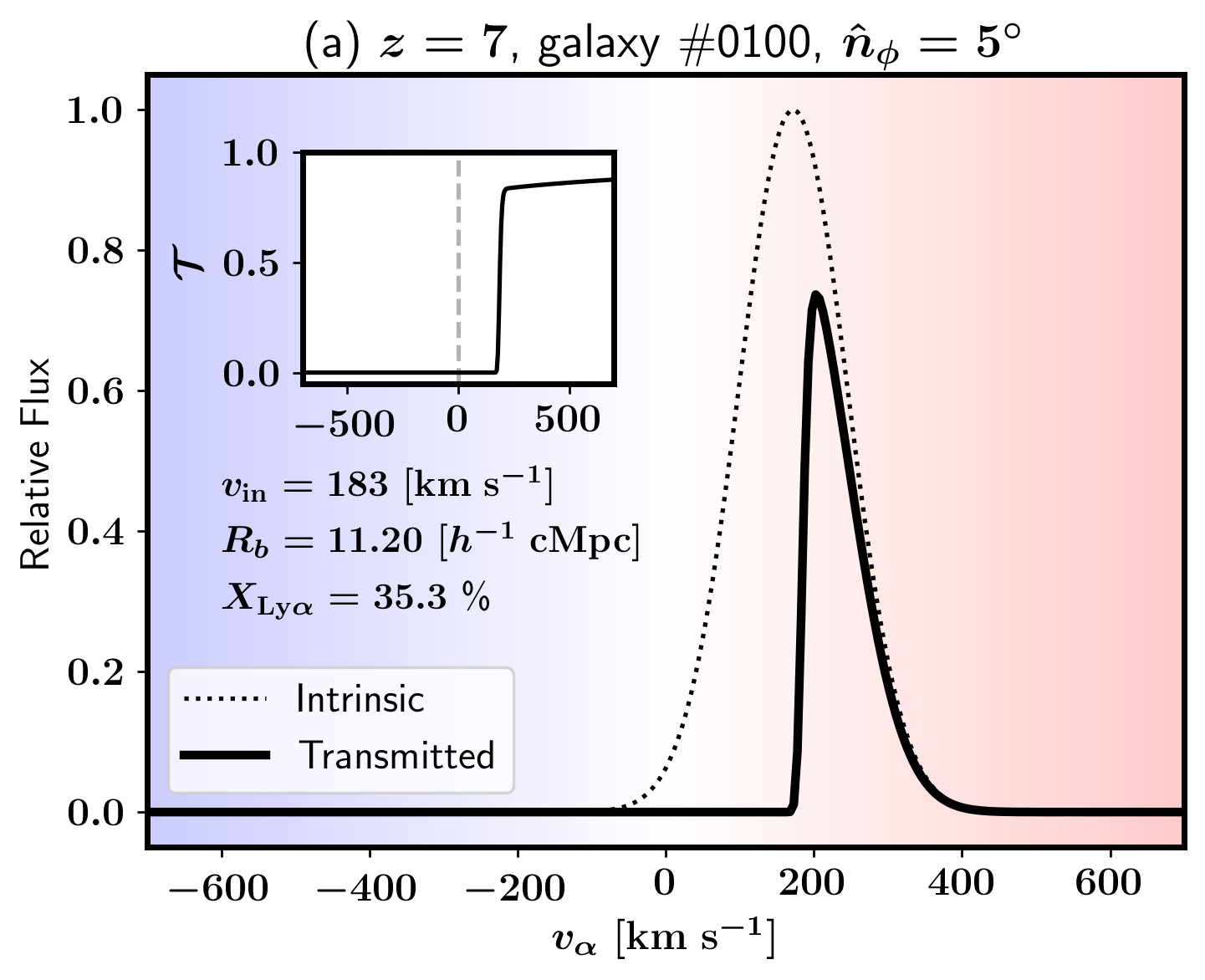}
\includegraphics[scale=0.5]{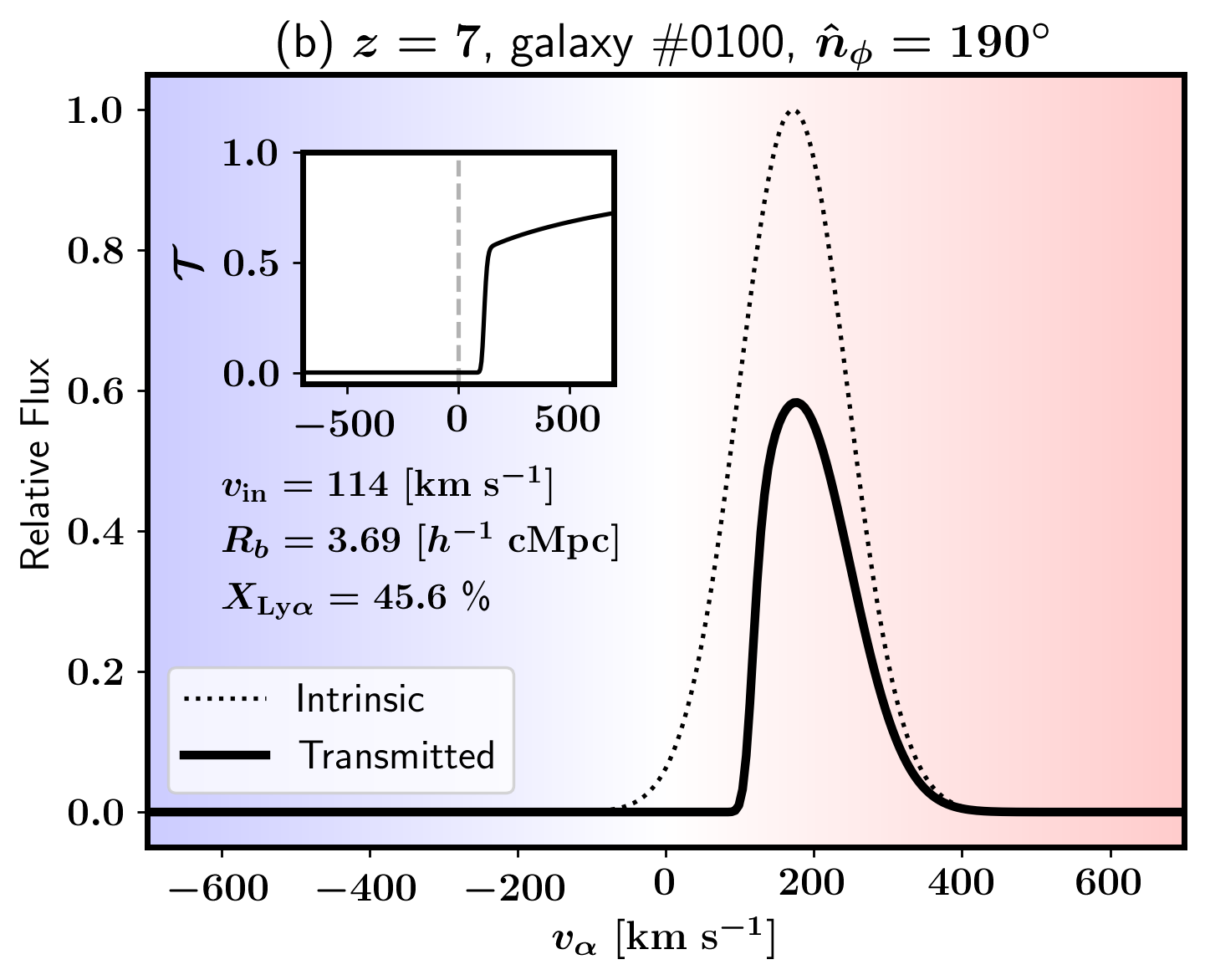}
\includegraphics[scale=0.5]{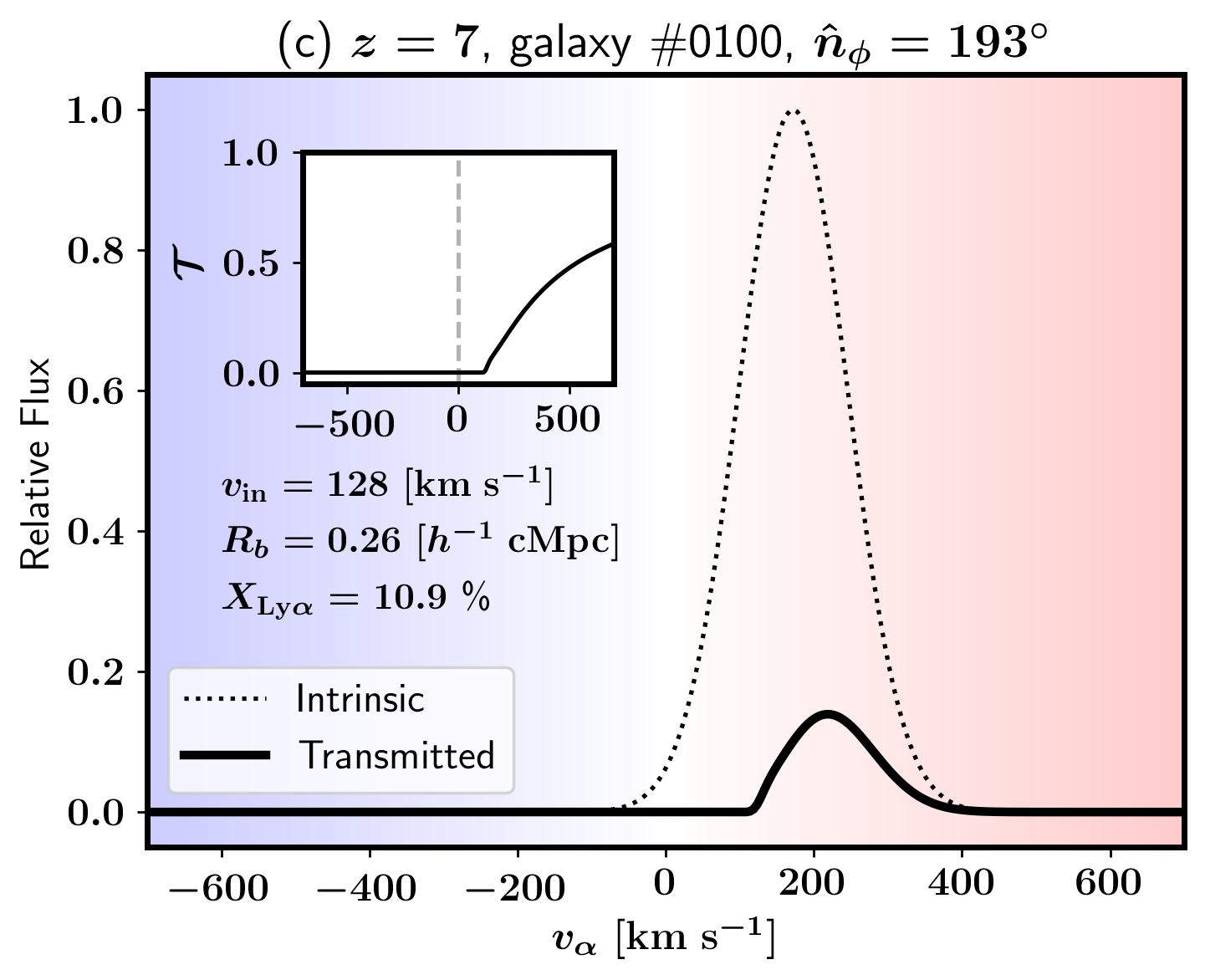}
\includegraphics[scale=0.5]{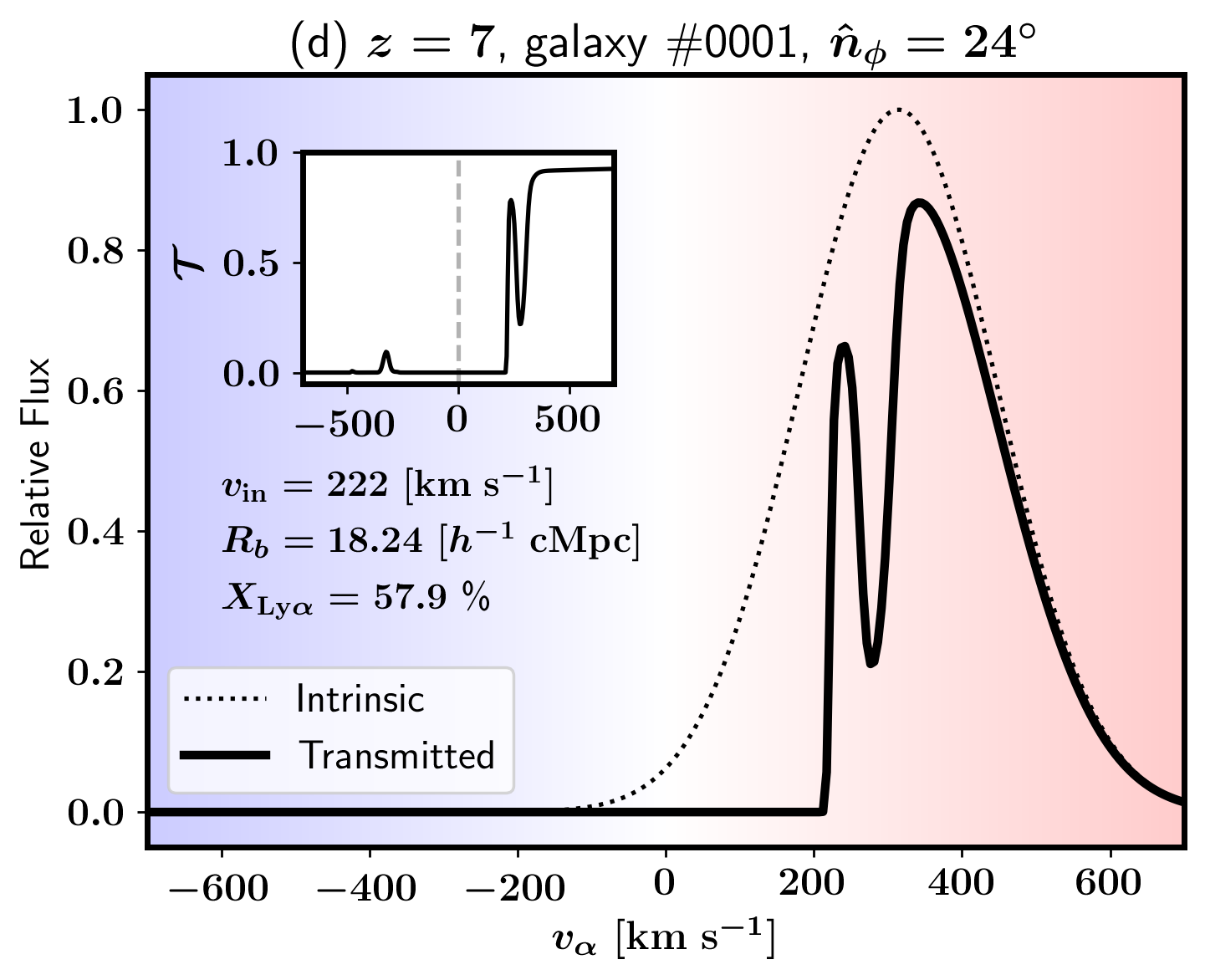}
\caption{Spectral shape of the model's intrinsic Ly$\alpha$ emission from the circumgalactic medium (dotted line) and transmitted emission through the intergalactic medium (solid line) along with the transmission curve in the inset panel for several sight lines on the $xy$ planes of galaxies \#0001 and \#0100 at $z=7$.  Panels $a$, $b$, and $c$ are for the azimuthal angles of $\phi=5^\circ,~190^\circ,$ and $193^\circ$ from galaxy \#0100, respectively, and Panel $d$ is for $24^\circ$ from galaxy \#0001, respectively (see Figure~\ref{fig:nHmap2}). The maximum infall velocity $v_{\rm in}$, HII bubble size $R_b$, and Ly$\alpha$ line transmissivity $X_{{\rm Ly}\alpha}$ for each sight line are given in the corresponding panel. } 
\label{fig:FluxH100}
\end{center}
\end{figure*}

We first visually inspect the transmission curves and pick several individuals that give useful insights into IGM transmissivity. In Figure~\ref{fig:FluxH100}, we plot the $F_{\rm in}$, $F_{\rm out}$, and $\mathcal{T}$ for several notable sight lines on the $xy$ planes of galaxies \#0001 and \#0100 at $z=7$. We show these sight lines as black dotted lines in Figure~\ref{fig:nHmap2}.

In Figures~\ref{fig:FluxH100}$a$ and $b$, We compare two sight lines in almost opposite directions ($\phi=5^\circ$ versus $190^\circ$) to galaxy \#0100 at $z=7$. Although the former case has a larger distance to the neutral region ($R_b=11$ versus $3.7~h^{-1}~{\rm cMpc}$), it has a lower transmitted fraction $X_{{\rm Ly}\alpha}$ (35\% versus 46\%) because it has a higher infall velocity (183 versus 114 ${\rm km}~{\rm s}^{-1}$). This comparison highlights that the infall velocity is another crucial factor for the IGM transmission and its sight-line variation. 

In Figure~\ref{fig:FluxH100}$b$ and $c$, we compare two sight lines that are only $3^\circ$ apart ($\phi=190^\circ$ versus $193^\circ$). Despite the small difference in  direction, the latter case has a much stronger damping-wing absorption ($R_b=3.7$ versus $0.26~h^{-1}~{\rm cMpc}$), which results in a more than four times lower transmission ($X_{{\rm Ly}\alpha}=46\%$ versus $11\%$). This dramatic difference is due to a self-shielded system that is located in the direction of $\phi=193^\circ$ from the galaxy (lower-right panel of Figure~\ref{fig:nHmap2}).

Figure~\ref{fig:FluxH100}$d$ shows an interesting case in which the transmitted spectrum is double peaked even though the intrinsic spectrum is single peaked. This "pseudo" blue peak occurs due to a leakage feature in the transmission curve, which appears as a sharp peak in the inset panel. This sight line is in the $\phi=24^\circ$ direction from galaxy \#0001 (upper-right panel of Figure~\ref{fig:nHmap2}), which goes through an extremely low HI density blob with $n_{\rm HI}\lesssim 10^{-10}~{\rm cm}^{-3}$ near the virial radius shown as a dark red spot on the HI density map. We find that the temperature of this blob is $\sim 10^6$ K, implying that the blob was created by a supernova explosion in a neighboring galaxy. This double-peak feature appears for other adjacent sight lines that go through the optically thin spot. We find similar features for the sight lines between $\phi=15^\circ$ and $30^\circ$ for this galaxy on the $xy$ plane. From visual inspections of several other galaxies, we find a few percent of sight lines exhibit similar features. We a leave more comprehensive analysis of this pseudo-double-peak feature for future studies.

\subsubsection{Global Statistics}

\begin{table}[]
\centering
\caption{Median and 68\% Range of IGM Transmissivity $X_{{\rm Ly}\alpha}$}
\label{tab:Xa}
\begin{tabular}{@{}c|c|c|c|c@{}}
\toprule
$M_{\rm UV}$ & -22  & -21 & -20 & -19 \\  \midrule
$z=6$ & $58.8^{+17.9}_{-9.9}\%$  & $59.3^{+19.8}_{-22.8}\%$ & $55.7^{+22.8}_{-23.1}\%$ & $49.8^{+26.1}_{-23.5}\%$ \\ \midrule
$z=7$ & $44.5^{+12.5}_{-12.1}\%$ & $38.4^{+19.3}_{-16.7}\%$ & $32.2^{+18.8}_{-14.8}\%$ & $27.5^{+19.9}_{-15.1}\%$ \\  \midrule
$z=8$ & N/A & $19.2^{+10.6}_{-8.8}\%$ & $12.3^{+10.3}_{-7.9}\%$ & $7.3^{+9.2}_{-5.2}\%$ \\ \bottomrule
\end{tabular}
\end{table}
%


\begin{table}[]
\centering
\caption{IGM Transmissivity Relative to $z=6$}
\label{tab:damping-wing}
\begin{tabular}{@{}c|c|c|c|c@{}}
\toprule
$M_{\rm UV}$ & -22   & -21   & -20 &  -19  \\ \midrule
$\left<X_{{\rm Ly}\alpha}\right>_{z=7}$/$\left<X_{{\rm Ly}\alpha}\right>_{z=6}$ & 77\% & 65\% & 58\% & 55\% \\ \midrule
$\left<X_{{\rm Ly}\alpha}\right>_{z=8}$/$\left<X_{{\rm Ly}\alpha}\right>_{z=6}$ & N/A   & 32\% & 22\% & 15\% \\ \bottomrule
\end{tabular}
\end{table}

\begin{figure*}
\begin{center}
\includegraphics[scale=0.39]{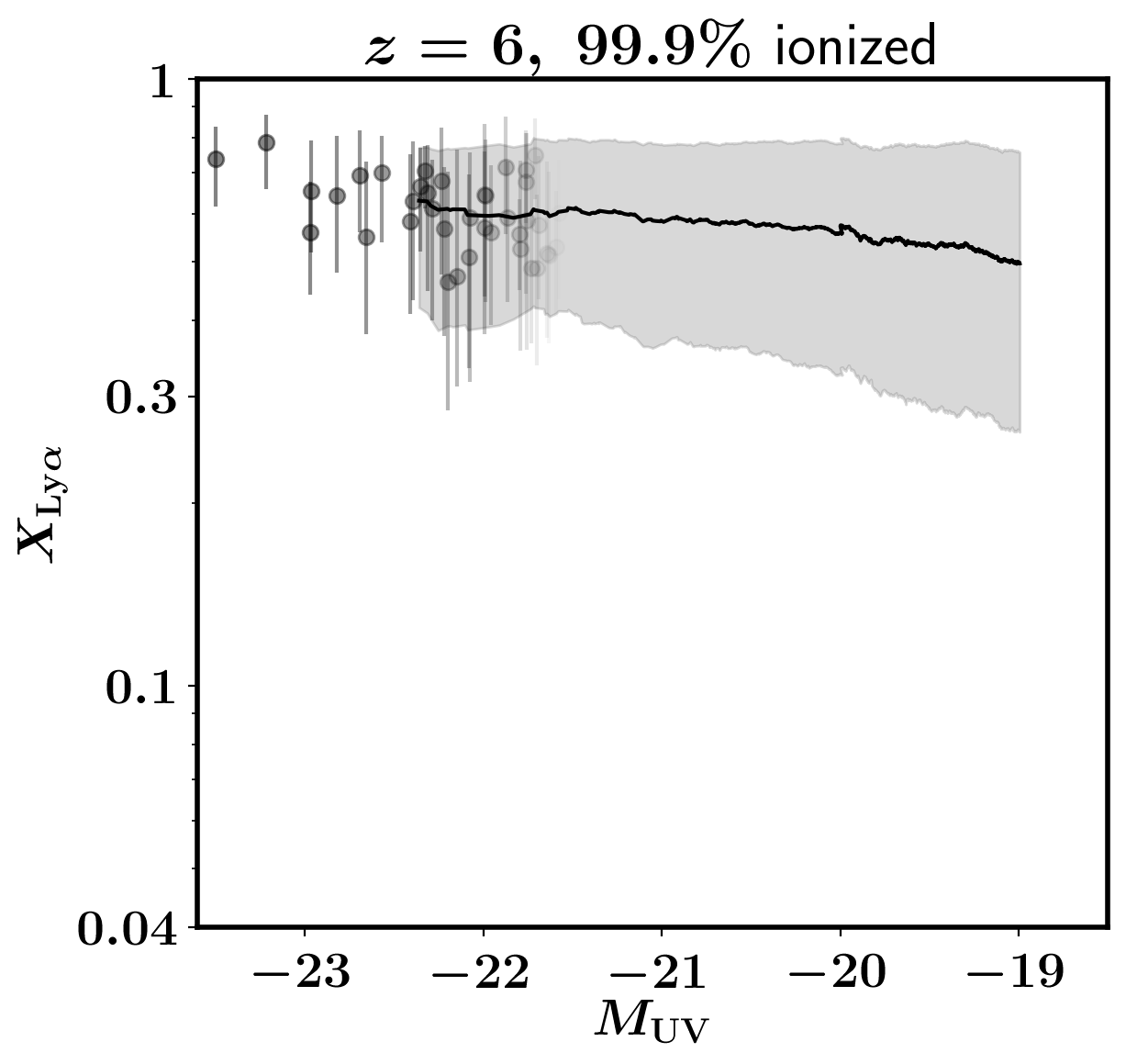}
\includegraphics[scale=0.39]{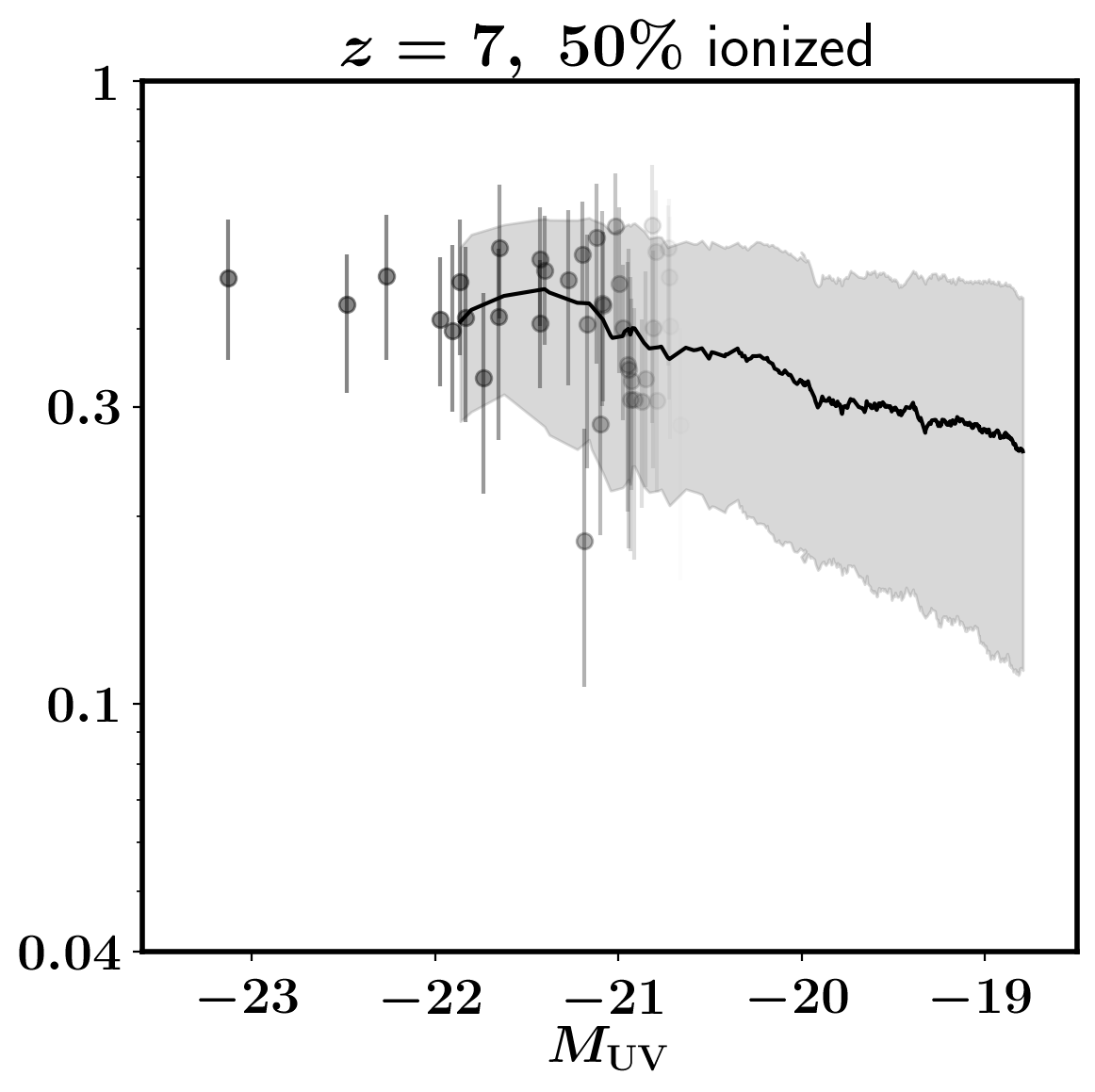}
\includegraphics[scale=0.39]{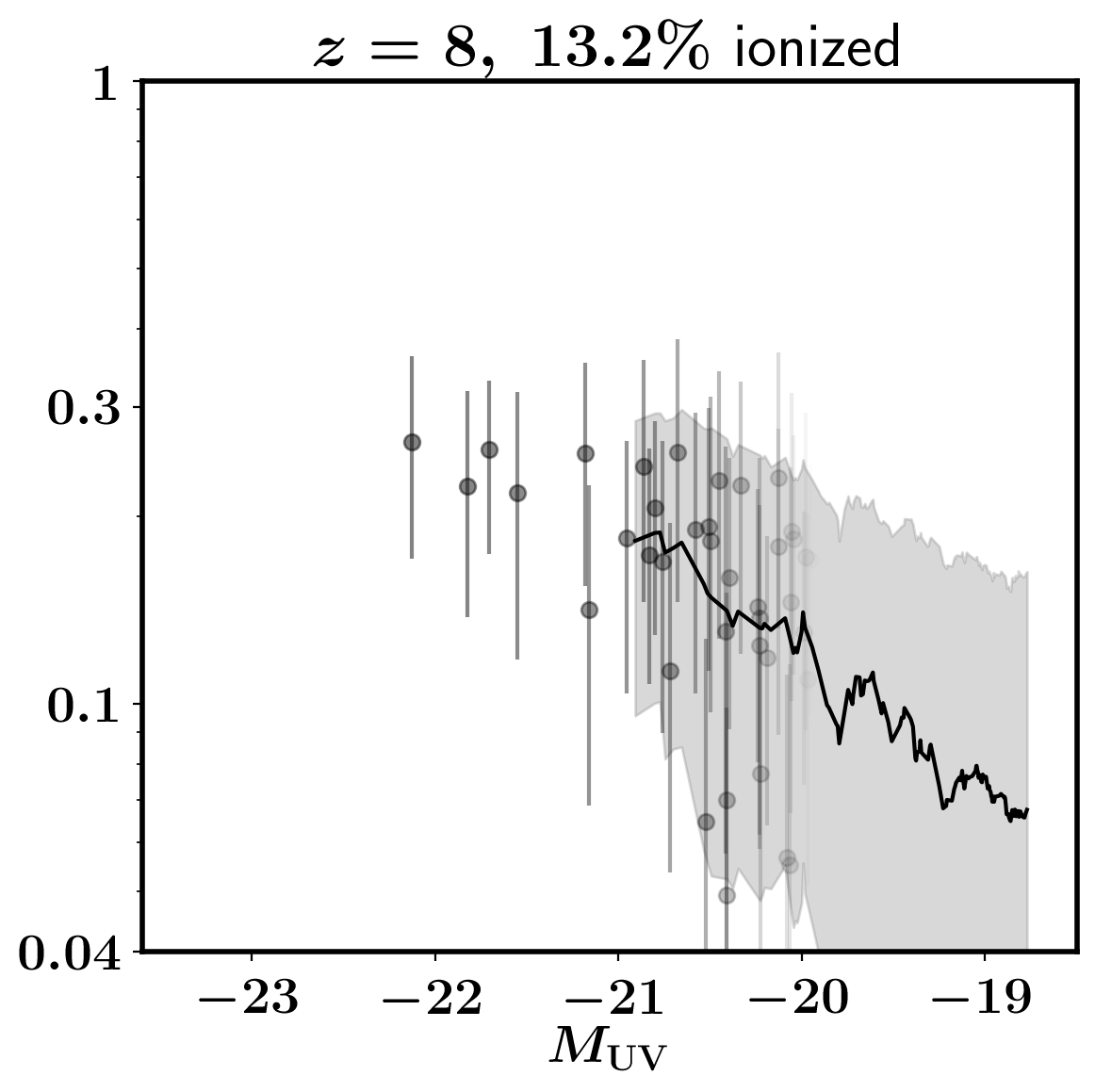}
\caption{Ly$\alpha$ transmissivity $X_{{\rm Ly}\alpha}$ for sample galaxies in the $z=6$ (left), $7$ (middle), and $8$ (right panel) snapshots. Each data point with an error describes the sight-line variation of an individual halo. The black solid line marks the median of $X_{{\rm Ly}\alpha}$ at given $M_{\rm UV}$. The gray shades bracket the 1$\sigma$ range of the combined sight-line-to-sight-line and galaxy-to-galaxy variation. Individual data points are shown only for relatively rare bright galaxies to avoid crowding the figure with too many data points.} 
\label{fig:Xa_vs_MUV}
\end{center}
\end{figure*}

We calculate $X_{{\rm Ly}\alpha}$ for 2000 sight lines for the galaxies with $M_{\rm UV}\lesssim -19$. We show the full statistics of the results in Figure~\ref{fig:Xa_vs_MUV}, where we show $X_{{\rm Ly}\alpha}$ as a function of $M_{\rm UV}$ with the galaxy-to-galaxy and sight-line-to-sight-line variations. Similarly to in Figure~\ref{fig:Rb_stat}, the error bars on the data points show the 68\% range for individual galaxies. The solid line marks the median value for the corresponding $M_{\rm UV}$, and the gray shade covers the 68\% range with the galaxy-to-galaxy and sight-line-to-sight-line variations combined. We list the $1\sigma$ range for $M_{\rm UV}=$ -22, -21, -20, and -19 at $z=6$, $7$, and $8$ in Table~\ref{tab:Xa}.

First, $X_{{\rm Ly}\alpha}$ decreases toward high $z$. For example, the median at $M_{\rm UV}=-21$ is $59.3$, $38.4$, and $19.2~\%$ at $z=6$, 7, and 8, where the IGM is 99.9\%, 50\%, and 13.2\% ionized, respectively. This is in accordance with the common expectation that $X_{{\rm Ly}\alpha}$ is a useful probe of the IGM neutral fraction.

It is also notable that $X_{{\rm Ly}\alpha}$ is well below 100\% at $z=6$, where the IGM is nearly 100\% ionized. The IGM absorption at $z=6$ should be entirely due to the resonance scattering by infalling gas given that the IGM is nearly fully ionized. Thus, we take the $z=6$ results as the fully ionized limit and calculate the relative change in $X_{{\rm Ly}\alpha}$ at higher redshifts. That is, we take the ratio of the median at $z=7$ or $8$, $\left<X_{{\rm Ly}\alpha}\right>_{z=7~{\rm or}~8}$, to that at $z=6$, $\left<X_{{\rm Ly}\alpha}\right>_{z=6}$, in order to separate out the damping-wing opacity from the neutral IGM. For example, the median transmission fraction of galaxies with $M_{\rm UV}=-21$ is suppressed to 65\% and 32\% at $z=7$ and $8$, respectively, compared to that at $z=6$. We list this ratio for $M_{\rm UV}=-22$, $-21$, $-20$, and $-19$ in Table~\ref{tab:damping-wing}. Both Figure~\ref{fig:Xa_vs_MUV} and Table~\ref{tab:damping-wing} show that UV-fainter galaxies show a stronger suppression in $X_{{\rm Ly}\alpha}$ than the brighter ones do. For example, the median transmission fraction of $M_{\rm UV}=-21$ galaxies drops to $35\%$ from $z=6$ to $8$, while that of $M_{\rm UV}=-19$ galaxies  drops to $15\%$.

\begin{figure*}
\begin{center}
\includegraphics[scale=0.4]{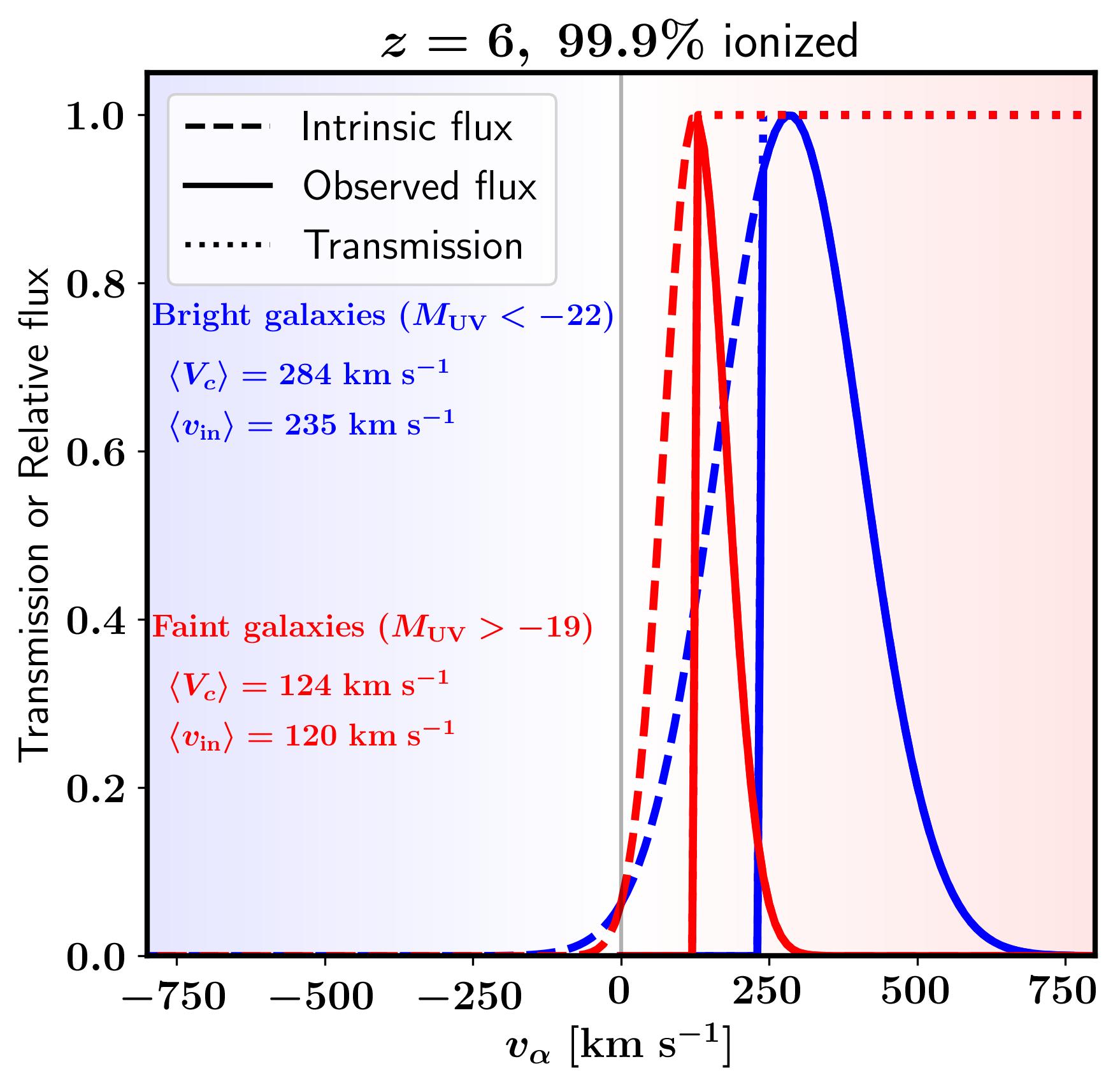}
\includegraphics[scale=0.4]{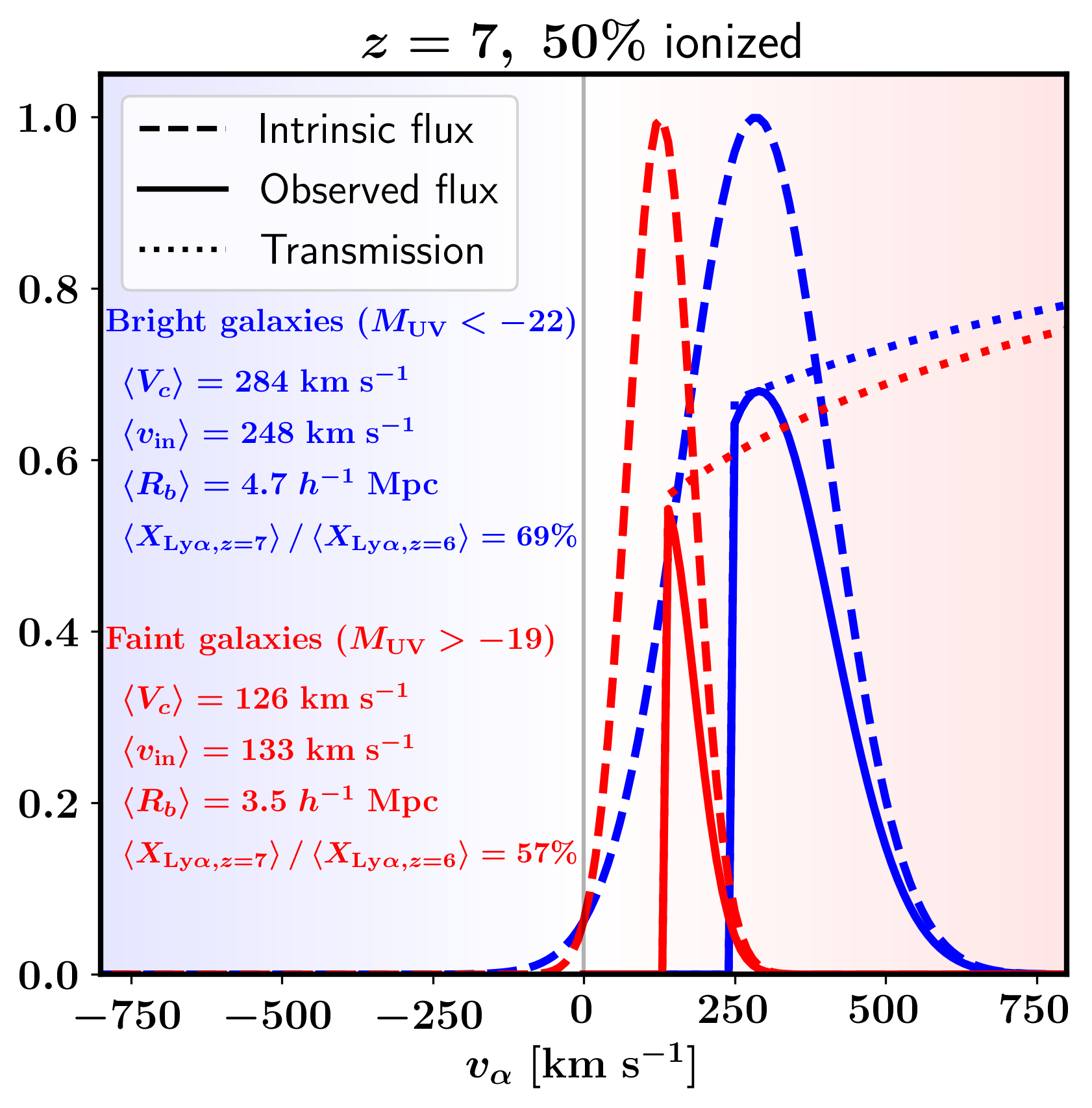}
\includegraphics[scale=0.4]{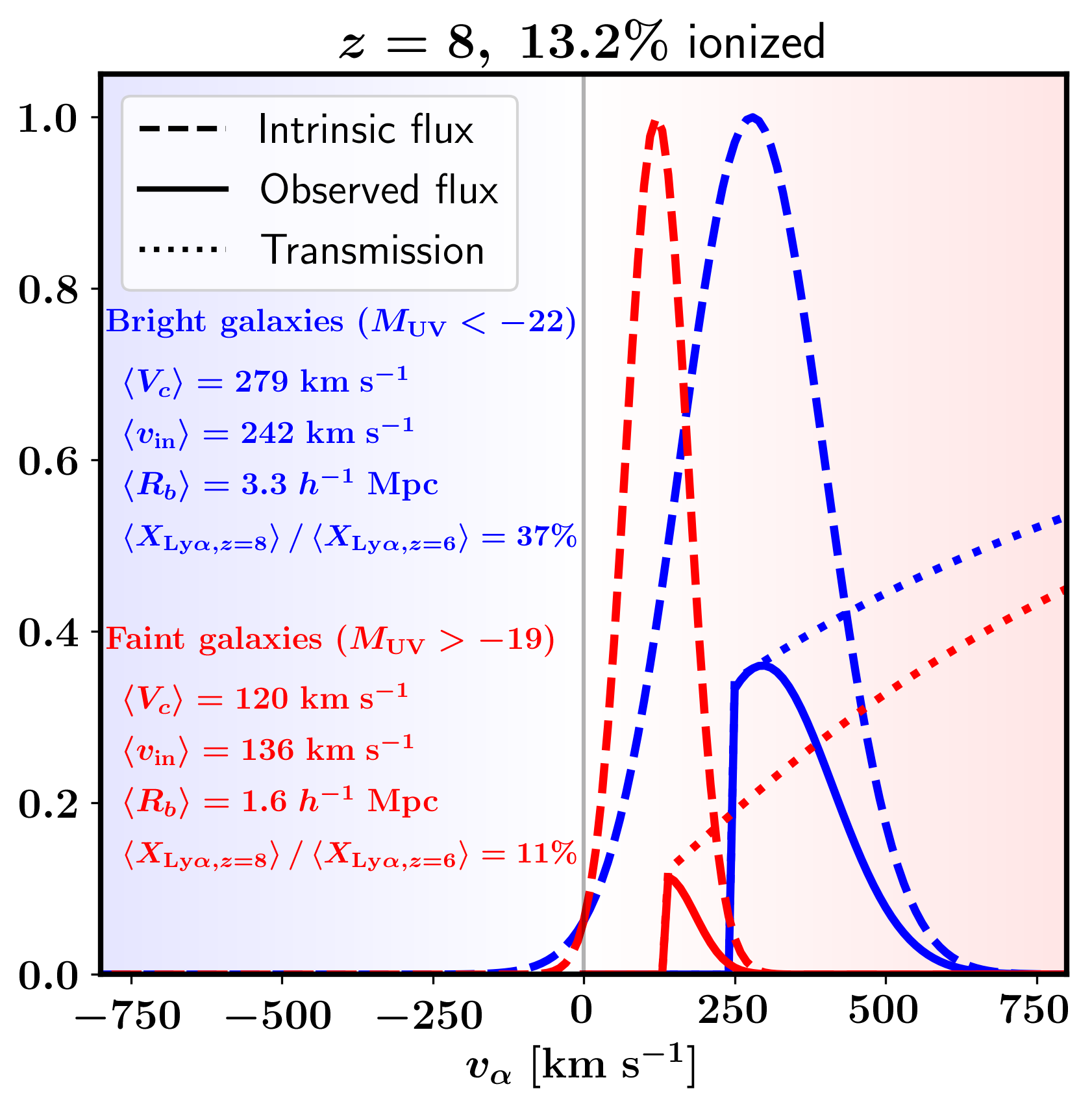}
\caption{The median emission-line profile (solid line) calculated by multiplying the median transmission (dotted line) to the intrinsic profile (dashed line). The red and blue lines show the results for the bright ($M_{\rm UV}<-22$) and faint ($M_{\rm UV}>-19$) galaxy groups, respectively. The left, middle, and right panels show the results for $z=6$, $7$, and $8$ galaxies, respectively. The median transmission curve is constructed using the median infall velocity $v_{\rm in}$ and effective HII bubble size $R_b$ for each group, the values of which are shown on the plot. The intrinsic profile is given by Eq.~\ref{eq:intrinsic}, where the circular velocity $V_c$ is given by the median $M_h$ of each group. The integrated line flux at $z=7$ or $8$ with respect to $z=6$ is also shown on the plot for each group in percent. } 
\label{fig:typical_Tr}
\end{center}
\end{figure*}

The $M_{\rm UV}$ dependence of $X_{{\rm Ly}\alpha}$ is often attributed to the difference in HII bubble size between bright and faint galaxies. Interestingly, we find that the bubble size only partially explains the dependence. According to Table~\ref{tab:Trdw}, the IGM transmissivity at a fixed frequency of $v_\alpha=400~{\rm km}~{\rm s}^{-1}$ changed from 73.1\% to 70.3\% between  $M_{\rm UV}=-21$ and $-19$ at $z=7$. This difference is smaller than the difference in $X_{{\rm Ly}\alpha}$ between the same $M_{\rm UV}$'s ($65\%$ versus $55\%$). This extra difference in $X_{{\rm Ly}\alpha}$ is attributed to the difference in the velocity offset of transmitted flux (Figure~\ref{fig:dv_MUV}), caused by the $M_{\rm UV}$ dependence of the IGM infall velocity $v_{\rm in}$ and the peak location of our intrinsic line profile. That is, brighter galaxies tend to transmit redder photons due to the stronger infall motion, and these redder photons are subject to smaller damping-wing opacity for a given HI density. 

We illustrate our understanding of the $M_{\rm UV}$ dependence of $X_{{\rm Ly}\alpha}$ in Figure~\ref{fig:typical_Tr}, where we show the median (transmitted) emission line profile for bright ($M_{\rm UV}<-22$) and faint ($M_{\rm UV}>-19$) galaxy groups at $z=6$, $7$, and $8$. We first construct the median transmission curve for each group using the median of $v_{\rm in}$ and $R_b$ for each group. Then, multiply the median transmission by the intrinsic emission profile obtained from Equation~(\ref{eq:intrinsic}) using the median circular velocity. The line flux relative to $z=6$ drops to 70\% (37\%) for the bright group while to 58\% (12\%) for the faint groups at $z=7~(8)$. The figure shows both bubble size and the wavelength of the transmitted photons create an $M_{\rm UV}$ dependence in the damping-wing opacity as we explained above.

We also highlight the large scatter in the $X_{{\rm Ly}\alpha}-M_{\rm UV}$ relation. Due to the complicated nature of Ly$\alpha$ radiative processes, galaxies at the same redshift with the same brightness can transmit a significantly different number of Ly$\alpha$ photons from sight line to sight line and from galaxy to galaxy. As $X_{{\rm Ly}\alpha}$ decreases toward low brightness and high redshift, the variation in $X_{{\rm Ly}\alpha}$ becomes larger relative to the median. The 1$\sigma$ range of $X_{{\rm Ly}\alpha}$ for $M_{\rm UV}=-21$ at $z=7$ is $59.3^{+19.8}_{-22.8}\%$, while that for $M_{\rm UV}=-20$ at $z=8$ is $X_{{\rm Ly}\alpha}=7.3^{+9.2}_{-5.2}\%$ (see Tab.~\ref{tab:Xa}). The ratio of the 1$\sigma$ range to the median is $43/59\approx 0.7$ in the former case, while $14.4/7.3\approx 2$ in the latter case. This means that one needs to measure Ly$\alpha$ line strengths of at least 50 (400) galaxies to constrain $\left< X_{{\rm Ly}\alpha} \right>$ to 10\% in the former (latter) case assuming perfectly accurate measurements.

\section{Summary and Discussion} \label{sec:discussion}

We have computed the IGM transmission for the Ly$\alpha$ emission line from star-forming galaxies during reionization from the Cosmic Dawn II simulation to understand what aspects of the IGM determine the transmission. Thanks to the star-formation physics and the subsequent radiative transfer implemented in a large box with high resolution, we were able to compute the IGM transmissivity for galaxies with  $M_{{\rm AB}1600}\gtrsim -22$ while accounting for the small-scale physics around the virial radii of the galaxies. We find that the IGM infall motion, HII bubble size, and self-shielded neutral gas systems are the critical factors. We discuss the implications of our results for the current understandings of LAEs below.

\subsection{Significance of Infall Motion for the IGM Transmissivity} 

Previous works have repeatedly reported the impact of the infalling IGM on the Ly$\alpha$ transmission \citep{2004MNRAS.349.1137S,2007MNRAS.377.1175D,2008MNRAS.391...63I,2011MNRAS.410..830D,2011ApJ...728...52L,2013MNRAS.429.1695B,2017ApJ...839...44S,2019MNRAS.485.1350W,2021MNRAS.504.1902G}. Although galactic outflow from supernova explosions may extend beyond the virial radius for some occasions, the outflow is confined within the virial radius in most sight lines \citep[e.g.,][]{2013ApJ...765...89S}. Thus, the peculiar motion of IGM outside the virial radius is dominated by the gravitational field of the source galaxy. The gravitationally infalling IGM suppresses a significant portion of the red-peak emission even after reionization has ended, requiring higher intrinsic Ly$\alpha$ luminosity to explain the observed Ly$\alpha$ luminosity function \citep{2011MNRAS.410..830D}. The absorption by infalling IGM extends to $v_\alpha\sim 100~{\rm km}~{\rm s}^{-1}$ in \cite{2011ApJ...728...52L} and \cite{2021MNRAS.504.1902G}, for example, while it goes to $v_\alpha\gtrsim 200~{\rm km}~{\rm s}^{-1}$ in our work and \citet{2019MNRAS.485.1350W}. This difference likely resulted from larger simulation box sizes of the latter two studies, which involve more massive galaxies with stronger infall motion. 

The truncation frequency of the IGM transmission is given by the maximum infall velocity along the line of sight. In most sight lines, the infall velocity peaks around the virial radius of the galaxy. If one uses a distance larger than the virial radius to define the IGM, the maximum infall velocity would decrease. In this case, some of the absorption by the infalling IGM in this work would be regarded as the absorption by the CGM. This is illustrated in Figure 6 of \citet{2019MNRAS.485.1350W}, where they define gas at $R_{\rm vir}<s<5 R_{\rm vir}$, $5R_{\rm vir}<s<10 R_{\rm vir}$, and $s>10R_{\rm vir}$ as the ``inner CGM'',  ``outer CGM'', and ``IGM'', respectively. All three components are regarded as the IGM in this work.

We note that AGNs, which the CoDa II simulation does not include, can result in a much higher transmission owing to much stronger outflow and ionizing radiation. Jets from AGNs can disrupt the infall motion well beyond the virial radius, and the HI density is low enough to be transparent at the resonance ($<10^{-10}~{\rm cm}^{-3}$) within the proximity zones. Indeed, low-$z$ observations show that luminous LAEs are often associated with AGNs \citep{2016ApJ...823...20K}.
Neighboring galaxies within the proximity zone can also receive a boost in the Ly$\alpha$ transmission as was reported by \citet{2020ApJ...896...49B}. Also, the X-ray from AGNs can change the reionization morphology by partially ionizing the surrounding HI region \citep{2017MNRAS.468.3718K}. 

The relation between $v_{\rm in}$ and the galaxy mass (or UV magnitude) does not evolve much with redshift as the infall motion is purely gravitational in nature. However, $v_{\rm in}$ can still affect the IGM transmission by setting the minimum wavelength of transmitted photons. Stronger infall motion results in the transmission of redder photons, which experience less damp-wing opacity. As a result, Ly$\alpha$ emission from brighter galaxies, which are likely more massive, are less subject to increasing neutral IGM fraction toward higher redshifts.

The infall motion varies depending on the viewing angle due to the complexity of density structures around the source galaxy. This variation adds to the statistical uncertainty of the transmitted flux from a single observation. We find the infall velocity typically has a scatter of $\sim 40~{\rm km}~{\rm s}^{-1}$ from sight line to sight line (Figure~\ref{fig:dv_MUV}). The sight-line difference in the infall velocity can overwhelm that in the damping-wing opacity in determining the total IGM transmissivity (e.g., Figure~\ref{fig:FluxH100}$a$ versus $b$).

\subsection{Observational Signature of the Infall Motion} 

Truncation of the red peak by the infalling IGM can make the spectral shape of the transmitted flux blue-skewed in frequency even if the intrinsic emission is symmetric (Figure~\ref{fig:FluxH100}). Such blue-skewed spectra are often observed from recent high-$z$ spectroscopic LAE surveys \citep[see, e.g., Figure 4 of][]{2020ApJ...904..144J}.

The infall motion is a potential explanation for the offset between the observed line and the systemic redshift of the source galaxy. The absorption by infalling opaque IGM will force the peak to be on the red side of $v_\alpha=v_{\rm in}\approx V_c$ regardless of the intrinsic emission. Given the correlation between the UV magnitude and the infall velocity (Figure~\ref{fig:dv_MUV}), the offset is expected to correlate with the UV magnitude, and this is supported by existing observations \citep[e.g., see Figure 10 of][]{2019PASJ...71...71H}.

The pseudo-blue-peak feature described in Figure~\ref{fig:FluxH100}$d$ is an interesting possible outcome of infalling IGM. Given that the IGM is unlikely to transmit photons from the blue side of Ly$\alpha$ at $z\gtrsim 6.5$ \citep[Figure~\ref{fig:TrH1}; also see][]{2021MNRAS.508.3697G}, the currently existing double-peak cases at $z\sim 6.5$ without exact systemic redshifts \citep{2018A&A...619A.136M, 2018ApJ...859...91S,  2021MNRAS.500..558M} may be this case. The line shape of our case is the most similar to NEPLA4 of \citet[][see their Figure 12]{2018ApJ...859...91S}, although the overall shape is about two times narrower in our case. The peak separation of double-peaked LAEs ranges from 100 to 300 ${\rm km}~{\rm s}^{-1}$, which is narrower than the reported value of $\sim 750~{\rm km}~{\rm s}^{-1}$ from $z=2-3$ LAEs by \citet{2012ApJ...745...33K}. The pseudo-blue-peak scenario can explain the narrow separation because the two peaks come from incomplete resonance absorption of a single red peak. We also note that more recent observation with improved spectral resolution by \citet{2021MNRAS.505.1382M} do find lots of narrowly separated double peaks at $z=2-3$. If any of the reported $z>6$ double-peaked LAE falls into the pseudo-double-peak case, its systemic redshift will be found blueward of both peaks in future observation\footnote{\citet{2018A&A...619A.136M} also suggested a model that produces a double-peak feature redward of the systemic redshift in their Figure 9. However, the suggested HI density configuration is different from our case.}.


\subsection{Observational Impacts of Self-shielded Systems} 

Before reionization, mini halos below $10^8~M_\odot$ could accrete gas from the cold IGM and develop dense gas clumps inside their potential wells \citep{2003AIPC..666...89S,2004MNRAS.348..753S}. Upon reionization, these clumps were exposed to the ionizing background radiation and started evaporating. However, the evaporation took up to several hundred megayears to finish for relatively massive ones \citep{2005MNRAS.361..405I,2016ApJ...831...86P,2020ApJ...898..149D} and, thus, some of the clumps are expected to have existed within HII regions and attenuated Ly$\alpha$ emission for incoming sight lines \citep{2013MNRAS.429.1695B,2015MNRAS.446..566M,2016MNRAS.463.4019K}. 
Previous works relied on analytical or subgrid models of these clumps to demonstrate their impact on large-scale Ly$\alpha$ transmission. 

We have revisited this subject with the CoDa II simulation, which directly resolves these dense clumps at (proper) kiloparsec scales while simultaneously reproducing the HII bubbles as large as several (proper) Mpc. We report that the self-shielded systems do not seem to have a large impact on the average transmission, but they seem to be responsible for a large fraction of sight lines with unusually low transmission. Thus, Ly$\alpha$ nondetections from UV-bright galaxies can be at least partially attributed to these dense clumps blocking the sight line. We note that the spatial resolution of CoDa II may have underestimated the opacity from these systems in this work given that high-resolution simulation studies \citep[e.g.,][]{2021arXiv210804837N} suggest there are a lot of self-shielded absorbers smaller than the cell size of CoDa II ($16~h^{-1}~{\rm ckpc}$).

The statistics of these clumps depend on multiple parameters including the ionizing background radiation intensity, local overdensity, timing of reionization, X-ray preheating, and baryon dark matter streaming motion \citep{2015MNRAS.446..566M,2016ApJ...831...86P,2020ApJ...898..149D,2020ApJ...898..168C,2021ApJ...908...96P}. The first three parameters are especially relevant to relatively massive ($\gtrsim 10^6~M_\odot$) systems that can survive the ionizing background radiation for more than $10^7$ yr. The IGM opacity would also depend on these parameters if the unresolved self-shielded systems add a significant amount of opacity to the IGM. Thus, it is worth exploring the Lyman-limit opacity of these absorbers in future studies dedicated to these small-scale structures.

\subsection{$M_{\rm UV}$ Dependence of IGM Transmissivity} \label{sec:Muv-dependence}

A number of past observations have reported that Ly$\alpha$ emission from relatively brighter galaxies tends to be less affected by the neutral IGM during reionization than fainter ones do \citep[e.g.,][Jung, I. et al. 2021 in preparation]{2010ApJ...725L.205F,2011ApJ...728L...2S,2011ApJ...743..132P,2012ApJ...744...83O,2012ApJ...760..128M, 2012MNRAS.422.1425C,2013ApJ...775L..29T,2014ApJ...794....5T,2021MNRAS.502.6044E}. We confirm this trend from our results in Figure~\ref{fig:Xa_vs_MUV} and Table~\ref{tab:damping-wing}). We illustrate our results by calculating the ``median emission profile'' for for bright ($M_{\rm UV}<-22$) and faint ($M_{\rm UV}>-19$) galaxies in Figure~\ref{fig:typical_Tr}, which is obtained by applying the median $v_{\rm in}$ and $R_b$ to the intrinsic emission profile created for the median $M_h$.

Based on Figure~\ref{fig:typical_Tr}, we conclude that the $M_{\rm UV}$ dependence of the transmission is driven by two factors. First, bright galaxies tend to reside in HII regions larger than where the fainter galaxies reside ($R_b=4.7$ versus $3.5~h^{-1}~{\rm cMpc}$ at $z=7$) because they can ionize larger volumes and are more likely to form in overdense regions, where numerous neighboring galaxies power the HII bubble together \citep{2019ApJ...879...36F,2021MNRAS.502.6044E}. Second, bright galaxies tend to transmit in longer wavelengths because they have a larger velocity offset in their intrinsic emission profile \citep[][ also compare $V_c$ in Figure~\ref{fig:typical_Tr}]{2018ApJ...857L..11M,2021MNRAS.504.1902G} and higher IGM infall velocity ($v_{\rm in}=248$ versus $138~{\rm km}~{\rm s}^{-1}$ at $z=7$) suppressing the transmission up to a longer wavelength. Often the former is considered to be the major reason for the $M_{\rm UV}$ dependence, but we find the latter is comparably responsible for the dependence. We also find that the $M_{\rm UV}$ dependence grows stronger toward high $z$. We find the median transmission relative to $z=6$ is $69\%$ versus $57\%$ at $z=7$ while it is 37\% versus 11\% at $z=8$. This highlights the importance of dealing with the dependence carefully in high-$z$ LAE surveys to avoid biasing the IGM neutral fraction estimation. 

The infall velocity $v_{\rm in}$ and the circular velocity $V_c$ stay nearly constant between $z=6$ and $8$, and thus do not directly contribute to the IGM transmission evolution. However, the strength of the $M_{\rm UV}$ dependence depends on the wavelength of the transmitted flux set by those two parameters. 
In the Appendix, we demonstrate the flux-weighted transmission and its $M_{\rm UV}$ dependence can be strengthened or weakened by modest amounts if we shrink or expand the intrinsic profiles, respectively. However, our main conclusion that the IGM transmission for brighter galaxies is less suppressed by the neutral IGM remains solid in this analysis.

\begin{table}[]
\centering
\caption{68\% Range to Median Ratio of $X_{{\rm Ly}\alpha}$}
\label{tab:68tomed}
\begin{tabular}{@{}c|c|c|c|c@{}}
\toprule
$M_{\rm UV}$ & -22  & -21 & -20 & -19 \\  \midrule
$z=6$ & $0.47$  & $0.72$ & $0.82$ & $1.00$ \\ \midrule
$z=7$ & $0.55$ & $0.94$ & $1.04$ & $1.27$ \\  \midrule
$z=8$ & N/A & $1.01$ & $1.48$ & $1.97$ \\ \bottomrule
\end{tabular}
\end{table}

\subsection{Variation in IGM Transmissivity and Its Implication for Observations} 

While an apparent decline in Ly$\alpha$ visibility at $z>6$ has been interpreted as a sign of an increasing IGM neutral fraction, the complex nature of Ly$\alpha$ radiative processes in the IGM limits the measurement accuracy of $\bar{x}_{\rm HI}$ based on the Ly$\alpha$ emission strength. The IGM infall velocity and size of HII regions vary from sight line to sight line and galaxy to galaxy even for galaxies with the same mass or brightness, and self-shielded systems occasionally block unfortunate sight lines. These uncertainties make it difficult to pin down the average IGM transmissivity at a given redshift with a small number of measurements. Moreover, the intrinsic Ly$\alpha$ emission from the CGM appears to also depend on not only the viewing angle but also various characteristics of galaxies \citep[e.g.,][]{2021MNRAS.505.1382M,2021MNRAS.503.4105T} as theoretically expected \citep[e.g.,][Song, H. et al. in preparation]{2012A&A...546A.111V,2019MNRAS.484...39S}.

We show in Table~\ref{tab:68tomed} the ratio between the 68\% range and the median value of $X_{{\rm Ly}\alpha}$ based on Table~\ref{tab:Xa}. The ratio is larger for fainter galaxies at higher redshifts. For galaxies with $M_{\rm UV}\ge-21$ at $z\ge7$, the ratio is $\gtrsim 1$, implying that we need at least a hundred measurements of Ly$\alpha$ emission strength to constrain the mean IGM transmissivity to $10\%$ accuracy with those galaxies. In practice, the required sample size is likely much larger because most of the measurements would yield only nondetections, and there will be extra uncertainty in the intrinsic emission from the CGM.

Recent state-of-art LAE surveys are starting to constrain the neutral fraction of the IGM at $z\sim 7.5$, but their constraints have not converged yet: \citet{2019ApJ...878...12H}, \citet{2020ApJ...904..144J}, and \citet{2020MNRAS.495.3602W} give $\bar{x}_{\rm HI}= 88^{+5\%}_{-10\%}$, $49^{+19\%}_{-19\%}$, and $55^{+11\%}_{-13\%}$, respectively. According to our findings, these studies are based on less than a hundred galaxies, which is not enough to suppress the uncertainty from the IGM transmission variation. 

For LAEs at $z\lesssim 7$, ground-based experiments are expected to provide a large-enough sample soon \citep[e.g., SILVERRUSH;][]{2021ApJ...911...78O}. For higher redshifts, future space experiments like the Nancy Grace Roman Space Telescope (NGRST), the James Webb Space Telescope (JWST), and Euclid will substantially improve the constraints with high-quality observations. 

Because brighter galaxies have a relatively smaller transmission variation, surveying bright LAEs in a wide field may be more effective than surveying faint galaxies in a narrow field. In this regard, NGRST and Euclid may be more efficient than JWST in constraining the IGM transmissivity at high redshifts, although a definitive conclusion requires accounting for the variation in the intrinsic emission from the CGM. 

\section*{Acknowledgements}
We thank A. Inoue, M. Ouchi, N. Yoshida, K. Moriwaki, J. Matthee, C. Mason, and A. Smith for helpful comments on this paper. H.P. was supported by the World Premier International Research Center Initiative (WPI), MEXT, Japan and JSPS KAKENHI grant No. 19K23455. H.S. was supported by the Basic Science Research Program through the National Research Foundation of Korea (NRF) funded by the Ministry of Education (2020R1I1A1A01069228). Numerical computations were carried out on the gfarm computing cluster of the Kavli Institute for Physics and Mathematics of the Universe and a high-performance computing cluster at the Korea Astronomy and Space Science Institute. T.D. was supported by the National Science Foundation Graduate Research Fellowship Program under Grant No. DGE-1610403. P.R.S. was supported in part by US NSF grant AST-1009799, NASA grant NNX11AE09G, NASA/JPL grant RSA Nos. 1492788 and 1515294, and supercomputer resources from NSF XSEDE grant TG-AST090005 and the Texas Advanced Computing Center (TACC) at the University of Texas at Austin. I.T.I. was supported by the Science and Technology Facilities Council (grant Nos. ST/I000976/1 and ST/T000473/1) and the Southeast Physics Network (SEPNet). K.A. was supported by NRF-2016R1D1A1B04935414 and NRF-2016R1A5A1013277.  I.J. acknowledges support from NASA under award number 80GSFC21M0002.

\appendix
 
\section{Dependence of IGM Transmissivity on Intrinsic Emission Profile}\label{sec:IFTest}

\begin{figure*}
\begin{center}
\includegraphics[scale=0.4]{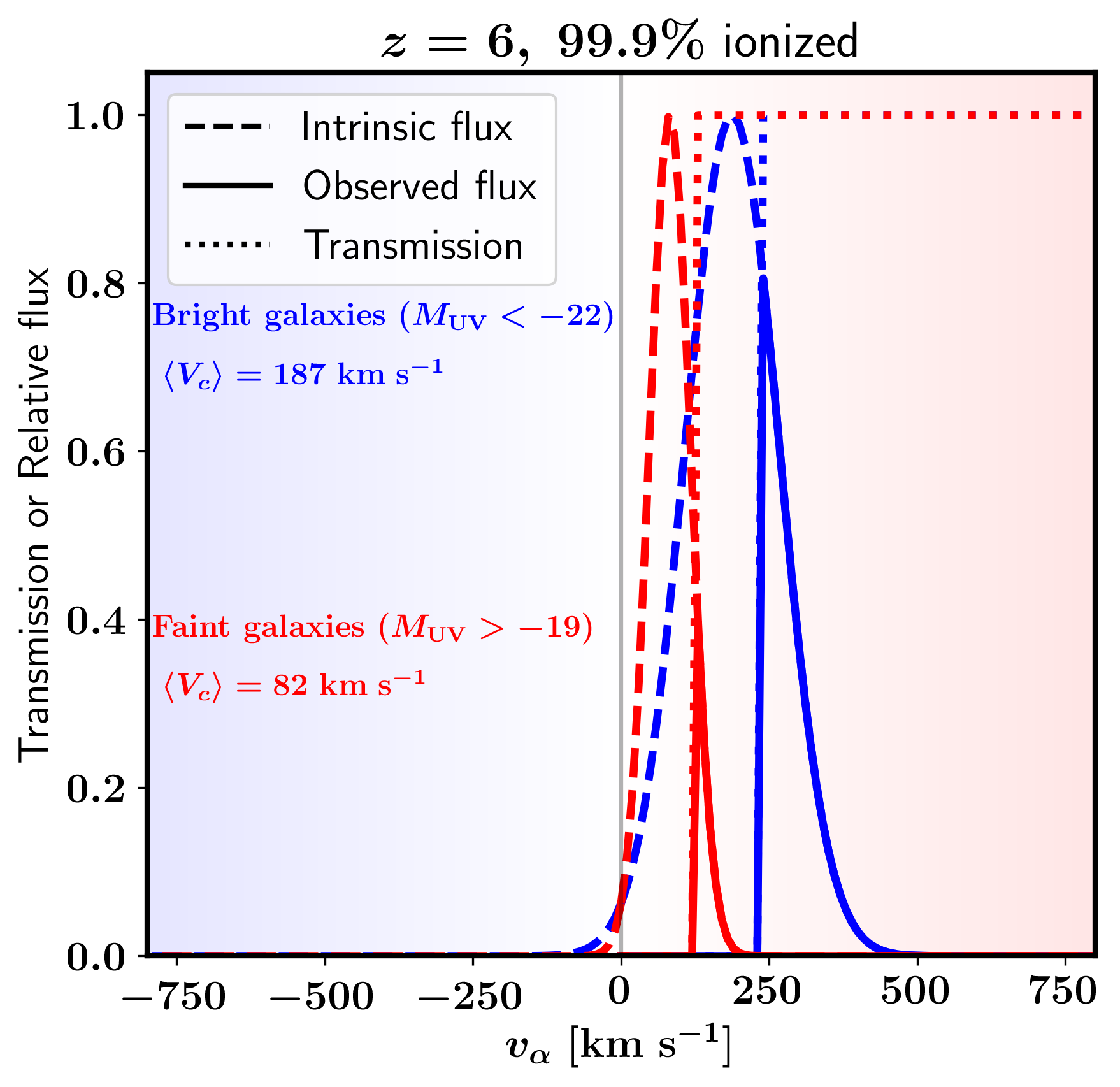}
\includegraphics[scale=0.4]{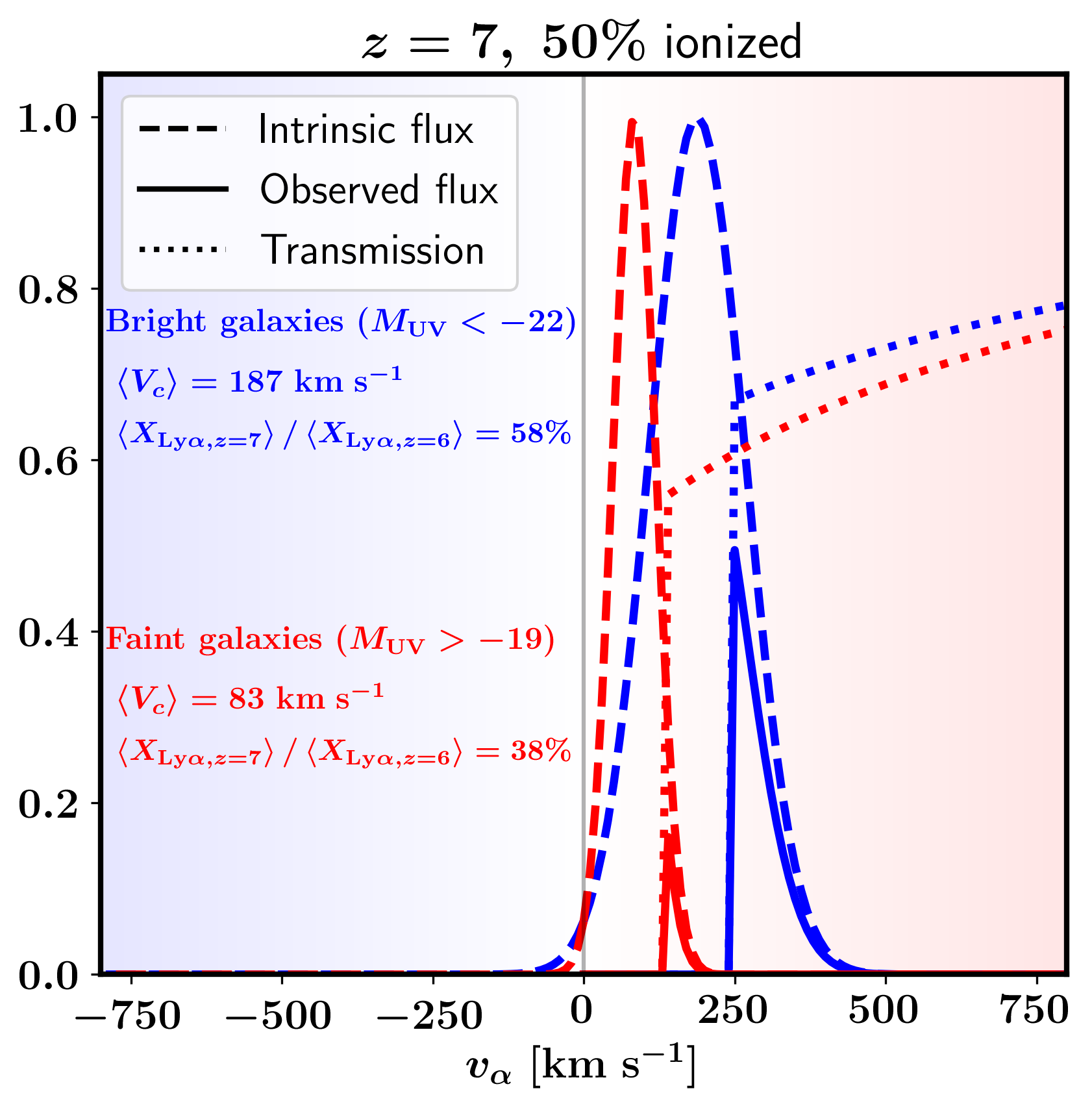}
\includegraphics[scale=0.4]{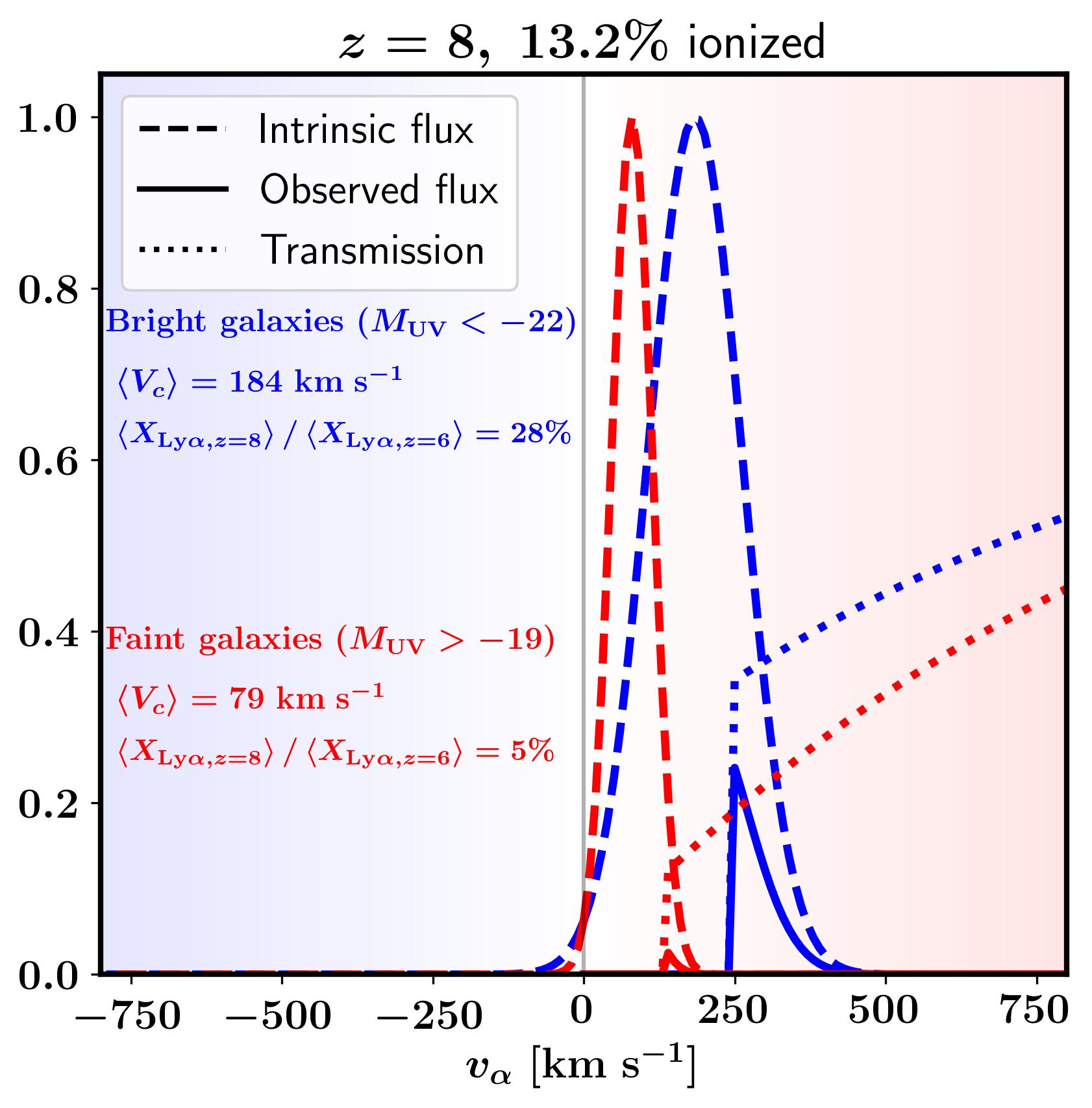}
\caption{Same as Figure~\ref{fig:typical_Tr} except that we reduce $V_c$ by 33\% in Eq.~\ref{eq:intrinsic} to shrink the intrinsic emission profile.} 
\label{fig:typical_Tr2}
\end{center}
\end{figure*}

\begin{figure*}
\begin{center}
\includegraphics[scale=0.4]{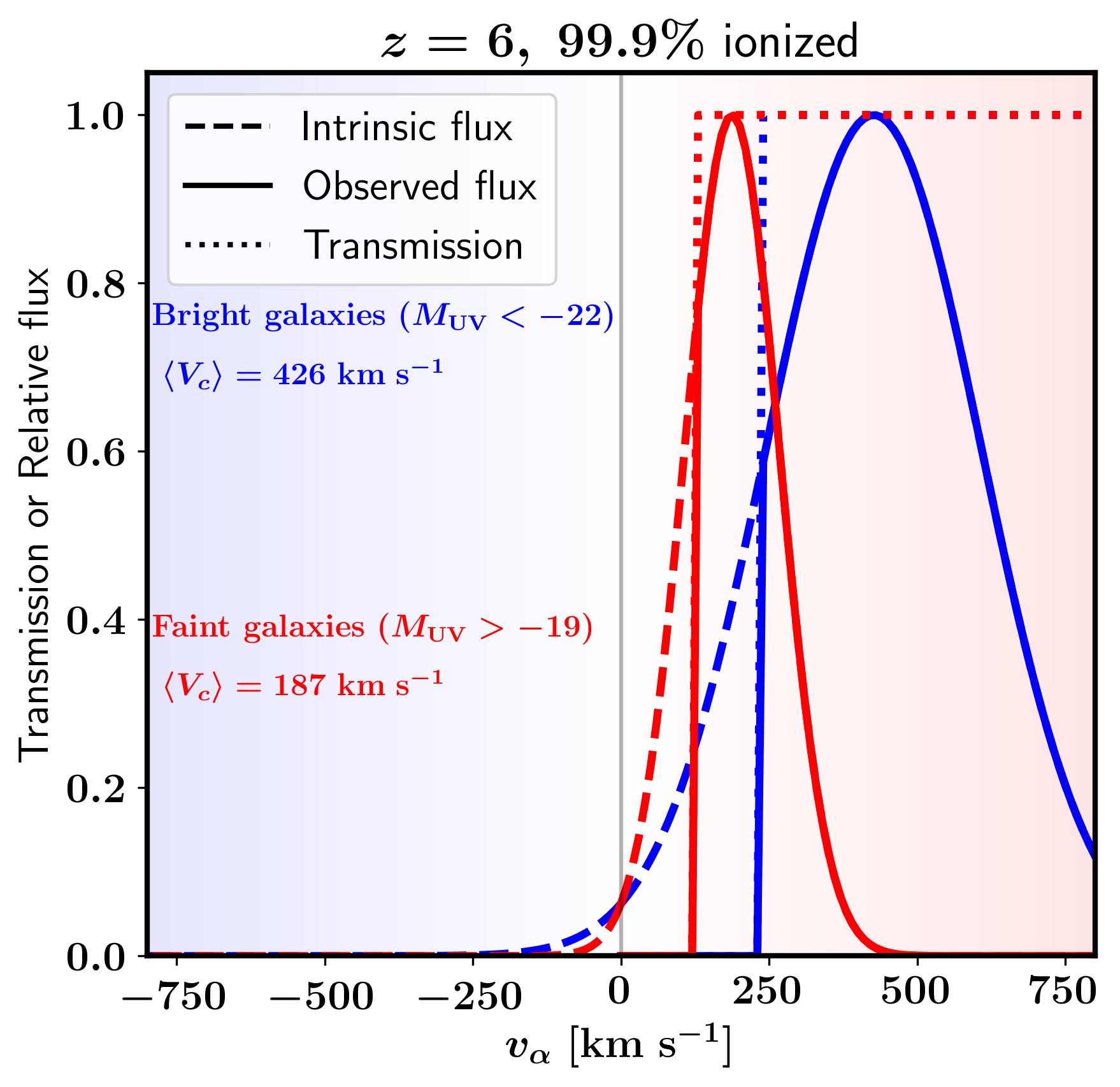}
\includegraphics[scale=0.4]{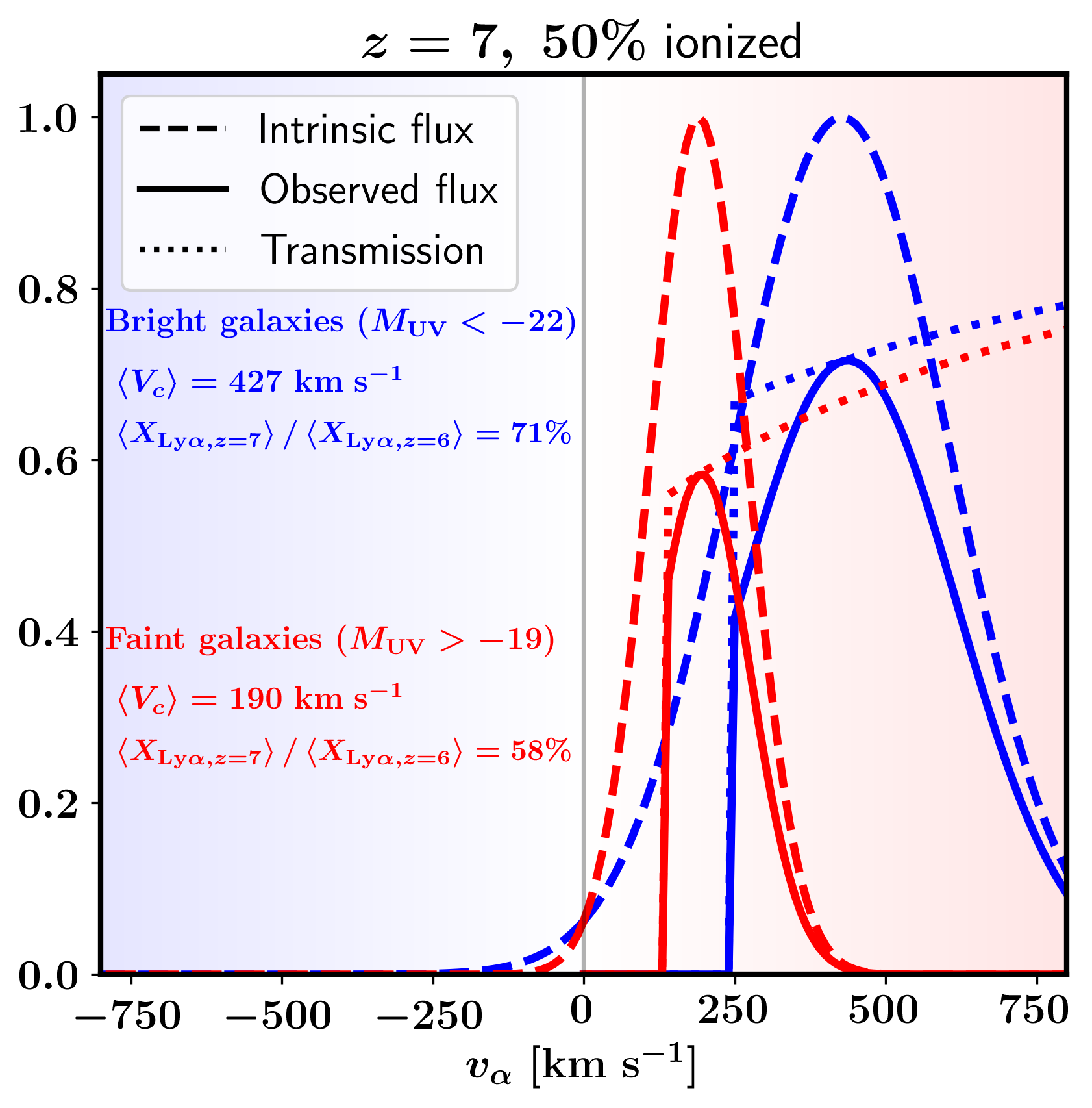}
\includegraphics[scale=0.4]{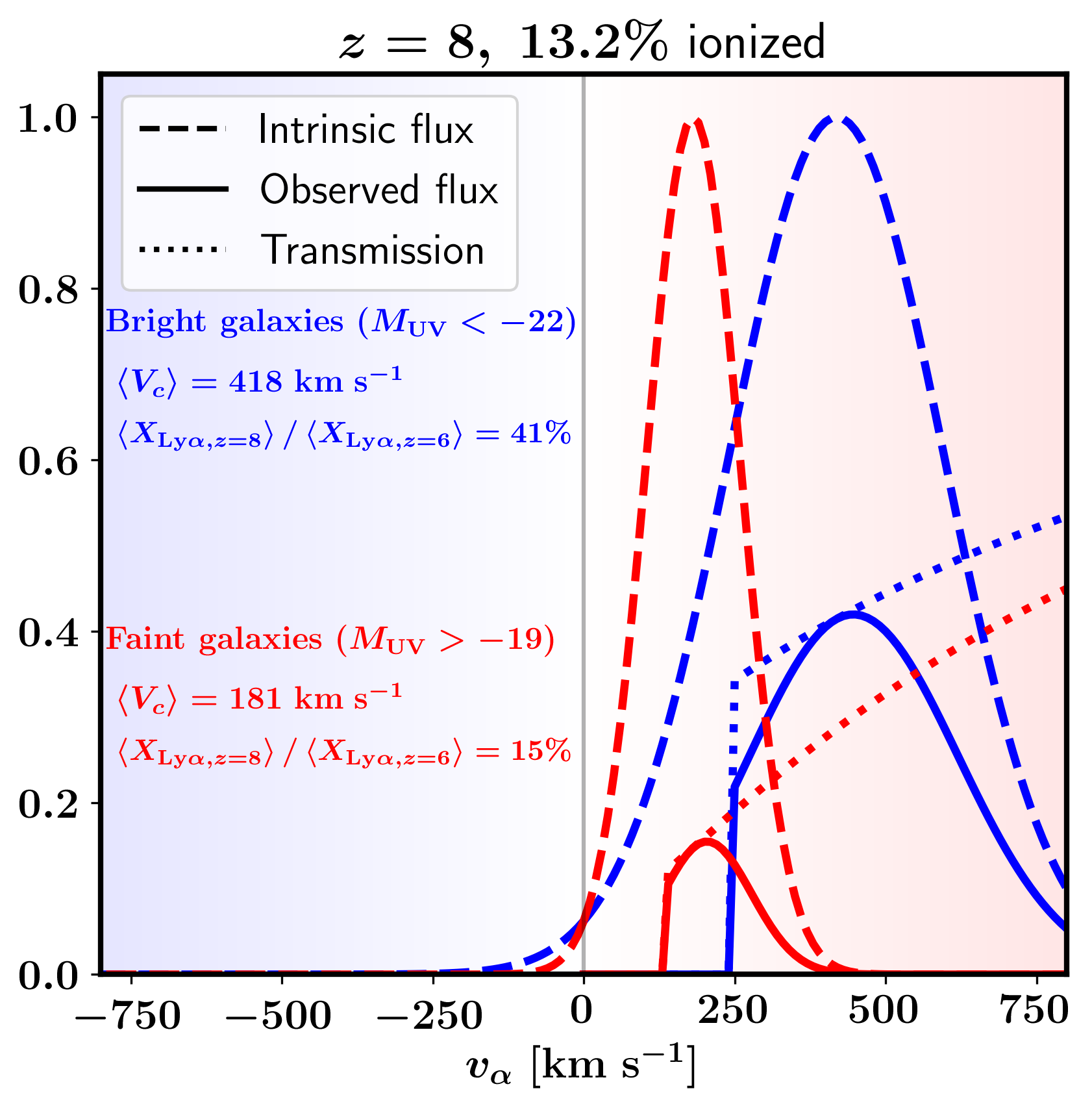}
\caption{Same as Figure~\ref{fig:typical_Tr} except that we increase $V_c$ by 50\% in Eq.~\ref{eq:intrinsic} to inflate the intrinsic emission profile.} 
\label{fig:typical_Tr3}
\end{center}
\end{figure*}

We test how sensitive our results are to our choice of the intrinsic emission profile (Eq.~\ref{eq:intrinsic}) by varying the profile in our calculation. We apply the median transmission to bright and faint galaxies in the same way as in Figure~\ref{fig:typical_Tr} in Section~\ref{sec:Muv-dependence} except that we shrink the profile by reducing the circular velocity term $V_c$ by 33\% in one case and inflating the profile by increasing $V_c$ by 50\% in the other case. The median infall velocity ($v_{\rm in}$) and effective bubble size ($R_b$) are kept same as in Figure~\ref{fig:typical_Tr}. The results for shrunk and inflated profiles are shown in Figure~\ref{fig:typical_Tr2} and \ref{fig:typical_Tr3}, respectively. 

The flux-weighted transmission, $\left<X_{{\rm Ly}\alpha}\right>$, with respect to $z=6$ is 69\% (37\%) and 57\% (11\%) for bright and faint galaxies at $z=7$ $(8)$, respectively, in the fiducial case. These fractions decrease to 58(27) and 38(5)\% at $z=7(8)$, respectively, for the shrunken intrinsic profile and increase to 71\% (41\%) and 58\% (15\%)  at $z=7$ $(8)$, respectively, for the inflated profile. Wider emission profiles result in the emission transmitted at longer wavelengths on average, which leads to higher flux-weighted transmission because the transmission increases toward longer wavelengths for a given $R_b$. We also note that the transmission from the faint galaxies is more sensitive to the profile shape, increasing three times from the shrunken to the inflated profile at $z=8$ while that from the bright galaxies increases 1.5 times. Despite these quantitative changes in the relative transmissivity values, the general trend of fainter galaxies being more suppressed by the damping-wing opacity remains throughout our experiment.

\bibliographystyle{apj}
\bibliography{reference}


\end{document}